%% file: HAQ_arXiv.tex
\documentclass[11pt,preprint,numberedappendix]{aastex}

\newcommand{\Av}{A(V)}

\newcommand{\noprint}[1]{}

\usepackage{graphicx,amsmath,amssymb,epsfig,psfig,arydshln}
\usepackage[T1]{fontenc}
\usepackage{multirow}
\usepackage{color}

\newcommand{\total}{$159$}

\shorttitle{The HAQ Survey: red QSOs from optical and NIR photometry - II}
\shortauthors{Krogager et al.}

\begin{document}


\title{The High $\mathrm{A_V}$ Quasar Survey: Reddened Quasi Stellar Objects selected from optical/near-infrared photometry - II}

\author{
J.-K. Krogager\altaffilmark{1,2},
S. Geier\altaffilmark{3,4},
J. P. U. Fynbo\altaffilmark{1},
B. P. Venemans\altaffilmark{5},
C. Ledoux\altaffilmark{2},
P. M\o ller\altaffilmark{6},\\
P. Noterdaeme\altaffilmark{7},
M. Vestergaard\altaffilmark{1,8},
T. Kangas\altaffilmark{4,9},
T. Pursimo \altaffilmark{4},
F. G. Saturni\altaffilmark{9,10},
O. Smirnova\altaffilmark{4,11}
}

\altaffiltext{1}{{\tiny Dark Cosmology Centre, Niels Bohr Institute, University of Copenhagen, Juliane Maries Vej 30, DK-2100 Copenhagen \O, Denmark}}
\altaffiltext{2}{{\tiny European Southern Observatory, Alonso de C\'ordova 3107, Vitacura, Casilla 19001, Santiago 19, Chile}}
\altaffiltext{3}{{\tiny Instituto de Astrof\'isica de Canarias (IAC), E-38205 La Laguna, Tenerife, Spain}}
\altaffiltext{4}{{\tiny Nordic Optical Telescope, Apartado 474, E-38700 Santa Cruz de La Palma, Spain}}
\altaffiltext{5}{\tiny Max-Planck Institute for Astronomy, K{\"o}nigstuhl 17, D-69117 Heidelberg, Germany}
\altaffiltext{6}{\tiny European Southern Observatory, Karl-Schwarzschildstrasse 2, D-85748 Garching bei M\"unchen, Germany}
\altaffiltext{7}{\tiny Institut d'Astrophysique de Paris, CNRS-UPMC, UMR7095, 98bis bd Arago, F-75014 Paris, France}
\altaffiltext{8}{\tiny Steward Observatory and Department of Astronomy, University of Arizona, 933 North Cherry Avenue, Tucson, AZ 85721, USA}
\altaffiltext{9}{\tiny Tuorla Observatory, Department of Physics and Astronomy, University of Turku, V\"ais\"al\"antie 20, 21500 Piikki\"o, Finland}
\altaffiltext{10}{\tiny University of Rome "La Sapienza," p.le A. Moro 5, I-00185 Rome, Italy}
\altaffiltext{11}{\tiny Institute of Astronomy, University of Latvia, Raina bulv. 19, Riga, 1586,Latvia}

\begin{abstract}
Quasi-stellar objects (QSOs) whose spectral energy distributions (SEDs) are reddened by dust either in their host galaxies or in intervening absorber galaxies are to a large degree missed by optical color
selection criteria like the one used by the Sloan Digital Sky Survey (SDSS). To overcome this bias against red QSOs, we employ a combined optical and near-infrared color selection.
In this paper, we present a spectroscopic follow-up campaign of a sample of red
candidate QSOs which were selected from the SDSS and the UKIRT Infrared Deep Sky Survey (UKIDSS).
The spectroscopic data and SDSS/UKIDSS photometry are
supplemented by mid-infrared photometry from the Wide-field Infrared Survey Explorer. In
our sample of 159 candidates, 154 (97\%) are confirmed to be QSOs.
We use a statistical algorithm to identify sightlines with plausible intervening absorption systems and identify nine such cases assuming dust in the absorber similar to Large Magellanic Cloud sightlines.
We find absorption systems toward 30 QSOs, 2 of which are consistent with the best-fit absorber redshift from the statistical modeling.
Furthermore, we observe a broad range in SED properties of the QSOs as probed by the rest-frame 2~$\mu$m flux. We find QSOs with a strong excess as well as QSOs with a large deficit at rest-frame 2~$\mu$m relative to a QSO template. Potential solutions to these discrepancies are discussed. Overall, our study demonstrates the
high efficiency of the optical/near-infrared selection of red QSOs.
\end{abstract}

\keywords{galaxies: active --- quasars: general --- quasars: absorption lines}

%------------------------Intro---------------------------------------
\section{Introduction}
Quasi-stellar objects (QSOs) are enigmatic objects in the Universe and due to their very high
intrinsic luminosities they can be seen out to very large cosmological
distances. Although great breakthroughs have developed since the first
detection of quasi-stellar radio sources \citep[][since then the term {\it
QSO} or {\it quasar} has gained prevalence]{Matthews1963}, many questions regarding their
physical nature remain unsolved. In order to draw robust conclusions concerning the
population of QSOs as a whole, it is important to have a representative
sample; however, most studies in the past have relied on color selections of
QSOs from large optical surveys, e.g., the Sloan Digital Sky Survey
\citep[SDSS,][]{York2000} and the 2dF QSO redshift survey \citep{Croom2004}. Although any
selection on color inherently biases the selected sample, the power of color
selection lies in the ability to quickly build a large sample without investing
large amounts of time on spectroscopic classification.

QSOs are thought to be transitional phenomena
originating in the environments of super-massive black holes
in the cores of galaxies. The QSO activity is triggered by mechanisms that 
cause material to be accreted onto the central back holes, e.g., mergers 
or hydrodynamical instabilities within the galaxies. Such major events
are also believed to induce strong star formation activity
leading to the formation of large amounts of dust.
Some part of the population is thus expected to be found in
dust-rich environments leading to red optical colors. The search for red QSOs
has a long history \citep[e.g.,][]{Benn98, Warren00, Gregg02, Richards03,
Hopkins04, Polletta2006, Glikman07, Glikman2012, Glikman2013, Lacy2007, Maddox2008, Urrutia09, Banerji2012, Maddox2012}. The detection of red QSOs in most of these works relied on either radio or X-ray detections \citep[see][for an
extensive discussion]{Warren00}. Recently, large area surveys in the
near-infrared have made it possible to select QSOs based on near-infrared
photometry \citep{Warren07, Peth11, Maddox2012} and this is the approach we
have adopted in this work.

The red colors of QSOs may be caused by dust in intervening absorption systems
as well. If the intervening absorber is very dust-rich, then it may cause reddening
and dimming of the background QSO to the point where optical selection will
fail to identify the source as a QSO. A bias may therefore exist against
very dusty and hence very metal-rich absorption systems, leading to the underestimation of the
cosmic chemical abundance in absorption systems \citep{Pontzen2009, Khare2012}.
In \citet[][hereafter Paper I]{Fynbo2013a}, we investigated the population of
red QSOs missing in the SDSS DR7 sample, motivated by the discoveries of
intervening absorbers causing reddening of the background QSOs to the point
where these QSOs were close to dropping out of the color criteria invoked by
SDSS-I/II \citep{Noterdaeme09b, Noterdaeme10, Kaplan10, Fynbo11, Noterdaeme12, Wang2012}. Selecting candidate QSOs on basis of their near-infrared colors
showed that QSOs were indeed missing in the SDSS DR7 sample of QSOs, see Paper I.
Any bias in QSO samples affects both the study of the QSOs themselves, absorption features
in the QSOs (e.g., broad absorption line (BAL) QSOs, see Saturni et al. 2015), and the samples of
intervening absorption systems \citep{Richards03, Maddox2012}.

In this work, we present revised criteria compared to those utilized in Paper I to target more reddened QSOs. The color criteria allow us to select a pure (though not complete) sample
of reddened QSO candidates. Here we present and discuss our spectroscopic follow-up campaign
The High A(V) Quasar (HAQ) Survey.
In Section~\ref{selection}, we present our selection criteria. In
Sections~\ref{observations} and \ref{results}, we describe our observations
and analysis of the HAQ sample.
In Section~\ref{discussion}, we discuss the implications of our work. Throughout this work, we assume a flat $\Lambda$CDM cosmology with $\Omega_m=0.3$, $\Omega_{\Lambda}=0.7$, and $H_0=70~{\rm km~s^{-1}~Mpc^{-1}}$.

\section{Photometric Data and Selection}
\label{selection}
\subsection{Photometric data}
The selection of candidate red QSOs is based purely on optical and near-infrared photometry.
The photometry was selected from the overlap region between the SDSS data release 7 ($u$, $g$, $r$, $i$, and $z$ bands) and the UKIRT Infrared Deep Sky Survey ($Y$, $J$, $H$, and $K_s$).

In our analysis, we also include data from the Wide-field Infrared Survey Explorer (WISE)
providing photometry in four bands in the mid-infrared at 3.4, 4.6, 12, and 22 $\mu$m.
In case of non-detections, we quote the flux in the given band as a 2\,$\sigma$ upper limit.

\subsection{Selection Criteria}

\begin{figure}
  \plotone{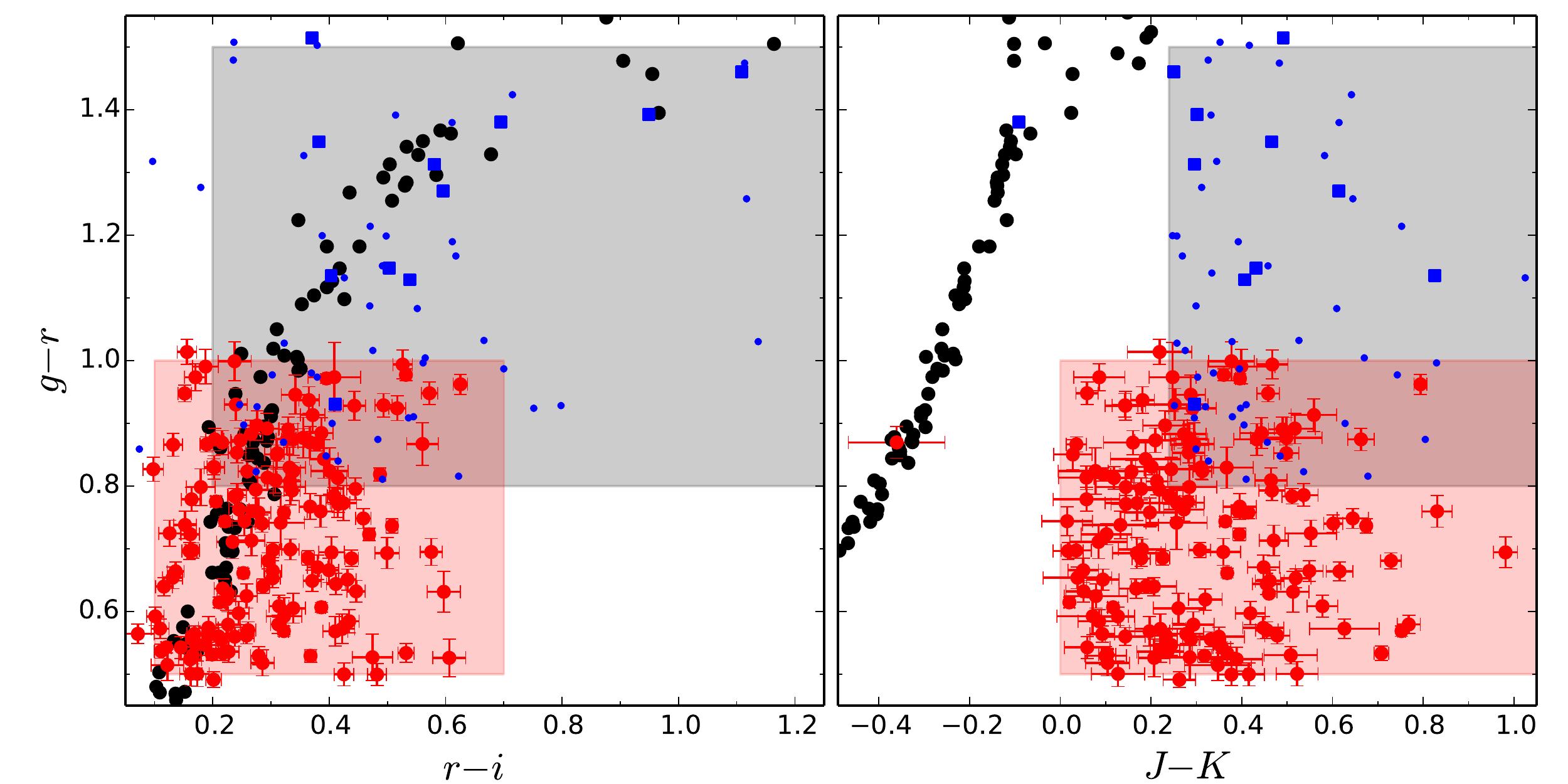}
  \caption{Comparison of the color criteria used in this work (red region) to those from Fynbo et al. 2013 (Paper I) (gray region).
  The left and right panels show $g-r$ versus $r-i$ and $g-r$ versus $J-K$, respectively. The big, red points with error-bars represent the sample that was observed in this work. Typical stellar colors are shown in black points. The blue points represent the sample from Paper I; QSOs are shown as small, blue points while contaminants are shown as big, blue squares. These contaminants (galaxies and stars) are primarily concentrated in the upper part of the color-color diagrams. This distribution is part of the motivation for the revised criteria resulting in a higher efficiency of QSO selection.
  \label{criteria}}		% <-- LABEL: criteria
\end{figure}

Our aim is to look for this population of red QSOs, which is missed in the
optical QSO samples, by using a set of color selection criteria that were
refined with respect to those of Paper I. By studying the distribution of
colors (see Fig.~\ref{criteria}) of the various identified targets in our
pilot study, we find that we can significantly reduce the small fraction of contaminating
galaxies and stars using these refined color criteria (all on the AB magnitude
system):

$J-K_s > 0$ ; $H-K_s > 0$ ; $J-H < 0.4$ ; $0.5 < g-r < 1.0$ ; $0.1 < r-i < 0.7$.\\

Moreover, the revised criteria also improve the selection of QSOs with
redshifts in the range $2.5<z<3.5$, which were missing in the sample presented in Paper I.
In total, we have selected 901 point sources common to the SDSS and UKIDSS
survey fields fulfilling these refined selection criteria down to a flux limit
of $J_\mathrm{AB}<19$. Of these, $\sim45\%$
had already been observed by SDSS (DR8) and found to be QSOs (either dust
reddened, BALs, or at $z\gtrsim3$). From the remaining 492 targets without
spectroscopy from SDSS, we selected our sample for spectroscopic follow-up.

In Fig.~\ref{criteria}, we compare the selection criteria from Paper I with the refined criteria described above. We only show the criteria for $g-r$, $r-i$, and $J-K$ as these are the colors that both sets of criteria have in common. As can be seen in Fig.~\ref{criteria}, the two selections overlap in these two color spaces. In total, 15 out of 58 objects from Paper I also fulfill all of the revised selection criteria. The targets from Paper I that were not spectroscopically classified as QSOs (i.e., stars and galaxies) are shown in Fig.~\ref{criteria} as large blue squares. Since these concentrate in a specific part of the color-color diagram shown in Fig.~\ref{criteria}, we can effectively remove them by excluding these regions of color space. For comparison, we indicate the colors of stars as black points. While the colors of stars coincide with the quasar colors in the optical, when we include the infrared colors the two populations separate more easily.

\section{Spectroscopic Observations and Data Reduction}
\label{observations}

During a range of observing runs in 2012, 2013, and 2014, {\total} candidate red QSOs were observed with the
Nordic Optical Telescope (NOT) on La Palma, using the Andalucia Faint
Object Spectrograph and Camera (ALFOSC). As in Paper I, we used grism \#4, which
covers the wavelength range from about 3200 \AA \ to 9100 \AA \ at a resolution
of about 300 with a slit width of 1.3 arcsec. Redwards of about 7500 \AA \ the spectra are strongly affected by
fringing, which was alleviated by dithering along the slit. In order to prevent 2$^{\rm nd}$ order contamination from wavelengths shorter than 3500~{\AA}, a blocking filter was used for the observations with grism \#4. We used filter no. \#94 which blocks out wavelengths shorter than $3560$~{\AA}. Given the red nature of the targets (by selection) and the reduced intensity of the second order spectrum the contamination from the remaining flux which is not blocked by the filter (from $3560-4500$~\AA) will be low. The overlap from the second order starts at $7120$~{\AA}, which is furthermore the spectral range affected by strong fringing. Moreover, we have the overall spectral shape from the broad-band photometry to make sure that we are not affected by significant second order contamination. Hence we consider second order contamination to be negligible.
The spectra were taken aligning the slit at the parallactic angle.
In some cases of bad seeing, we observed with a 1.8 arcsec slit.
Four sources (HAQ0047+0826, HAQ0151+1453, HAQ1115+0333, and HAQ2300+0914) were also observed with grism \#6 which covers the wavelength range from about 3200 \AA \ to 5500 \AA \ at a resolution of about 500 with the 1.0 arcsec slit. Two targets (HAQ1115+0333 and HAQ2225+0527) were observed with grism \#7 to look for intervening absorption. Grism \#7 covers wavelengths from about 3850 {\AA} to 6850 {\AA} at a resolution of about 650 with the 1.0 arcsec slit.
We binned the CCD pixels by a factor of 2 along the wavelength axis.

After the release of SDSS-DR9 in August 2012 we noted that 11 of the candidates, which we followed up, had been observed by SDSS. With the new release of DR10, the number of candidates with spectra from SDSS is now 25.
For targets that were observed by the Baryon Oscillation Spectroscopic Survey (BOSS), we present the spectra from the NOT along with BOSS spectra. However, we preferentially use the BOSS spectra in our analysis if available, since they have larger wavelength coverage, higher signal to noise and better resolution ($R\sim2000$) \citep{BOSS}.
In Table~\ref{sample}, we give a full list of all the observed targets.

The spectra were reduced using a combination of IRAF\footnote{IRAF is distributed by the
National Optical Astronomy Observatory, which is operated by the Association of
Universities for Research in Astronomy (AURA) under cooperative agreement with
the National Science Foundation.} and MIDAS\footnote{ESO-MIDAS is a copyright protected software product of the European Southern Observatory. The software is available under the GNU General Public License.} tasks for low-resolution spectroscopy.
Cosmic rays were rejected using the software written by \citet{vanDokkum01}.
In case of photometric observing conditions the spectrophotometric standard star observed on the same night as the science spectra was used for the flux calibration. Otherwise we used a standard response curve to calibrate the spectra. The spectra and photometry were corrected for Galactic extinction using the extinction maps from \citet{Schlegel98}. In order to improve the absolute flux-calibration, we scaled the spectra to the $r$-band photometry from SDSS.

%--------------------------TABLE----------------------------------------
\input{./redQSOs_table1}
%LABEL: sample
%-------------------------------------------------------------------------

\section{Results}
\label{results}

\begin{figure}[h]
\plotone{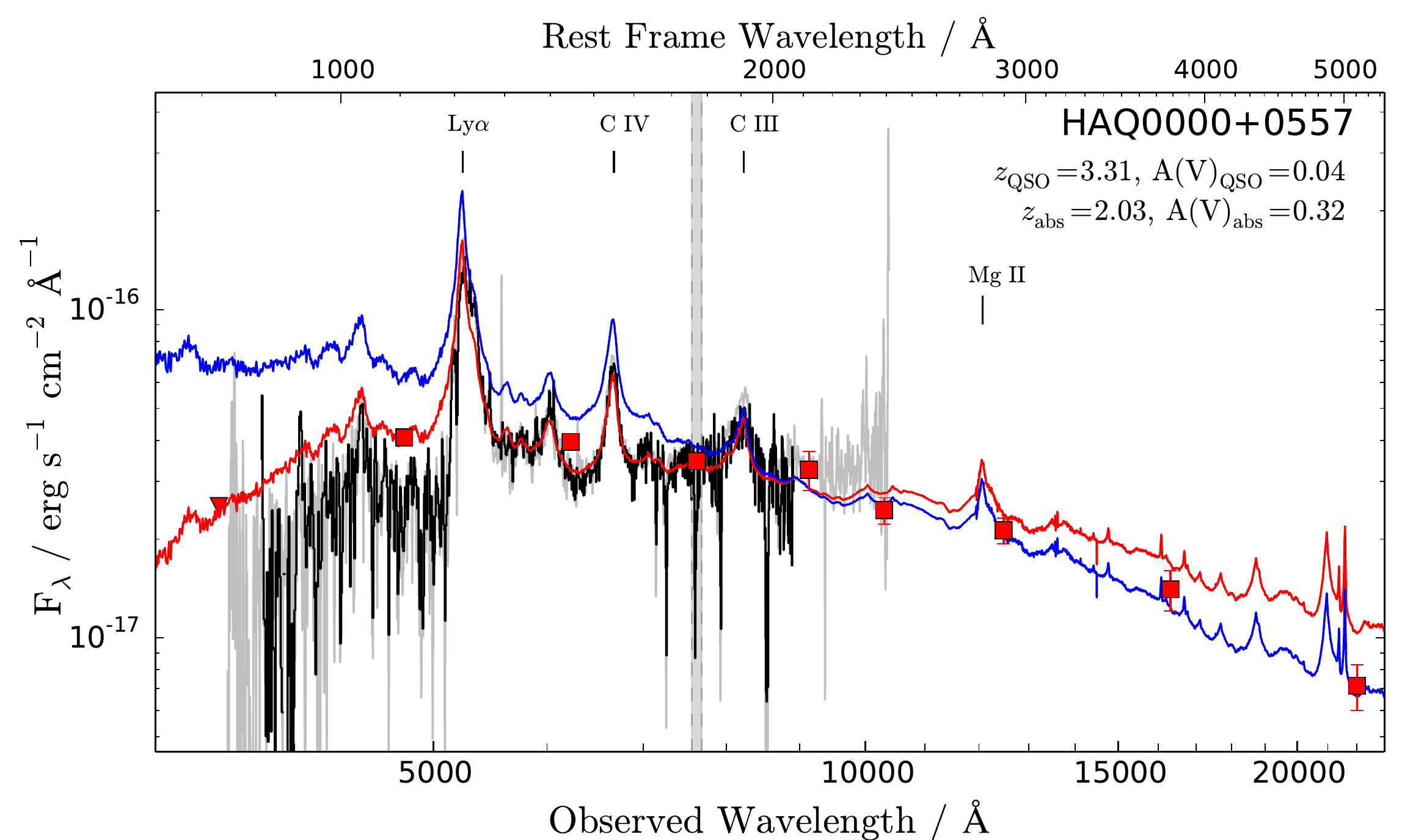}
\caption{The observed spectrum is plotted as a solid black line. For targets observed with both grism 4 and 6, we show both the grism 4 and grism 6 spectra. In case of available SDSS data the spectrum from SDSS is shown in gray.
In the upper right corner, the estimated emission redshift and rest-frame $V$-band extinction are provided.
The unreddened composite QSO spectrum is shown in blue, redshifted to the spectroscopic redshift, and in red we show the redshifted composite spectrum reddened by the indicated amount of extinction. Overplotted with filled squares are the SDSS and UKIDSS photometric data points.
The NOT spectra have been scaled to match the $r$-band photometric data point from SDSS. Unless
otherwise noted we have assumed an SMC-like extinction curve. Note that the
spectra have not been corrected for telluric absorption (marked with a gray band at $\sim7600$~\AA).
(The full set of figures is available on the survey webpage http://www.dark-cosmology.dk/$\sim$krogager/redQSOs/data.html)}
\label{spectra}
\end{figure}

In Fig.~\ref{spectra}, we show the 1-dimensional spectra of all targets in our sample along with the photometry from SDSS and UKIDSS. We are able to securely identify 154 out of {\total} targets as QSOs.
We have a tentative identification of the source HAQ2247+0146 which is probably a highly absorbed BAL QSO at redshift $z\approx 2.1$. Of the remaining four objects one is unidentified, one is most probably a BL Lacertae object \citep{Stein1976}, and two objects are identified as a stars. We thus have a purity (or efficiency) of $P\sim97$\%, defined as the ratio of QSOs to total objects in the sample. One of the stars in our sample (a M dwarf) was observed by mistake\footnote{The correct target ID for this star in SDSS is: J225430.23+063833.8} and is the result of a wrong object in the field being put on the slit. The real candidate for HAQ2254+0638 has therefore not been observed. The other star (a G dwarf) has entered our selection due to an error in the $K_s$-band photometry for this candidate. In the right panel of Fig.~\ref{criteria}, the single outlying red point (overlapping with the stellar sources) corresponds to this G dwarf and reflects the erroneous photometry in the $K_s$-band. The star is located at high galactic latitude toward the Galactic center with high foreground extinction ($A(V)=0.52$) due to the so-called North Galactic Spur. In Table~\ref{followuptab}, we present the identification of all objects, their radio flux from FIRST, and an estimate of the extinction assuming Small Magellanic Cloud (SMC) type dust at the redshift of the QSO unless specified otherwise. Only three QSOs in our sample were flagged as QSO candidates by SDSS. These are marked with the SDSS-flag `QSO\_HIZ' or `QSO\_CAP' in Table~\ref{followuptab}. This low number of QSO identifications in SDSS is caused by the overlap of the sources with the stellar locus in color-color space. Hence SDSS flags these sources as stars and not QSOs. By expanding the color-space with the near-infrared photometry we are able to break the degeneracy and separate stars from QSOs. More of our targets were identified by the BOSS survey, since BOSS relies on many other QSO selection algorithms than just color criteria \citep[][and references therein]{Ross2012}. In the notes of Table~\ref{followuptab}, we give the redshift inferred by BOSS for the cases with BOSS spectra, along with remarks on available spectra observed with other grisms, e.g., Grism \#6 or \#7. We note, that the targets marked as BAL QSO do not follow a stringent classification of BAL QSOs; however, this is to indicate that the QSO has intrinsic absorption. For a more detailed analysis and classification, see Saturni et al. (2014).

\subsection{Confirmed QSOs}
\label{quasars}
\begin{figure}
  \plotone{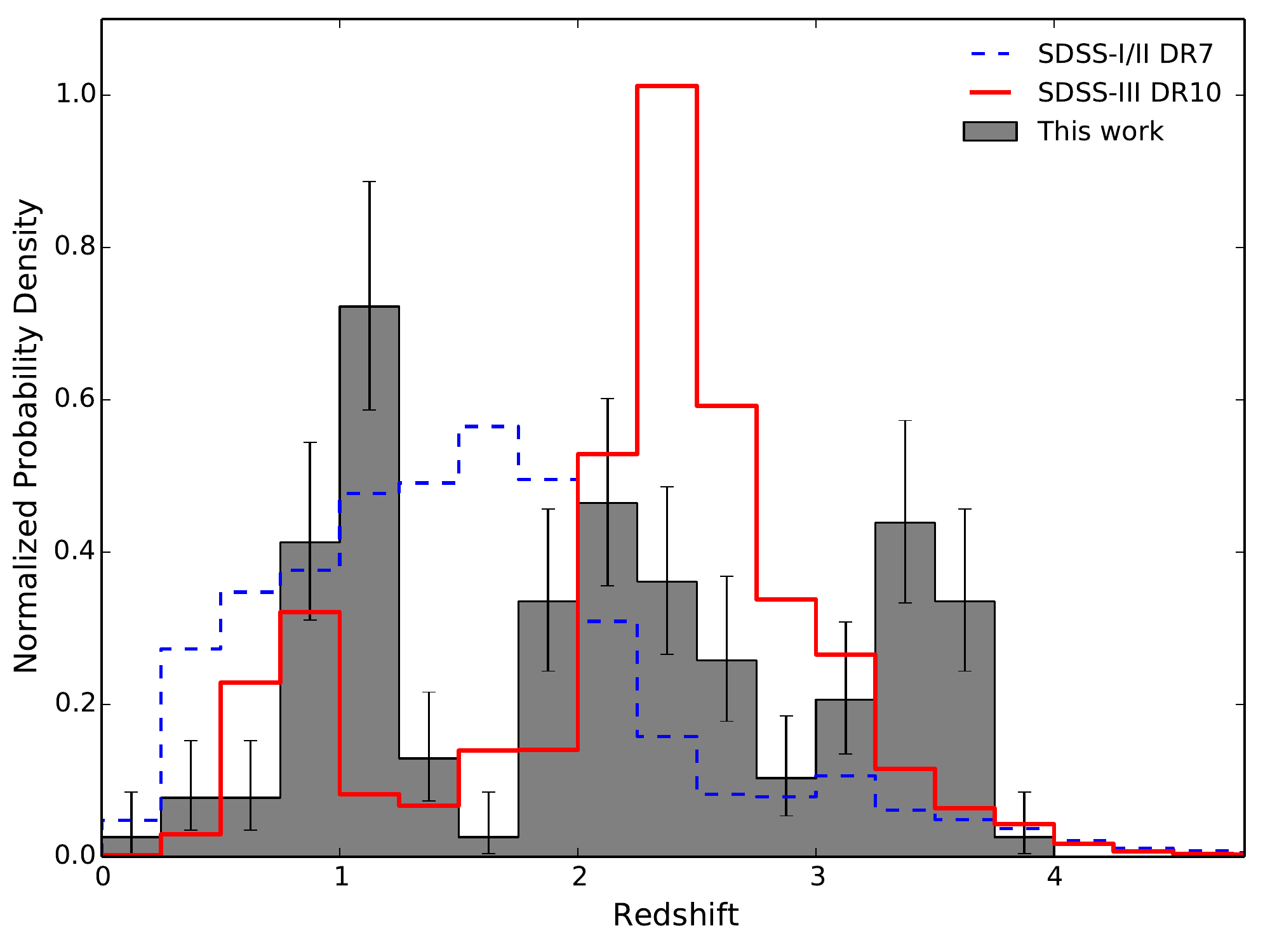}
  \caption{Probability density distribution of QSO redshifts for the HAQ sample (gray filled histogram) compared to the distributions of SDSS-I/II data release 7 (blue dashed histogram) and SDSS-III/BOSS data release 10 (red line histogram). The errorbars on the gray histogram indicate the 1$\sigma$ Poisson errors for each bin. The errors for the SDSS data are negligible due to the much larger sample sizes ($\sim1000$ times larger).
  \label{redshifts}}		% <-- LABEL: criteria
\end{figure}

As in Paper I, the redshifts have been determined by the emission lines visible in the spectra. In Fig.~\ref{redshifts}, we show the redshift distribution of the HAQ sample. The redshift distribution shows a peak around $z_{\rm QSO}\sim1$, and is more even for redshifts $z_{\rm QSO}>2$, however, not completely uniform. For comparison, we show the distributions of QSO redshifts from the SDSS-I/II (DR7) and SDSS-III/BOSS (DR10). The distribution of the SDSS-I/II sample peaks around $z_{\rm QSO}\sim1.5$ and quickly drops off for higher redshifts where the QSO colors begin to resemble stellar colors. The more complex selection algorithms utilized for the SDSS-III/BOSS were designed to increase the number of QSOs with $z_{\rm QSO}>2.15$. This explains the strong peak in the SDSS-III/BOSS redshift distribution.

We estimate the extinction of each QSO by fitting a QSO template to the photometry and spectrum combined. Spectral regions that are influenced by strong emission and absorption lines are masked out in the fit. We use, as in Paper I, a combined template from \citet{vandenBerk01} and \citet{Glikman06}. We disregard photometric points on the blue side of the Ly-$\alpha$ emission line, upper limits, and data points influenced by strong absorption (especially in case of broad absorption lines). Furthermore, we always exclude the $r$-band from the fit since this band has been used to scale the spectra to match the photometry. In order to test whether we primarily observe signs of dust in the QSO itself or dust along the line-of-sight, we fit two sets of models: one model assuming that the dust is located in the QSO host galaxy (the null hypothesis) and another model allowing both dust in the QSO and in an intervening absorption system (the general model). Furthermore, we run each set of models for two different reddening laws, Small Magellanic Cloud (SMC) and Large Magellanic Cloud (LMC) as parametrized by \citet{Gordon03} with a modification for wavelengths greater than 4400~{\AA} \citep[see][]{FM2005}. The model we fit can be summarized as follows:

$$ F_{\rm obs} = C\,\cdot\,F_0\, \cdot \exp \left(-\frac{1}{2.5 \log(e)} \left[ A({\rm V}) \cdot k_{\rm QSO} + A({\rm V)_{abs}} \cdot k_{\rm abs} \right] \right) ,$$

\noindent where $F_0$ denotes the rest-frame QSO template before reddening is applied, $F_{\rm obs}$ refers to the reddened template, and {\Av} is the amount of extinction applied in the QSO's rest frame given the reddening law, $k_{\rm QSO}$. We assume that the dust in the QSO is SMC type \citep[e.g.,][]{Hopkins04}. Likewise, {\Av$_{\rm abs}$} denotes the amount of extinction applied in the putative absorber's rest-frame at $z_{\rm abs}$ given the reddening law for the absorber, $k_{\rm abs}$ (SMC or LMC). $C$ is an arbitrary scale factor, since we do not know the intrinsic flux of the QSO before reddening is applied. The null hypothesis can then be constructed by restricting the extinction and redshift for the absorber in the general model: A(V)$_{\rm abs} = z_{\rm abs} = 0$. This nested nature of the models allows us to use a likelihood ratio test to compare the two models.

For a given reddened template, $F_{\rm obs}$, we calculate synthetic fluxes for the template in each photometric band, weighted by the appropriate filter transmission curve. Next, we interpolate the template onto the wavelength grid of the observed spectrum thereby creating a "model spectrum", which can be directly compared to the observed spectrum. For the fitting, we re-bin the spectra by a factor of two to decrease the influence of noise. We calculate the residuals using both the spectroscopic and photometric data available for each given object.
The fit is then performed using $\chi^2$ minimization utilizing a Levenburg--Marquardt algorithm as implemented in the Python package {\tt lmfit}\footnote{Written by Matthew Newville. Full documentation available at: http://cars9.uchicago.edu/software/python/lmfit/}. For the null hypothesis, we keep the absorber parameters fixed (A(V)$_{\rm abs} = z_{\rm abs}=0$) and only fit the remaining two parameters: the extinction at the QSO redshift, A(V), and the scale factor, $C$. The parameter A(V) is restricted to values larger than A(V)$>-0.2$. The inclusion of negative values of A(V) takes into account any possible variation in the intrinsic slope of the QSO spectra. A negative value of A(V) occurs if the QSO slope is intrinsically steeper than the template and no or little dust is present in the QSO. The limit of A(V)$>-0.2$ is motivated by the spread in $g-r$ color distribution of QSOs from \citet[][see their fig.~8]{Richards2001}. The assumption that steeper QSO UV slopes can be approximated by including negative A(V) values relies on the fact that, to first order, the SMC extinction law itself is a power-law (valid to within an error of $\la5$\%).

For the general model, we allow $z_{\rm abs}$ and A(V)$_{\rm abs}$ to vary as well, however, the extinction at the absorber redshift is restricted to ${\rm A(V)_{abs}}>0$, since a negative extinction in the absorber is non-physical. Moreover, the absorber redshift is restricted to be $0<z_{\rm abs}<z_{\rm QSO}$.
We stress that when referring to the null hypothesis we quote the amount of extinction in the QSO rest-frame as just A(V); however, in order to avoid confusion when referring to the general model, we quote the amount of extinction in the QSO rest-frame and absorber rest-frame as A(V)$_{\rm QSO}$ and A(V)$_{\rm abs}$, respectively.

\subsubsection{Model Comparison}
\label{model_comparison}
In order to evaluate whether the general model with dust in an intervening system provides a significantly better fit we use a likelihood ratio test. This test quantifies how likely the observed improvement is, given the added free parameters. The logarithmic likelihood ratio is defined as $L=-2 \ln (\Lambda_0 / \Lambda_{\rm G})$, where $\Lambda_0$ and $\Lambda_{\rm G}$ are the likelihoods for the null hypothesis and the general model, respectively. This can be simplified in our case since we assume that our uncertainties are Gaussian. In this case, $\chi^2=-2 \ln(\Lambda)$ is a direct measure of the logarithmic likelihood. We thus get: $L=\chi^2_0-\chi^2_{\rm G}$, where $\chi^2_0$ and $\chi^2_{\rm G}$ are the $\chi^2$ of the best fit for the null model and the general model, respectively. The ratio, $L$, will be distributed approximately as a $\chi^2$ distribution with number of degrees of freedom, $\nu$, given by the change in degrees of freedom between the two models, which is equal to the change in number of parameters ($\Delta \nu$ = 2). For each set of models for a given data set, we can then calculate the chance probability of encountering the calculated change in $\chi^2$ given the addition of two extra free parameters. This is the so-called $p$-value. In order to reject the null hypothesis, i.e., preferring a model with dust in the intervening system, we use a 5$\sigma$ level significance, hence the $p$-value must be less than $p<5.7\cdot 10^{-7}$. For a $\chi^2$ distribution with $\nu = 2$ this corresponds to a threshold of $L>28.75$. The change in $\chi^2$ must therefore be larger than 28.75 in order to reject the null hypothesis. We use such a strict criterion since the reddening estimates are very degenerate when introducing a second reddening system. Moreover, we require that the fit with intervening absorption (the general model) provides a \emph{good} fit, since a model which to begin with provides a bad fit easily can improve significantly in terms of $\chi^2$ when adding two free parameters but still provide a bad fit. 

For a model to provide a good fit to the data, the distribution of normalized residuals\footnote{Also sometimes referred to as standardized residuals: $(\mu-x)/\sigma$, where $\mu$, $x$, and $\sigma$ denote, respectively, model, data, and uncertainty.} should be normally distributed, i.e., follow a Gaussian distribution of $\mu=0$ and $\sigma^2=1$. We subsequently use a Kolmogorov--Smirnov (KS) test to check the departure of the normalized residuals from normality. The one-sample KS test measures the difference between the cumulative distribution of the test sample and that of a normal distribution. One can then assign a probability ($P_{\rm KS}$) of encountering at least this difference given the hypothesis that the test sample is drawn from a normal distribution. Hence a high value of $P_{\rm KS}$ means that the test sample is consistent with being drawn from a normal distribution.
For our purpose, we require that the distribution of normalized residuals have $P_{\rm KS}>0.1$, i.e., at the 10\% confidence level, we have no probabilistic evidence against the hypothesis that the residuals are normally distributed.
Finally, we disregard fits where one or more parameters have reached the limit of the allowed range, thus not giving a fully converged fit.

\begin{figure}
\plotone{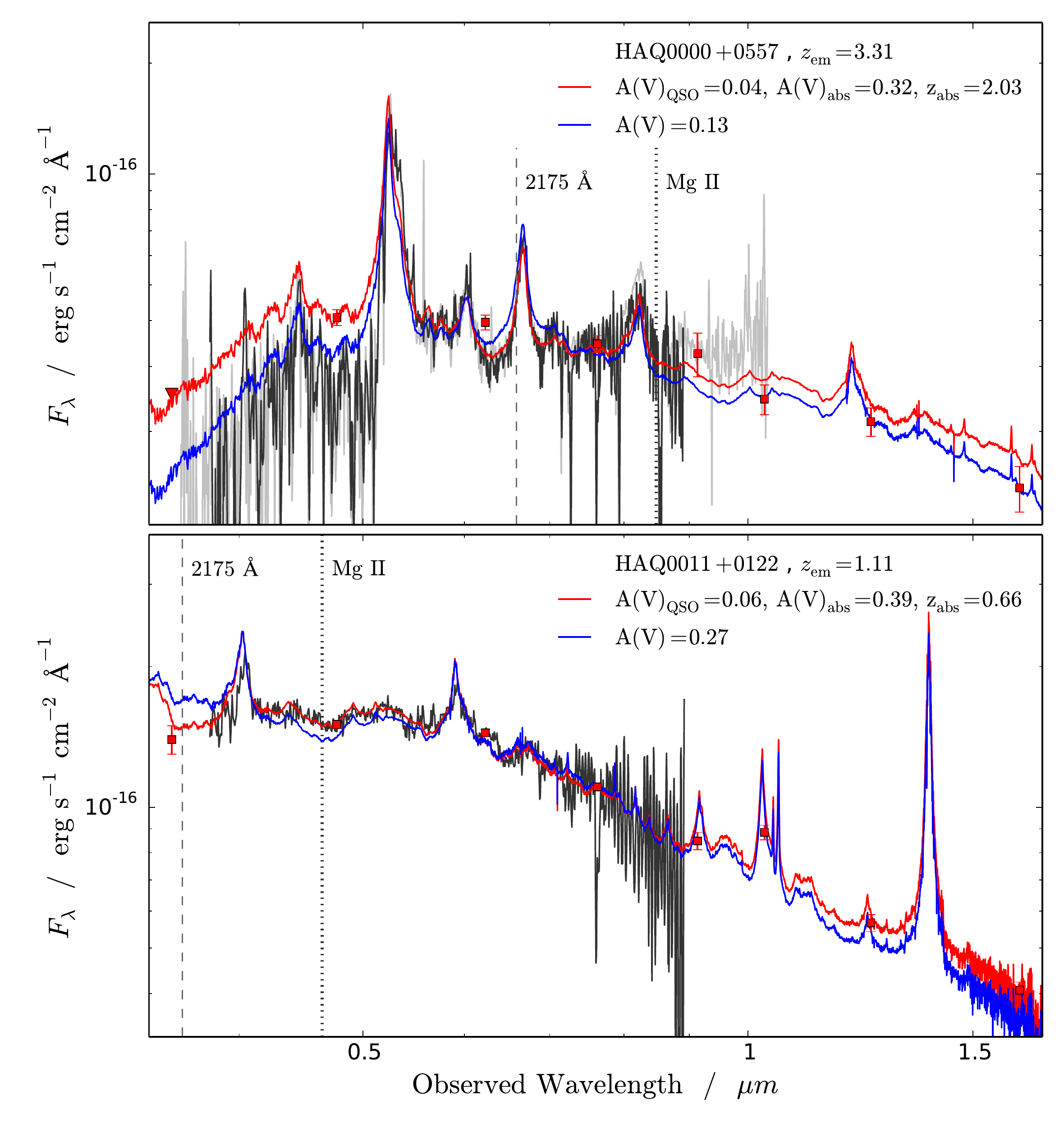}
\caption{Spectra and photometry for two of the QSOs with evidence for dust in an intervening absorption system. Each panel shows the NOT spectrum in black and the SDSS and UKIDSS photometry as red squares. The blue and red templates show the best-fit null model and general model, respectively (see text for definition). In the top panel, the underlying gray spectrum is the SDSS spectrum which has been smoothed with a three pixel Gaussian kernel for presentation purposes. The dashed and dotted vertical lines indicate the locations of the 2175\,{\AA} bump and Mg\,{\sc ii} at the best-fit redshift for the absorber.}
\label{LMC_fits} 	% <-- LABEL: LMC_fits
\end{figure}

For the model fits assuming SMC type dust in the absorber, we find no evidence for improved fits when including intervening dusty absorbers. However, when assuming that the dust in the absorber is LMC type we find 9 QSOs for which the model with intervening dust is preferred. The best-fit parameters for these 9 QSOs are given in Table~\ref{table:LMC_dust}. In all other cases where the null hypothesis is preferred, we list the extinction, {\Av}, at the QSO redshift assuming SMC type dust in Table~\ref{followuptab}.
In Appendix~\ref{appendix:table}, we present a table with all the details for the cases with significant evidence for dust in an intervening system. The table shows the best-fit parameters to both the null model and the general model along with the resulting $\chi^2$ for each model. Furthermore, we give the $P_{\rm KS}$ value from the KS test to the normalized residuals for each model. Finally, we give the overall $p$-value, i.e., the chance probability of the observed improvement (in terms of $\chi^2$) given the extra free parameters.
The uncertainties on the fit parameters for the general model have been obtained by using a Markov chain Monte Carlo method in order to provide a robust evaluation of the confidence intervals. Details on the Monte Carlo simulations are given in Appendix~\ref{appendix:table}.
Two of these nine cases are shown in Fig.~\ref{LMC_fits} to demonstrate the improvement of the fit when including intervening dust. The remaining cases are shown in Appendix~\ref{appendix:intervening}.

%--------------------------TABLE-------------------------------------------
\input{./redQSOs_table2}
% LABEL: followuptab
%----------------------------------------------------------------------------

\begin{deluxetable}{lcccr}
\tablecaption{Dust in Intervening Absorbers
\label{table:LMC_dust}}
\tablewidth{0pt}
\tablehead{
\colhead{Target} & \colhead{$z_{\rm QSO}$}  &
\colhead{$z_{\rm abs}$} & \colhead{$A({\rm V})_{\rm abs}$} & \colhead{$A({\rm V})_{\rm QSO}$}\\
}
\startdata

HAQ0000+0557 & 3.31 & 2.03 $^{+0.03}_{-0.03}$ & 0.32 $^{+0.03}_{-0.03}$ & 0.04 $^{+0.01}_{-0.01}$ \\[3pt] 
HAQ0011+0122 & 1.11 & 0.66 $^{+0.02}_{-0.02}$ & 0.39 $^{+0.04}_{-0.05}$ & 0.06 $^{+0.02}_{-0.02}$ \\[3pt] 
HAQ0031+1328 & 1.02 & 0.79 $^{+0.01}_{-0.01}$ & 0.35 $^{+0.02}_{-0.02}$ & 0.08 $^{+0.01}_{-0.01}$ \\[3pt] 
HAQ0051+1542 & 1.90 & 1.24 $^{+0.03}_{-0.03}$ & 0.39 $^{+0.04}_{-0.03}$ & $-$0.15 $^{+0.02}_{-0.03}$ \\[3pt]
HAQ0339+0420 & 1.80 & 1.10 $^{+0.06}_{-0.07}$ & 0.44 $^{+0.07}_{-0.06}$ & 0.01 $^{+0.03}_{-0.04}$ \\[3pt] 
HAQ1248+2951 & 3.55 & 2.36 $^{+0.10}_{-0.09}$ & 0.20 $^{+0.02}_{-0.02}$ & $-$0.03 $^{+0.01}_{-0.01}$ \\[3pt] 
HAQ1509+1214 & 2.80 & 1.71 $^{+0.06}_{-0.04}$ & 0.45 $^{+0.13}_{-0.12}$ & $-$0.15 $^{+0.05}_{-0.06}$ \\[3pt] 
HAQ2203$-$0052 & 1.24 & 0.65 $^{+0.04}_{-0.03}$ & 0.73 $^{+0.07}_{-0.07}$ & $-$0.11 $^{+0.02}_{-0.03}$ \\[3pt]
HAQ2343+0615 & 2.16 & 1.48 $^{+0.03}_{-0.03}$ & 0.29 $^{+0.04}_{-0.04}$ & $-$0.05 $^{+0.03}_{-0.03}$ \\

\enddata
\tablecomments{The estimates of A(V)$_{\rm abs}$ are calculated assuming the LMC extinction curve by Gordon et al. (2003).}
\end{deluxetable}

\subsubsection{Robustness and precision of the likelihood ratio test}
\label{robustness}
When determining whether a given QSO spectrum was better fitted by the general model than the null model, we assigned a probability of observing the given increase in likelihood assuming that the null model is true. We rejected the null model when this probability was below 5\,$\sigma$. Hence, we might have rejected the null model in cases where this was indeed the true model, and vice versa. In order to quantify how robust our method is, we therefore generate a set of model QSOs with varying amounts of both intrinsic and intervening dust. We then quantify how many absorption systems are recovered using the likelihood ratio test at different signal-to-noise ratios, and how well the input parameters are estimated by the fit.
Each QSO model is described by four parameters: the QSO redshift ($z_{\rm QSO}$), the extinction at the QSO redshift (${\rm A(V)_{QSO}}$), the absorber redshift ($z_{\rm abs}$), and the extinction at the absorber redshift (${\rm A(V)_{abs}}$). We assume that the dust in the QSO is of SMC type in all cases and generate two sets of models with the dust in the absorber being, respectively, of SMC or LMC type. For each model, we then generate synthetic data sets mimicking the SDSS and UKIDSS photometric bands and the spectral coverage of the NOT observations. We generate 500 models of synthetic data sets for each of the absorber extinction curves (LMC or SMC) and for both high and low signal-to-noise ratios: in total 2000 models. 
High and low signal-to-noise ratios here refer to ${\rm SNR_{spec}=15,~SNR_{phot}=20}$ and
${\rm SNR_{spec}=5}$, ${\rm SNR_{phot}=10}$, respectively, where ${\rm SNR_{spec}}$ refers to the average SNR per pixel in the synthetic spectral data and ${\rm SNR_{phot}}$ refers to the SNR of each synthetic photometric band. For the spectral noise model, we further add a noise component mimicking the fringing in the red part of the CCD of the ALFOSC spectrograph. Specifically this is $\sim20$\,\% (peak-to-peak fringe level) for wavelengths greater than $8000$\,{\AA}.

We use the same fitting method as described in Sect.~\ref{quasars} to fit both the null model and the general model. Hereafter we apply the same criteria for rejecting the null model as were used in the analysis of our data. That way we quantify how well we were able to identify intervening systems and with what precision the parameters are recovered. For models with LMC extinction in the absorber, we recover 39.8\,\% of the intervening systems down to ${\rm A(V)_{abs}}>0.2$ with high signal-to-noise ratio. While we at low SNR only recover 15.0\,\% and only with ${\rm A(V)_{abs}}>0.4$. For models with SMC type extinction, we recover 18.2\,\% of the absorbers down to ${\rm A(V)_{abs}}\ga0.5$ in the high SNR data and 0.0\,\% in the low SNR data. 
The precision on the best-fit parameters for the recovered intervening systems are given in Table~\ref{table:model_precision} for each of the extinction curves (LMC or SMC) and for both high and low SNR. Here we show the 16th, 50th, and 84th percentiles of the residuals for the three fit parameters. In Appendix~\ref{appendix:simulations}, we present the details of the recovered intervening systems along with details about the initial parameter distributions and noise models.

\begin{deluxetable}{lcccccc}
\tablecaption{Precision of recovered parameters from the likelihood ratio test
\label{table:model_precision}}
\tablewidth{0pt}
\tablehead{
	&	\multicolumn{3}{c}{High SNR} & \multicolumn{3}{c}{Low SNR}  \\[3pt]
  & \colhead{$\Delta A({\rm V})_{\rm QSO}$} & \colhead{$\Delta A({\rm V})_{\rm abs}$} & \colhead{$\Delta z_{\rm abs}$} &
\colhead{$\Delta A({\rm V})_{\rm QSO}$} & \colhead{$\Delta A({\rm V})_{\rm abs}$} & \colhead{$\Delta z_{\rm abs}$}
}
\startdata
LMC & \phs$0.00^{+0.04}_{-0.04}$ & $0.01^{+0.08}_{-0.09}$ & $0.00^{+0.05}_{-0.03}$ &
	$-0.02^{+0.07}_{-0.07}$ & $0.03^{+0.14}_{-0.10}$ & $0.00^{+0.06}_{-0.05}$ \\[10pt]

SMC & $-0.08^{+0.12}_{-0.16}$ & $0.09^{+0.24}_{-0.15}$ & $0.04^{+0.17}_{-0.18}$ & \nodata & \nodata & \nodata \\
\enddata
\tablecomments{For each input extinction law, we show the residuals (input -- output) of the various model parameters.}
\end{deluxetable}

In order to estimate the robustness of the likelihood ratio test, we also perform a sanity check of the null hypothesis. For this we generate 2$\times$500 spectra with dust only at the QSO redshift at high and low SNR and fit these with both the null model and the general model. In {\it all} cases, we reject the general model; in other words, the preferred model is the model with dust only in the QSO itself. The agreement between input and best-fit parameter allows us to quantify the precision in the A(V) estimates using this method of fitting templates to the data. We observe no bias in the estimated parameters and find a statistical uncertainty of $\pm0.01$ on the A(V) estimates for high SNR, while we get $\pm0.02$ at low SNR. These are only statistical uncertainties; the actual error on a single measurement in {\it real} data may be larger due to the template not being a perfect description for every QSO.

\subsubsection{Intervening Absorption Systems}

Our next step is to go through all of the spectra manually to look for intervening absorption systems such as DLAs or Mg\,{\sc ii} systems that would provide a host for dust at lower redshift.
Due to the limited signal-to-noise ratio and low resolution we are not able to detect weak absorption systems. Also, the limited wavelength coverage means that we might miss absorption systems if the Ly$\alpha$ or Mg\,{\sc ii} lines fall outside the spectral range for the given absorption system.
We identify 27 QSOs with clearly visible absorption systems and three tentative detections (due to low
signal-to-noise ratio). The systems are mainly identified by the characteristic
Mg\,{\sc ii}\,$\lambda\lambda\,2796,2803$ doublet, one target is identified by the Ca\,{\sc ii}\,$\lambda\lambda\,3934,3969$, and eight systems are identified by Ly$\alpha$, C\,{\sc iv}, or both. The identifications are secured by detections of other absorption lines (such as Fe\,{\sc ii} lines) at the same redshift. All absorption systems
are listed in Table~\ref{tab:absorbers}. Out of the nine QSOs that have preferred solutions with intervening dust in the model comparison, two have Mg\,{\sc ii} systems at redshifts close to the best-fit absorption redshift, one target has inconsistent redshifts between the absorption system and the best-fit,
and one target (HAQ2343+0615) has slight indications of a Mg\,{\sc ii} absorption system at the best-fit absorber redshift $z_{\rm abs}=1.49$. However, the quality of the data does not allow a firm detection.
In the rest of the cases where the absorber dust model was preferred, we either do not have good enough SNR to securely detect any intervening systems (HAQ2203-0052 and HAQ2343+0615), the Mg\,{\sc ii} doublet is not covered by the spectral range (HAQ1248+2951 and HAQ1509+1214 where Mg\,{\sc ii} falls on top of the telluric absorption) or the algorithm picks up spectra that differ significantly from the template. Such cases could be caused by differences in the QSO continuum or variations in the intensity of weak and blended iron emission lines \citep{Pitman2000}. Nevertheless, many of these quasars selected by our analysis may still be reddened by foreground absorbers; we are just not able to confirm the existence of the absorbing galaxy with this modeling approach.

The target, HAQ0000+0557, shows a very strong Mg\,{\sc ii} (EW$_{\rm rest}\,(2796)=4.5$\AA) absorber at $z_{\,_{\rm Mg\,II}}=2.048$. This QSO is one of the QSOs for which the general model with dust in an intervening absorber was preferred, see Sect.~\ref{model_comparison}. The preferred redshift for the absorber from the fitting was $z_{\rm abs}=2.03\pm0.03$ in agreement with the Mg\,{\sc ii} system. This strongly suggests that the dust reddening (with A(V)$_{\rm abs}=0.32$) be caused by this Mg\,{\sc ii} system. Although the spectrum is well fit by the intervening dust model, the IR photometry is matched better by the model with only QSO dust; see top panel of Fig.~\ref{LMC_fits}. Further analysis of the extinction curve is needed in order to fit all the data. However, for this we would need a longer spectroscopic wavelength range observed simultaneously to rule out offsets in the photometry due to intrinsic QSO variability.\\
\indent
The spectrum of HAQ0339+0420 has a Mg\,{\sc ii} absorption system at a redshift of $z_{\,_{\rm Mg\,II}} = 1.21$ coinciding with the preferred redshift for an intervening system from the fit at $z_{\rm abs}=1.10$ with ${\rm A(V)_{abs}}=0.44$. Although the best-fit redshift is lower than the redshift for the Mg\,{\sc ii} absorber, the two are consistent when considering the $3\,\sigma$ confidence interval for the absorber redshift $0.92 < z_{\rm abs} < 1.24$.\\
\indent
The spectrum of HAQ1248+2951 shows a Mg\,{\sc ii} system at $z_{\,_{\rm Mg\,II}}=1.56$, however, the best-fit absorption redshift from the statistical modeling is $z_{\rm abs} = 2.36\pm0.10$. At this absorbing redshift, the Mg\,{\sc ii} absorption lines would fall outside the spectral coverage of our setup. Hence, the system at $z_{\,_{\rm Mg\,II}}=1.56$ may simply be an intervening system, which hosts no dust. However, with the current data it is not possible to constrain the dust properties further, since we do not have coverage of the full spectral range. Moreover, the addition of a second absorber in the fit will increase the degeneracy of the fit parameters, which is not possible to break with the data at hand.

The QSO, HAQ1115+0333, was observed with grisms \#6 and \#7 to provide higher-resolution spectroscopy of the absorber which was identified in the spectrum taken with grism \#4. The QSO shows a strong Mg\,{\sc ii} absorber at $z=1.18$ and a weaker (Ly$\alpha$, C{\small IV}) system at $z=2.57$. However, the spectrum of this system is consistent with the dust being at the QSO redshift $z_{\rm QSO}=3.10$.
The rest of the identified absorption systems in Table~\ref{tab:absorbers} are all either consistent with no dust or with the dust residing in the QSO system only.
However, one target (HAQ\,2225+0527) is the exception to the rule. This system has a Damped Ly$\alpha$ absorption system at redshift $z_{\rm DLA}=2.131$. The extinction toward the QSO is quite high (${\rm A(V)}=0.29$), and is consistent with the dust extinction being caused by SMC-type dust in the DLA when taking into account the added constraint on $z_{\rm abs}$. The reason why this system was not identified in our statistical analysis is due to the fact that SMC-type dust is extremely difficult to identify with this approach, since the featureless extinction curve does not provide any redshift-dependent features. Hence, the degeneracy between the various fit parameters is very large. In Sect.~\ref{robustness}, we discussed the limitations of the statistical approach and we found that for SMC-type dust only absorption systems with ${\rm A(V)_{abs}}\ga0.5$ are recovered. Given the value of A(V)=0.29 for HAQ2225+0527 we therefore do not expect to detect the intervening reddening with this particular modeling approach.

\begin{deluxetable}{lllcccl}
\tabletypesize{\small}
\tablecaption{Intervening absorption systems.
\label{tab:absorbers}}
\tablewidth{0pt}
\tablehead{
\colhead{Target} & \colhead{$z_{\rm QSO}$} & \colhead{$z_{\rm spec}$\tablenotemark{(a)}} & \colhead{$z_{\rm abs}$\tablenotemark{(b)}} & \colhead{${\rm A(V)_{abs}}$\tablenotemark{(b)}} & \colhead{${\rm A(V)_{QSO}}$} & \colhead{Notes}
}
\startdata

HAQ0000+0557  &  3.31  &  2.05  &  2.03$\pm$0.03  &  0.32$\pm$0.03  &  0.04$\pm$0.01  & strong Mg\,{\sc ii} absorber \\
HAQ0008+0835  &  1.19  &  0.85  &  \nodata  &  \nodata  &  0.00  & Mg\,{\sc ii} absorber \\
HAQ0008+0846  &  1.23  &  1.07  &  \nodata  &  \nodata  &  0.04  & Mg\,{\sc ii} absorber \\
HAQ0015+0736  &  3.63  &  2.47  &  \nodata  &  \nodata  &  $-$0.05  & strong Mg\,{\sc ii} absorber with associated Ly$\alpha$ \\
HAQ0015+1129  &  0.87  &  0.81  &  \nodata  &  \nodata  &  0.78  & Mg\,{\sc ii} absorber \\
HAQ0024+1037  &  1.22  &  0.29  &  \nodata  &  \nodata  &  0.27  & Ca\,{\sc ii} absorber \\
HAQ0044+0817  &  3.35  &  2.90  &  \nodata  &  \nodata  &  0.00  & Damped Ly-$\alpha$ absorber\\
HAQ0053+0216  &  0.99  &  0.90: &  \nodata  &  \nodata  &  0.71  & Mg\,{\sc ii} absorber, low SNR \\
HAQ0059+1238  &  3.50  &  3.21  &  \nodata  &  \nodata  &  0.00  & Ly$\alpha$ and C\,{\sc iv} absorber \\
HAQ0143+1509  &  3.76  &  1.24  &  \nodata  &  \nodata  &  0.00  & Mg\,{\sc ii} absorber \\
HAQ0211+1214  &  2.11  &  1.02  &  \nodata  &  \nodata  &  0.05  & Mg\,{\sc ii} absorber \\
HAQ0337+0539  &  3.28  &  1.20  &  \nodata  &  \nodata  &  0.03  & Mg\,{\sc ii} absorber \\
HAQ0339+0420  &  1.80  &  1.21  &  1.10$\pm$0.07  &  0.44$\pm$0.07  &  0.01$\pm$0.03  & Mg\,{\sc ii} absorber \\
HAQ0340+0408  &  1.62  &  1.03: &  \nodata  &  \nodata  &  0.13  & Mg\,{\sc ii} absorber, low SNR \\
HAQ0355$-$0053 & 2.15  &  1.24  &  \nodata  &  \nodata  &  0.00  & Mg\,{\sc ii} absorber \\
HAQ1115+0333  &  3.10  &  1.18  &  \nodata  &  \nodata  &  0.15  & Mg\,{\sc ii} absorber \\
HAQ1115+0333  &  3.10  &  2.57  &  \nodata  &  \nodata  &  0.15  & Weak Ly$\alpha$ and C\,{\sc iv} absorber \\
HAQ1248+2951  &  3.55  &  1.56  &  2.36$\pm$0.10  &  0.20$\pm$0.02  &  $-$0.03$\pm$0.01  & Mg\,{\sc ii} absorber \\
HAQ1327+3206  &  2.48  &  0.63  &  \nodata  &  \nodata  &  0.01  & Mg\,{\sc ii} absorber \\
HAQ1355+3407  &  3.16  &  1.08  &  \nodata  &  \nodata  &  0.00  & Mg\,{\sc ii} absorber \\
HAQ1534+0013  &  3.44  &  2.86  &  \nodata  &  \nodata  &  0.00  & Damped Ly$\alpha$ absorber \\
HAQ1633+2851  &  1.14  &  0.71: &  \nodata  &  \nodata  &  0.52  & Mg\,{\sc ii} absorber, low SNR \\
HAQ2159+0212  &  1.26  &  1.04  &  \nodata  &  \nodata  &  0.10  & Mg\,{\sc ii} absorber \\
HAQ2222+0604  &  2.45  &  1.29  &  \nodata  &  \nodata  &  0.25  & Mg\,{\sc ii} absorber \\
HAQ2225+0527  &  2.32  &  2.13  &  \nodata  &  \nodata  &  0.29  & Damped Ly$\alpha$ absorber \\
HAQ2231+0509  &  1.76  &  1.21  &  \nodata  &  \nodata  &  0.35  & Mg\,{\sc ii} absorber \\
HAQ2241+0818  &  2.43  &  2.32  &  \nodata  &  \nodata  &  0.04  & Damped Ly$\alpha$ absorber \\
HAQ2244+0335  &  3.37  &  2.89  &  \nodata  &  \nodata  &  0.00  & Damped Ly$\alpha$ absorber \\
HAQ2303+0238  &  2.22  &  1.47  &  \nodata  &  \nodata  &  0.11  & Mg\,{\sc ii} absorber \\
HAQ2305+0117  &  2.67  &  1.49  &  \nodata  &  \nodata  &  0.13  & Mg\,{\sc ii} absorber\\
HAQ2305+0117  &  2.67  &  2.67  &  \nodata  &  \nodata  &  0.13  & Damped Ly$\alpha$ absorber at QSO redshift\\
HAQ2311+1444  &  3.31  &  1.67  &  \nodata  &  \nodata  &  0.00  & Mg\,{\sc ii} absorber \\

\enddata
\tablenotetext{(a)}{$z_{\rm spec}$ here denotes the spectroscopic redshift of the absorption system. Redshifts followed by a colon (:) denote tentative detections due to low SNR.}
\tablenotetext{(b)}{Redshift and extinction values are only given in cases where the absorber model was preferred in the likelihood ratio test described in Sect.~\ref{model_comparison}. The values reported here assume LMC type dust at the absorber redshift from the fit.}
\end{deluxetable}

\subsubsection{Spectral Energy Distributions}
\label{SEDs}
\begin{figure}
\plotone{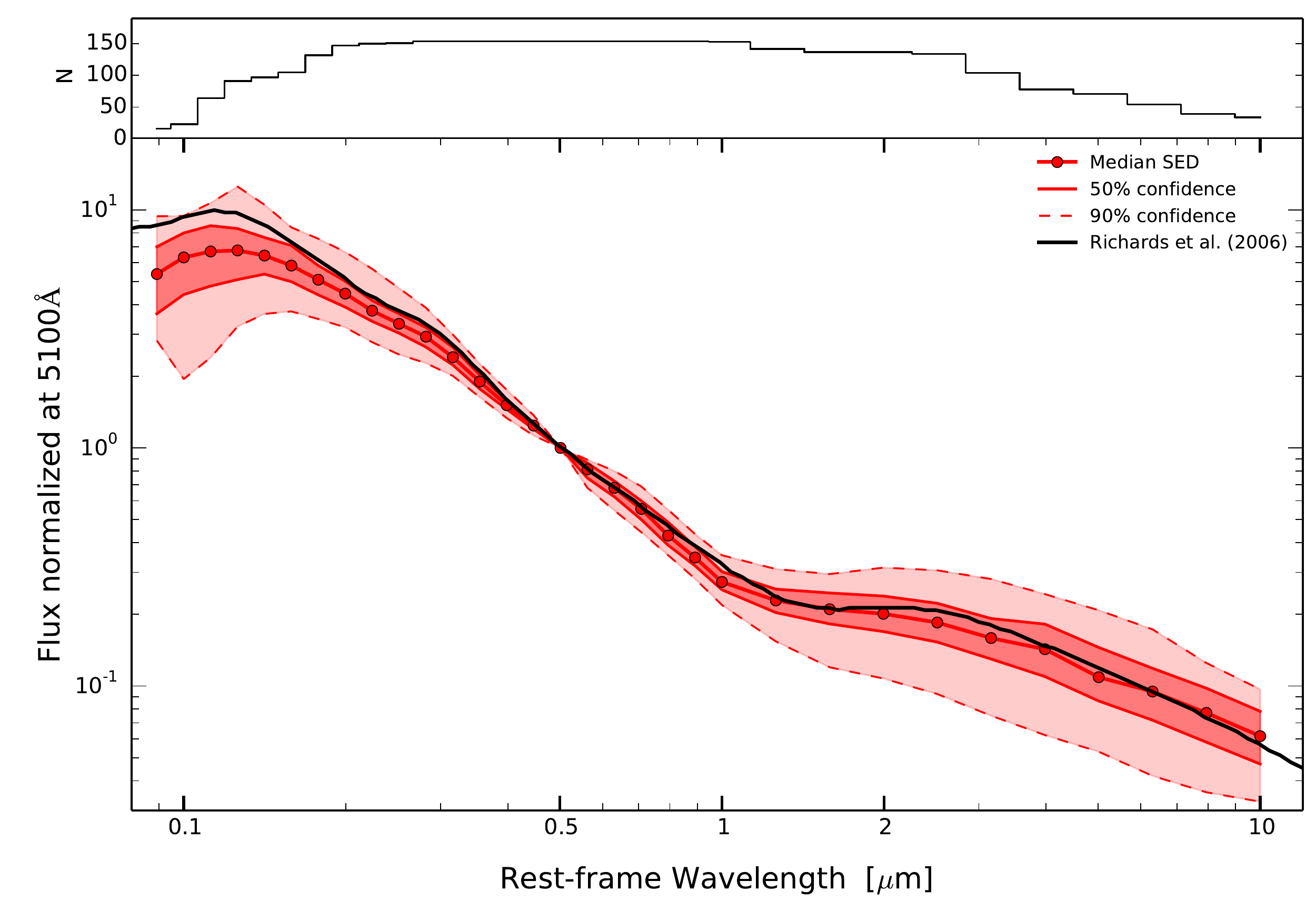}
\caption{Median rest-frame stacked spectral energy distribution (SED) for all QSOs in our sample (red points). The individual SEDs have been scaled at 5100 {\AA} before stacking. The dark and light red shaded areas show the 50\% and 90\% confidence intervals, respectively. The overplotted black line shows the continuum template of \citet{Richards06}. The top panel shows the number of individual SEDs that contribute to a given wavelength bin.} \label{fig:medianSED} 	% <-- LABEL: Avplot
\end{figure}

We investigate the average behavior of the QSOs in our sample by stacking their rest-frame, dust-corrected energy distributions. We correct for dust assuming the best-fit reddening obtained as described above.
Hereafter, we linearly interpolated all the rest-frame SEDs onto a common wavelength grid. We then normalize each SED at rest-frame wavelength 5100\,{\AA}. The average SED is computed using the median as well as the 50 and 90\,\% confidence intervals. The confidence intervals only take into account the variance in the sample because this, along with the interpolation, is the dominant source of uncertainty. For reference, the typical uncertainty on the photometric measurements is of the order of 5\,\%. In Fig.~\ref{fig:medianSED}, we show the median SED and its confidence intervals. We show the continuum model from \citet{Richards06} for comparison. In the top panel of the figure, we show the number of individual SEDs that contribute to a given wavelength bin.

In the blue part of the SED, for $\lambda<1500$\,{\AA}, we observe the largest discrepancy between our data and the template. This is, however, expected due to the low number of sources contributing to the stack in this range and due to the fact that we have not corrected for BAL QSOs in our sample, which significantly redden the overall SED blue-wards of C\,{\sc iv}. Apart from this discrepancy at short wavelengths, the sample is very homogenous for wavelengths less than rest-frame 1\,$\mu$m. At larger wavelengths, the sample exhibits a higher degree of variance. We therefore study the properties of the deviant SEDs at rest-frame 2\,$\mu$m, since at this wavelength the number of objects contributing to the stack is still high ($>80$\,\%). Above 3$\mu$m the large variance is dominated by the fact that only a small number of objects contribute to the stack.
We classify outliers as SEDs whose flux in the rest-frame 2\,$\mu$m data point differs by more than 5\,$\sigma$ from the template at the given wavelength.

\begin{figure}
\plotone{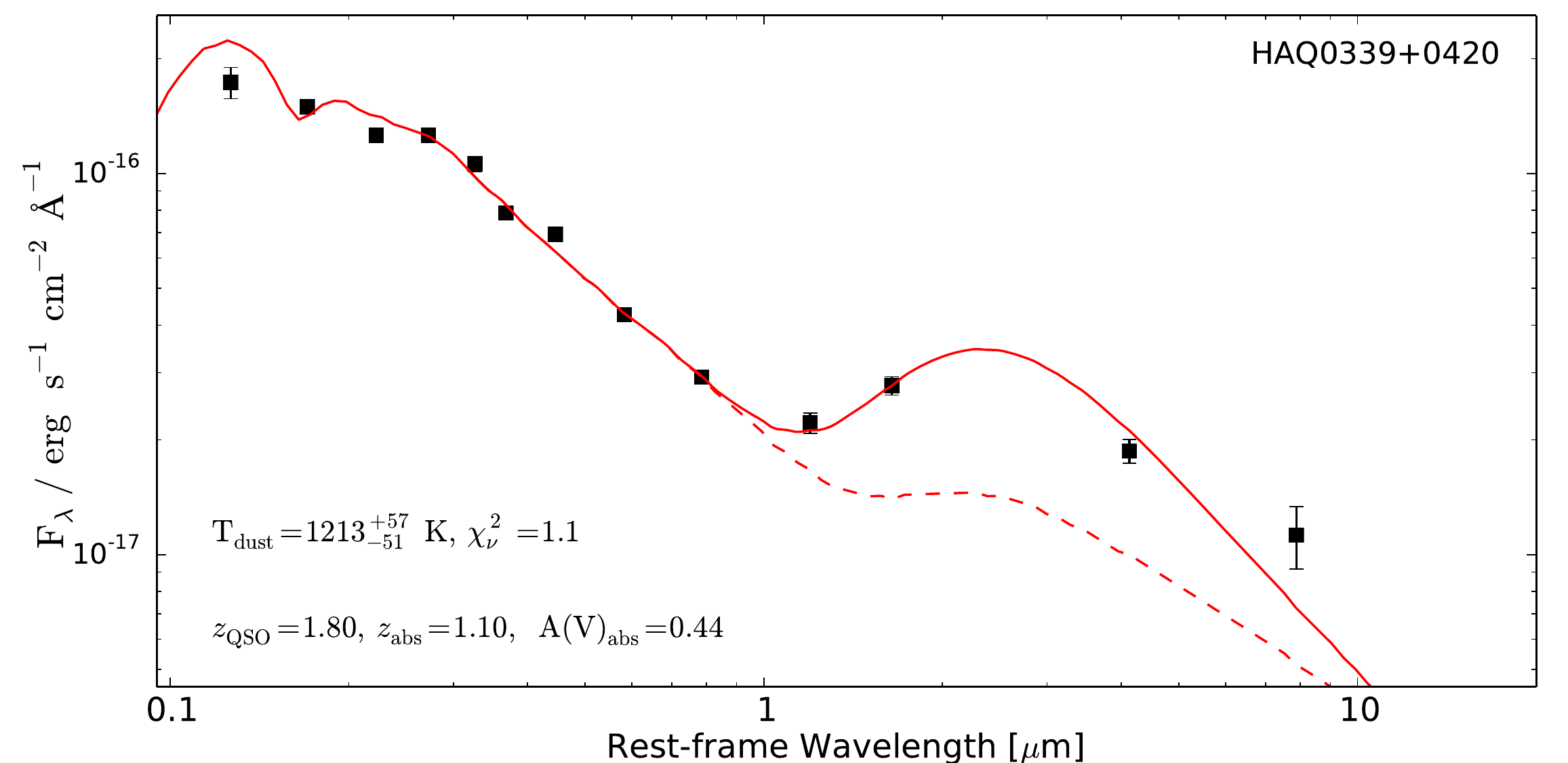}
\caption{A QSO with rest-frame infrared excess at $2~\mu$m (black points). The dashed line shows the best-fit template from fits to the spectra and SDSS and UKIDSS photometry as described in Sect.~\ref{quasars}. The solid line shows the same template with an extra dust component fitted to the $H$, $K$ and WISE bands. In the bottom left corner, we give the best-fit dust temperature, $T_{\rm dust}$ and the reduced $\chi^2$ from the dust component fit. This particular QSO was identified as a system with intervening absorption (see Section~\ref{model_comparison}), we therefore indicate the QSO redshift, the inferred absorption redshift and the best-fit extinction at the absorption redshift. We emphasize that the reddening law in the absorber's rest-frame is assumed to be of LMC type.
\label{fig:dustfit1}} 	% <-- LABEL: Avplot
\end{figure}

\begin{figure}
\plotone{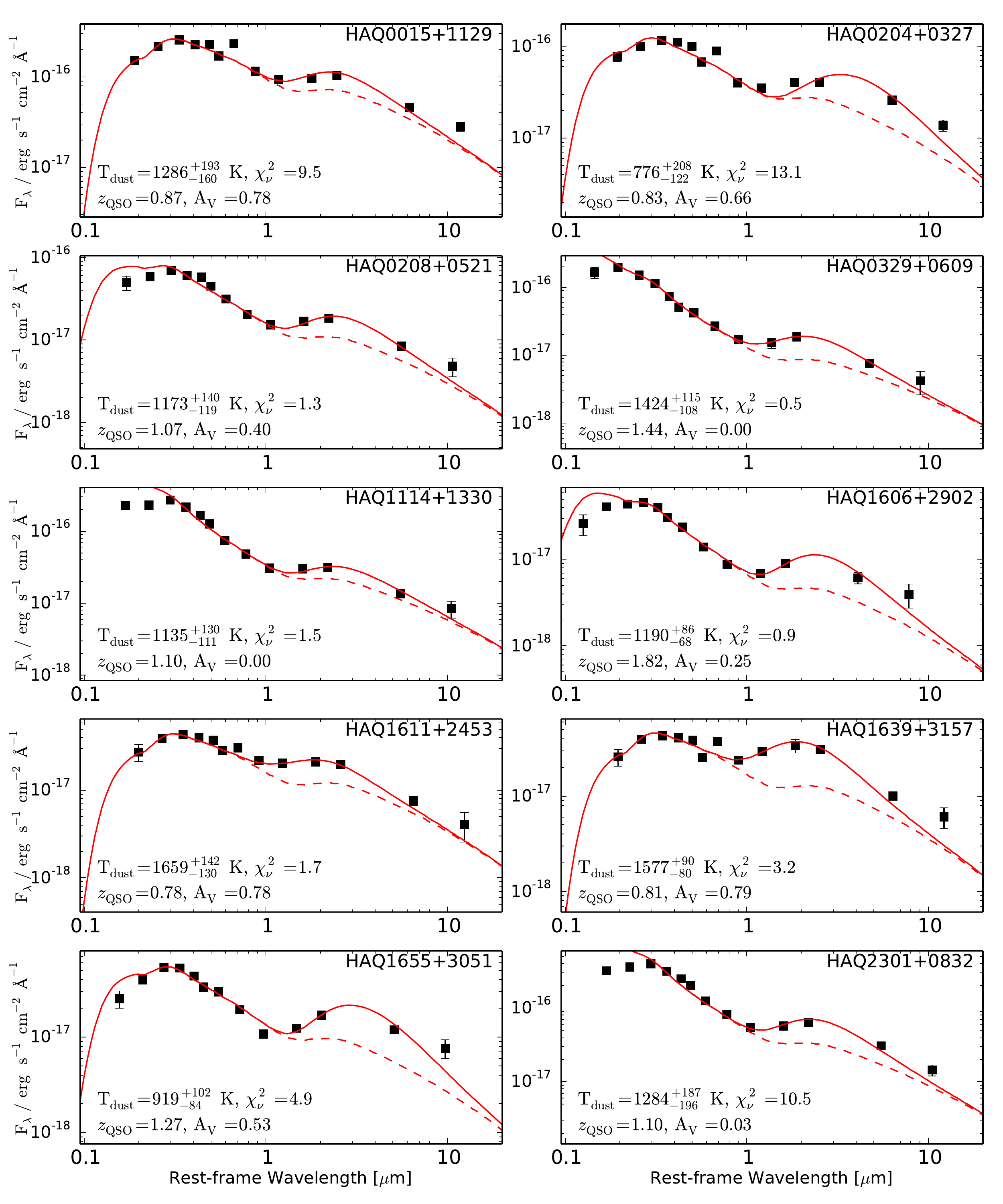}
\caption{Individual QSOs with rest-frame infrared excess at $2~\mu$m (black points). The dashed line shows the best-fit template from fits to the spectra and SDSS and UKIDSS photometry as described in Sect.~\ref{quasars}. The solid line shows the same template with an extra dust component fitted to the $H$, $K$ and WISE bands. In each panel, we give the best-fit dust temperature, $T_{\rm dust}$, the QSO redshift, the inferred extinction at the QSO redshift, and the reduced $\chi^2$ from the dust component fit.}
\label{fig:dustfit}
\end{figure}

In total, we find 27 QSOs that differ more than 5\,$\sigma$ from the template at rest-frame 2\,$\mu$m, 11 of which have an apparent excess compared to the template, and 16 have a deficit compared to the template. We included one additional target (HAQ1639+3157) that was not selected by this criterion because the large error on the WISE band 1 in this case falsely enhanced the uncertainty in the interpolated flux at rest-frame 2\,$\mu$m. 
The QSOs with rest-frame infrared excess are shown in Fig.~\ref{fig:dustfit1} and \ref{fig:dustfit}. We construct a simple model in which the excess emission is caused by re-emission from hot dust. We approximate the emission by a black-body emission with a single temperature, $T_{\rm dust}$. By adding this component to the reddened template (here from Richards et al.\,2006) and fitting the reddened template plus dust emission to the $H$ and $K$ bands along with the four WISE bands, we are able to get a very good agreement in
most cases with dust-temperatures in the range of 800~K$<T_{\rm dust}<$1600~K. The best-fit dust emission model is shown in Figs.~\ref{fig:dustfit1} and \ref{fig:dustfit} together with the best-fit template without extra dust emission.\\
The QSOs with rest-frame infrared deficit are summarized in Fig.~\ref{fig:IRdeficit}. These QSOs (except three: HAQ0011+0122, HAQ1233+1304, and HAQ1527+0250; see Sect.~\ref{dust}) all have rest-frame infrared SEDs consistent with no or very little reddening; however, from our fits to the rest-frame UV and optical data, we infer a significant amount of reddening. This seems to indicate a problem with the assumed extinction law. In order to test this, we re-fit the overall SED while changing the slope of the extinction law. We do this in terms of the parameter, $R_V\equiv A(V)/E(B-V)$, where $E(B-V)=A(B)-A(V)$. A large value of $R_V$ thus gives rise to a flat extinction curve, whereas small values indicate steeper extinction curves; for comparison, the SMC extinction curve from \citet{Gordon03} has $R_V=2.74$. Only for two QSOs (HAQ1106+0300, HAQ0206+0624) did the fit converge, in these cases yielding $R_V=0.8\pm0.2$ and $R_V=0.6\pm0.2$. The other fits failed to converge or failed to reproduce the IR fluxes, however, this does not rule out the fact that the extinction law may differ from the assumed SMC law. Instead, this reflects that simply varying $R_V$ is not enough to get the shape of the extinction curve to match the data. A detailed analysis of these extinction curves for a subset of our sample is currently underway (Zafar et al. 2014, in preparation). All the individual SEDs are shown in Appendix~\ref{appendix:SEDs}.

\begin{figure}
\plotone{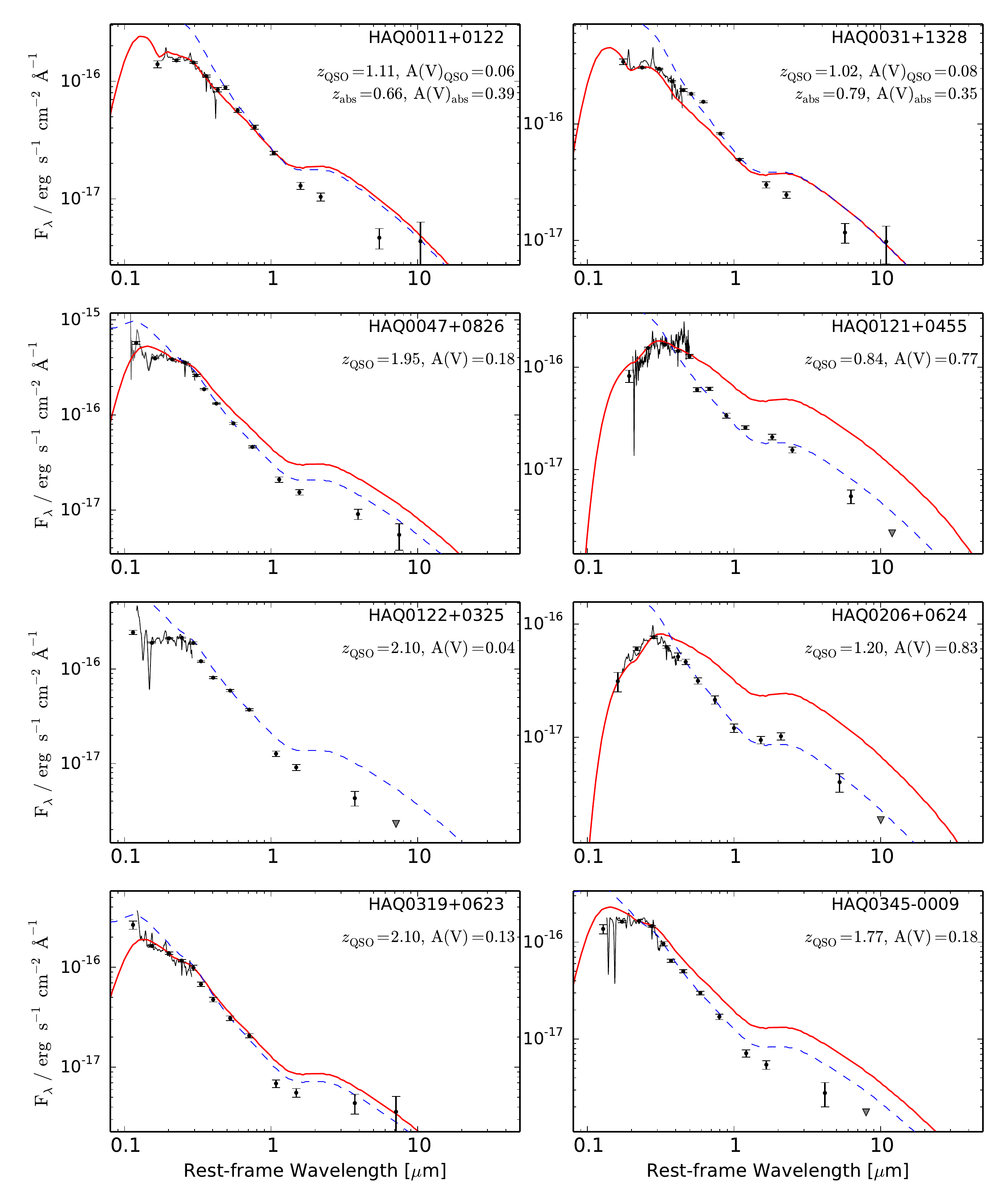}
\caption{Individual QSOs with rest-frame infrared deficit at $2~\mu$m (black points). In each panel, the dashed, blue line shows the unreddened template from \cite{Richards06}. The red, solid line shows the same template reddened by the amount given in each panel. In two cases the reddening is very low (consistent with zero), therefore only the dashed line is shown.} \label{fig:IRdeficit} 	% <-- LABEL: Avplot
\end{figure}

\begin{figure}
\plotone{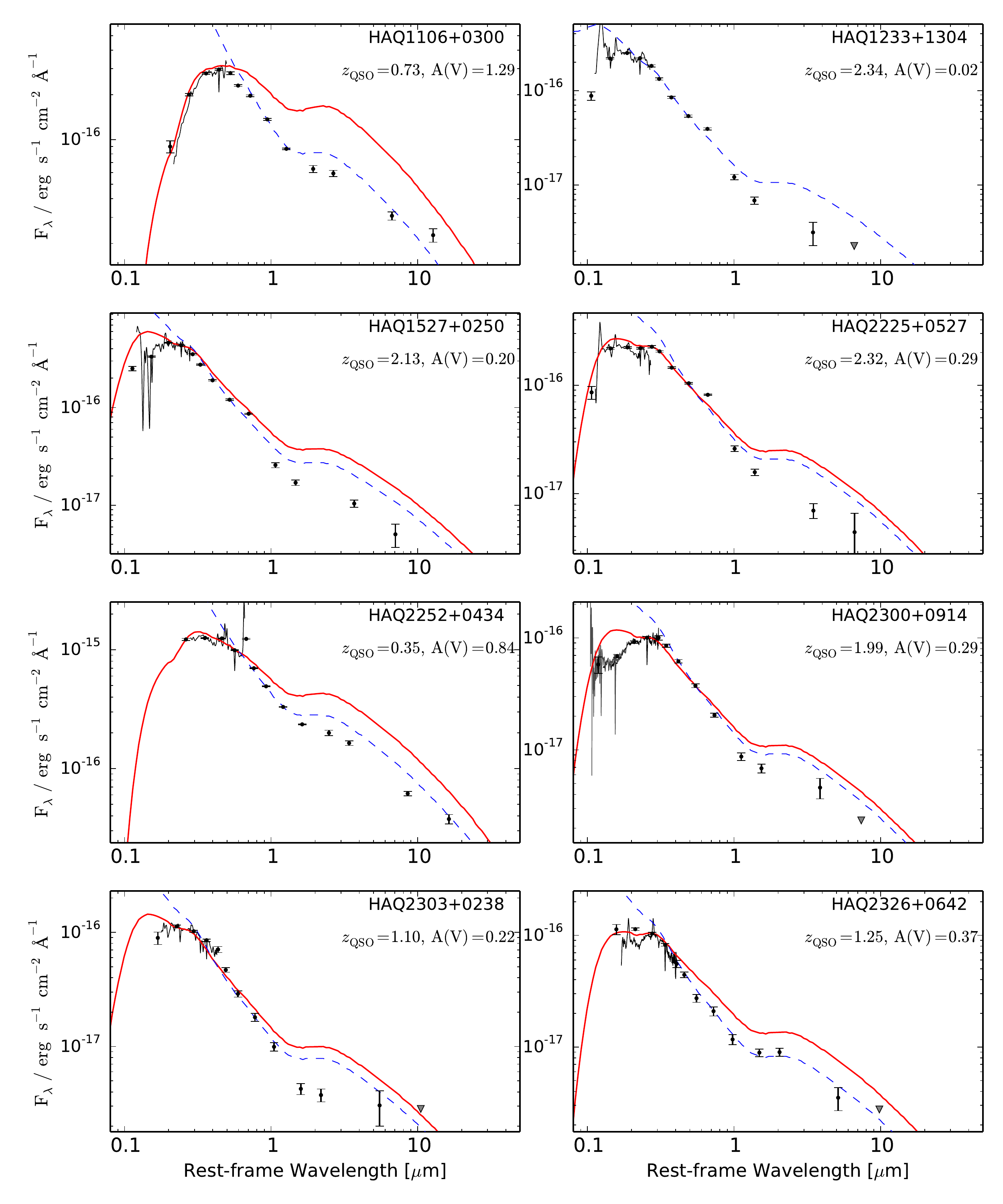}
\figurenum{8}
\caption{Continued.} 	% <-- LABEL: Avplot
\end{figure}

\subsubsection{Radio Properties}

\begin{figure}
\plotone{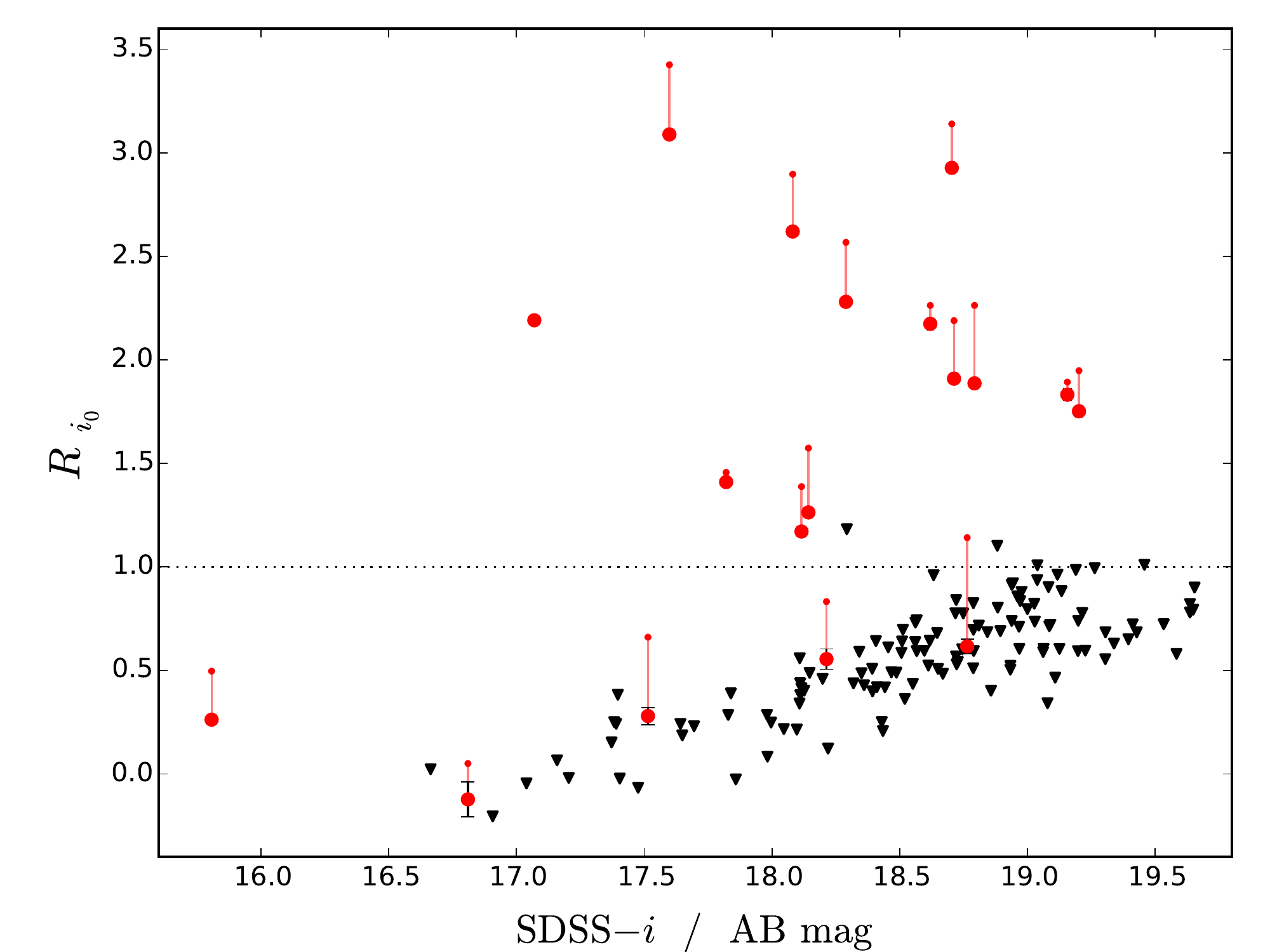}
\caption[]{Dust corrected radio-to-optical index, $R_{i_0}$, as function of observed $i$ band magnitude are shown as red circles with error bars. The small red dots connected by thin lines to each data point mark the values of $R_i$ before correcting for dust.
The dotted horizontal line marks the proposed boundary between radio loud ($R>1$) and radio quiet ($R<1$).
Non-detections from our sample are shown as 5~$\sigma$ upper limits (black triangles).
} \label{fig:radio} 	% <-- LABEL: Avplot
\end{figure}

We have matched our catalog to the VLA FIRST Survey \citep{Becker1995} to check for radio detections in our sample. Although our selection has not been based on radio detections, we have compiled data from the VLA/FIRST survey to compare the radio properties in our sample with other studies, e.g., \citet{Glikman2012, Glikman2013}. The survey provides measurements at 1.4 GHz ($\lambda=20$~cm) with a typical rms level of 0.15 mJy. A search through the survey's publicly available database\footnote{\url{http://sundog.stsci.edu/index.html}} reveals that 18 (out of \total) sources have radio detections at 1.4 GHz within a search radius of 10 arcsec. Thirteen targets in our sample lie outside the FIRST survey area. The radio fluxes (or 3\,$\sigma$ upper limits for non-detections) are given in Table~\ref{followuptab}.
We characterize the radio properties in terms of {\it radio loudness}. The term radio loudness has been defined in various ways, e.g., using the rest-frame radio luminosity or the ratio of radio-to-optical flux. We have chosen to use the ratio of radio-to-optical flux since this is independent of redshift and cosmological assumptions.
We use the definition of the radio-to-optical ratio, $R_m$ from \citet{Ivezic2002} to quantify the radio-loudness of QSOs. The authors define $R_m$ as the ratio of the radio flux to the optical flux: 
$$R_m = \log \left( \frac{F_{\rm radio}}{F_{\rm optical}}\right) = 0.4\,(m - t)~ , $$
where $m$ refers to the AB magnitude in any of the SDSS bands and $t$ is the \emph{``radio AB magnitude''} at 1.4 GHz: $t=-2.5\,\log\left( \frac{F_{\rm int}}{3631 {\rm Jy}} \right)$. Following this definition, a radio-loud source will have $R_m>1$. For our analysis we use the $i$ band of SDSS in order to compare with the work of \citet{Ivezic2002}; switching to $g$ or $r$ band does not change the inferred radio-loud fraction. In Fig.~\ref{fig:radio}, we show the dust-corrected radio-to-optical ratio, $R_{i_0}$, for the detected sources in our sample using the {\Av} inferred from the combined fit to spectroscopy and photometry to de-redden the $i$-band magnitude. We also show the non-detected sources as conservative 5~$\sigma$ upper limits. As seen in Fig.~\ref{fig:radio}, all non-detections except for four are consistent with being radio-quiet. Hence, we are able to put very firm constraints on the fraction of radio-loud QSOs. If we assume that \emph{all} non-detections are radio-quiet (consistent at the 3$\sigma$ level), then we find a radio-loud fraction ranging from 9 to 12\%, including the uncertainty of the dust-correction.

\citet{Ivezic2002} and \citet{Balokovic2012} find a radio-loud fraction of $8\pm1$\% and $10\pm1$\%, respectively, for QSOs with $i<18.5$. For the same $i$-band magnitude limit, we find a radio-loud fraction of 12$-$14\%. This is slightly higher but consistent with the previous results.
A higher fraction of radio-loud QSOs would be expected when targeting red QSOs as the radio flux has been shown to correlate with optical ($g-r$) color \citep{White2007}. We further note that due to the color selection of our sample, the estimate of radio-loud fraction is not representative of the overall QSO population. 

\subsection{Contaminants}

Two objects in our sample are identified as stars (marked in Table~\ref{followuptab}): one M-dwarf and one G-dwarf.
Their spectra are plotted together with stellar templates from SDSS\footnote{http://www.sdss.org/dr5/algorithms/spectemplates} in Appendix~\ref{appendix:contaminants}.
The G-dwarf template has been reddened (\Av$=0.6$) to match the observed spectrum by assuming MW type dust from \citet{Gordon03}.

We find two objects (HAQ1607+2611 and HAQ2247+0146) that are not securely identified. The object HAQ1607+2611 is probably a QSO with very weak emission lines, whereas HAQ2247+0146 is most likely a special subclass of BAL QSOs (so-called FeLoBALs) with very strong and broad absorption on the blue side of Mg\,{\sc ii}. However, the spectra do not provide enough clues to firmly secure the identification and redshift.

%-------------------------DISCUSSION-------------------------------------
\section{Discussion}
\label{discussion}

Our primary goal was to search for intervening absorption systems hosting metals and dust,
which would go undetected in optically selected samples of QSOs (e.g., SDSS-I/II). We found 30 such absorption systems, however, with varying amounts of dust. Moreover, we found interesting dust features most likely caused by dust within the QSO system. In the following, we take a closer look at the intervening systems in our sample and at the dust properties in the environment of the QSOs.

\subsection{Dust in Intervening Absorbers}
\label{absorbers}
In order to check whether any of the QSOs showed signs of intervening absorption, we analyzed all the spectra with an automated likelihood algorithm.
In Sect.~\ref{model_comparison}, we found nine cases where the code returned a preferred intervening dusty system. These are listed in Table~\ref{table:LMC_dust}. For all nine systems, we found that the extinction was caused by dust in the absorber only; the extinction at the QSO redshift was consistent with no reddening. All the other systems in our sample were fully consistent with either no reddening or dust at the QSO redshift, but note that our method is mainly sensitive to intervening absorbers with extinction curves displaying the 2175\,{\AA} extinction feature (see below).

Most statistical studies of absorption systems (Mg\,{\sc ii} or DLAs) find that the average reddening, though small, is consistent with SMC-type extinction \citep{York2006, Vladilo08, Khare2012}. However, indications of the 2175\,{\AA} bump, typical in Milky Way and LMC sightlines, are observed in some cases both statistically using stacking \citep{Malhotra1997} and in individual absorbers \citep{Junkkarinen2004, Wang2004, Ellison2006,  Srianand08, Eliasdottir2009, Noterdaeme09a, Jiang2010, Zhou2010, Wang2012, Zafar2012}.
Interestingly, we found {\it no} systems with preferred SMC extinction in an intervening system from our algorithm although the overall population seems to indicate SMC-type dust. This is, however, to be expected when taking into account the limitations of the method we utilized. In Sect.~\ref{robustness}, we found that absorbers needed to cause very high extinction (A(V)$\ga0.5$) with the SMC law in order to be detected, even at high SNR. This is due to the high degeneracy between dust at the absorber redshift and intrinsic dust in the QSO because the SMC extinction curve is mostly featureless. Our simulations showed that we were much more sensitive to LMC-type extinction due to the very characteristic 2175\,{\AA} bump which provided a rather unique imprint in the spectra. This allowed for a much more secure determination of the redshift of the absorber if this feature was available in the spectrum. Still, we are limited to absorbers with extinction higher than A(V)$>0.2$ at high SNR (an average SNR per spectral pixel of 15).\\
\indent
Since these estimates were purely based on statistical modeling, the results could be improved if an absorption system at a corresponding redshift could be identified. This would then serve as an anchor for $z_{\rm abs}$, which allows us to measure the extinction curve in the absorber with greater precision. For the nine cases reported here, two of them had a detection of a Mg\,{\sc ii} absorption system close to the best-fit redshift and one sightline showed a tentative detection. This indicates that the dust model was indeed detecting physical features and not just artifacts in the spectra. Furthermore, one DLA was found to be a plausible host of SMC type dust causing reddening of the background QSO when restricting the dust absorber redshift. This system, which will be discussed in more detail in a future work, is the kind of system we originally looked for with this survey.

Our results suggest that previous studies of dust extinction caused by intervening absorbers might have been biased against LMC type dust. This is plausibly caused by two effects. On one hand, for absorbers causing low amounts of extinction (as is usually the case for intervening absorbers found to date) the 2175~{\AA} feature characteristic of the LMC and MW extinction curves may be very weak. This will generally lead an observer to classify the extinction as SMC-type, as the 2175~{\AA} bump disappears in the noise \citep[see discussion in][]{Khare2012}. On the other hand, for dust-rich absorbers causing high amounts of extinction, where a distinction between various extinction curves is more easily done, the reddening of the background QSO may cause the QSO to drop out of the optical color selection utilized in previous samples. It has indeed been shown that such highly reddened QSOs are underrepresented in previous samples of QSOs from optical color selection \citep{Fynbo2013a, Glikman2012, Glikman2013}.
However, with the advance in near- and far-infrared photometric surveys
and various new QSO classification algorithms populations of highly reddened QSOs will be identified in larger numbers \citep{Maddox2012, Graham2014}.

\subsection{Dust in the QSOs}
\label{dust}

For our entire sample, we find a median attenuation of {\Av} = 0.13, corresponding to a median
$E(B-V)$ of 0.047, assuming the value of $R_V=2.74$ from \citet{Gordon03}.
The individual dust-corrected SEDs show a very homogeneous behavior at wavelengths $\lambda<1\mu$m. However, at larger wavelengths we see some variance.
We find that the QSOs with excess IR flux can be explained well by a simple single-temperature blackbody dust component. The dust temperatures inferred in these cases lie in the range from $800\lesssim T_{\rm dust} \lesssim 1600$~K with typical uncertainties of $100-200$~K. This is in good agreement with dust temperatures found in active galactic nuclei \citep{Sanders1989,Rodriguez-Ardila2006}.
In the cases of very bad fits (e.g., HAQ0204+0327), the discrepancy might be explained by adding a host galaxy
component or a model with multiple dust temperatures. However, such modeling is beyond the scope of this discussion and is not possible due to the few data points currently available in this wavelength range.

As mentioned in Sect.~\ref{results}, we find 16 cases where the flux at
rest-frame 2~$\mu$m is significantly lower than what is predicted by the
template. There may be more such QSOs, but we chose a conservative cut at
5$\sigma$ to limit the influence of template variance. In these cases, the
mismatch seems to hint at a problem with the assumed extinction law, since the
unreddened template in all but two cases provide a good fit in the rest-frame
infrared and optical, while a large amount of reddening is required to fit the
rest-frame ultraviolet data. Although uncommon, similarly steep extinction curves have been observed in previous works \citep[e.g.,]{Welty1992, Larson1996, Hall2002, Fynbo2013a, Fynbo2014, Jiang2013, Leighly2014}. For the two targets where this is not the case (HAQ0011+0122 and HAQ1233+1304), the lack in flux may be explained by intrinsic differences in the QSO energy distribution relative to the template. A lack of dust emission compared to the template would explain the lower rest-frame IR fluxes in these cases. This was also noted by \citet{Richards06}. Internal variations in the dust emission (both deficit and excess) may be linked to physical properties in the obscuring material, such as temperature, geometry, and viewing effects (inclination).

In targets where we see indications of non-SMC-type dust, we need an extinction
curve that is \emph{steeper} or with different curvature than SMC in order to explain the strong UV
reddening and little or no reddening in the infrared. In Sect.~\ref{SEDs}, we modeled this behavior of the extinction law by varying the value of $R_V$. This analysis only converged to meaningful values (non-negative values for $A(V)$ and $R_V$) for two QSO, which resulted in best-fit values of $R_V=0.8\pm0.2$ and $R_V=0.6\pm0.2$.
This indeed hints at the need for a very steep extinction curve, but this should only be taken as a preliminary analysis. In Paper I, we reported similar indications of extinction curve mismatches. Furthermore, many of these targets exhibit extremely weak emission lines, indicating that the central region, which is emitting these lines, may be somewhat obscured. These QSOs are similar to the peculiar red QSOs discussed by \citet{Hall2002}, \citet{Meusinger2012} and \citet{Jiang2013}.
The QSOs found by \citet{Hall2002} show similar characteristics, i.e., missing strong emission lines and a continuum mismatch at wavelengths shorter than $\lambda < 3000$~\AA . Many plausible explanations for such a continuum shape are discussed by these authors, however, they conclude that a steeper extinction curve (steeper than SMC or with different shape, e.g., a break) is the most likely explanation given the apparent lack of BAL features. Similar conclusions are reached by \citet{Meusinger2012}.

In contrast to the QSOs in the studies of \citet{Meusinger2012} and \citet{Hall2002} four of the
peculiar QSOs in this work show BAL features, two of which (HAQ0122+0325 and HAQ1233+1304) show very little apparent reddening, consistent with A(V)=0. The remaining BAL QSOs (HAQ0047+0826, HAQ0345-0009, and HAQ1527+0250), which show signs of non-SMC type dust, have very weak emission lines and their SEDs at $\lambda >
3000$~\AA \ are well represented by the unreddened QSO template. Further work
on larger samples over a larger wavelength range is needed to shed light on the
nature of these peculiar targets and to investigate the actual shape of the
extinction curve needed to explain the SEDs of these QSOs.
Similar steep extinction curves have been observed toward
the Milky Way centre \citep{Nishiyama2008, Nishiyama2009, Sumi2004}. It is thus
not clear whether the detections toward these reddened QSOs are caused by the
central engine in the QSOs or simply by the fact that we observe the central
parts of the host galaxy. \citet{Leighly2014} argue that by modeling the dust around the QSO with a spherical geometry and dust grain properties similar to those of the LMC or the Milky Way they can reproduce the steep extinction curve in the rest-frame UV observed in Mrk 231. In this case the dust in the central region is provided by a central star-burst. The scattering and absorption properties of the dust in a spherical distribution then gives rise to multiple scatterings, which causes the UV photons to suffer from higher extinction than the redder wavelengths. In contrast to the results of Leighly et al., \citet{Jiang2013} use a different grain size distribution to reproduce a similarly steep extinction curve toward a high-redshift QSO. In the work of \citet{Jiang2013}, the steep extinction curve is caused by a lack of larger grains, i.e., their grain size distribution is truncated at a maximum size of 70~nm. The authors invoke dust destruction in the QSO environment as a plausible mechanism, or differences in dust production and growth in QSO outflows \citep*[as argued by][]{Elvis2002}. This lack of large grains causes extinction in the UV to be relatively higher than in the optical and near-infrared. The question about the origin of the steep extinction curve is thus still open for debate. Furthermore, as is noted by Leighly et al., it is curious why this anomalous type of extinction is only observed toward a few reddened QSOs and not seen in the overall population of reddened QSOs.

\section{Conclusions}

In this paper, we present a clean selection method that only relies on optical
and near-infrared photometry down to a flux limit of $J_\mathrm{AB} < 19$,
i.e., we do not require detections from radio or X-ray surveys. The method is a modification of the criteria applied in Paper I rejecting the small contamination of cool stars and evolved galaxies found in Paper I. The refined method allows us to effectively select reddened QSOs regardless of their X-ray and radio properties over a large area of the sky. In our sample of {\total} objects only 2 turn out not to be QSOs and 2 remain unidentified.
Our primary goal with this selection was not to compile a complete sample of red QSOs, but instead we designed the criteria to provide a high purity to investigate any hidden population of absorption systems toward heavily reddened QSOs. 
We used a statistical algorithm to identify whether the SEDs were better represented by dust in the QSO or in an intervening system. This way we identified nine QSOs where dust in an intervening system was preferred. All of these were identified as having LMC-type dust showing signs of the 2175~{\AA} feature. Two (tentatively three) of these systems have Mg\,{\sc ii} absorption at a corresponding redshift to the best-fit redshift indicating that the statistical modeling was capable of correctly identifying dust in absorption systems. Moreover, we discovered a DLA toward a highly reddened QSO. Although this system was not selected in our likelihood ratio test, the data are consistent with the dust reddening being caused by SMC-type dust in the DLA. This system will be studied in more detail in a forthcoming paper.

Complementary to our primary search for dusty intervening systems, our selection serendipitously discovered QSOs with a wide range of dust properties. More work on expanding the sample and quantifying the physical dust properties toward these obscured QSOs is currently underway.
A complete list of all the candidate QSOs in our survey is given in Table~\ref{candidatetab} in Appendix~\ref{candidates}.

\acknowledgements
The authors thank the anonymous referee whose constructive feedback
improved the quality of this work.
The authors thank the many students who dedicated their time to observing
QSOs during the summer schools at the NOT in the years
2012, 2013, and 2014. JK acknowledges support from a studentship at the
European Southern Observatory in Chile.
JPUF acknowledges support from the ERC-StG grant EGGS-278202. The Dark Cosmology Centre is funded by the DNRF. BPV acknowledges funding through the ERC grant "Cosmic Dawn."
Based on observations made with the NOT, operated
on the island of La Palma jointly by Denmark, Finland, Iceland,
Norway, and Sweden, in the Spanish Observatorio del Roque de los
Muchachos of the Instituto de Astrof\'isica de Canarias.
Funding for the SDSS and SDSS-II has been
provided by the Alfred P. Sloan Foundation, the Participating Institutions, the
National Science Foundation, the U.S. Department of Energy, the National
Aeronautics and Space Administration, the Japanese Monbukagakusho, the Max
Planck Society, and the Higher Education Funding Council for England. The SDSS
Web Site is http://www.sdss.org/. The SDSS is managed by the Astrophysical
Research Consortium for the Participating Institutions. The Participating
Institutions are the American Museum of Natural History, Astrophysical
Institute Potsdam, University of Basel, University of Cambridge, Case Western
Reserve University, University of Chicago, Drexel University, Fermilab, the
Institute for Advanced Study, the Japan Participation Group, Johns Hopkins
University, the Joint Institute for Nuclear Astrophysics, the Kavli Institute
for Particle Astrophysics and Cosmology, the Korean Scientist Group, the
Chinese Academy of Sciences (LAMOST), Los Alamos National Laboratory, the
Max-Planck-Institute for Astronomy (MPIA), the Max-Planck-Institute for
Astrophysics (MPA), New Mexico State University, Ohio State University,
University of Pittsburgh, University of Portsmouth, Princeton University, the
United States Naval Observatory, and the University of Washington. We
acknowledge the use of UKIDSS data.  The United Kingdom Infrared Telescope is
operated by the Joint Astronomy Centre on behalf of the Science and Technology
Facilities Council of the U.K. 
This publication makes use of data products from WISE , which is a joint project of the University of California, Los Angeles, and the Jet Propulsion Laboratory/California Institute of Technology, funded by the National Aeronautics and Space Administration.

%%%%%%%%%%%%%%%%%%%%%%%%%%%%%%
%% BIBLIOGRAPHY
%%%%%%%%%%%%%%%%%%%%%%%%%%%%%%

\bibliographystyle{apj}

%%%%%%%%%%%%%%%%%%%%%%%%%%%%%%%
%% APPENDICES
%%%%%%%%%%%%%%%%%%%%%%%%%%%%%%%

\appendix
\section{Uncertainty Estimates}
\label{appendix:table}

In order to estimate the confidence intervals of the best-fit parameters, we used a Markov chain Monte Carlo method. The simulations were performed using the {\tt emcee} package for Python \citep{emcee}. We used flat priors on all parameters and ran 100 walkers for 700 iterations of which the first 100 were discarded as the "burn in" phase. In Table~\ref{tab:dust_fit}, we give the 68\% and 99.7\% confidence intervals of the parameter distributions.

\begin{deluxetable}{lcccccccccc}
\rotate
\tabletypesize{\scriptsize}
\setlength{\tabcolsep}{4pt}
\tablecaption{Dust Model Comparison
\label{tab:dust_fit}}
\tablewidth{0pt}
\tablehead{
  &   &  & \multicolumn{2}{c}{Null Model} & & & & \multicolumn{2}{c}{General Model} & \\
\colhead{Target} & \colhead{$z_{\rm QSO}$}  &  \colhead{$A({\rm V})$}  &
\colhead{$\chi^2_0$} & \colhead{$P_{\rm KS,\,_0}$}   &
\colhead{$z_{\rm abs}$} & \colhead{$A({\rm V})_{\rm abs}$} & \colhead{$A({\rm V})_{\rm QSO}$} & \colhead{$\chi^2_{\rm G}$} &
\colhead{$P_{\rm KS,\,_{\rm G}}$}  & \colhead{$p$} \\
}
\startdata
HAQ0000+0557 & 3.31 & 0.13 $\pm$ 0.01 & 380.1 & 0.002 & 2.03$^{+0.03}_{-0.03}(1\,\sigma)^{+0.08}_{-0.08}(3\,\sigma)$ & 0.32$^{+0.03}_{-0.03}(1\,\sigma)^{+0.11}_{-0.09}(3\,\sigma)$ & 0.04$^{+0.01}_{-0.01}(1\,\sigma)^{+0.04}_{-0.04}(3\,\sigma)$ & 257.2 & 0.914 & $2.1 \cdot 10^{-27}$ \\[5pt] 
HAQ0011+0122 & 1.11 & 0.27 $\pm$ 0.01 & 345.6 & 0.135 & 0.66$^{+0.02}_{-0.02}(1\,\sigma)^{+0.05}_{-0.05}(3\,\sigma)$ & 0.39$^{+0.04}_{-0.05}(1\,\sigma)^{+0.14}_{-0.13}(3\,\sigma)$ & 0.06$^{+0.02}_{-0.02}(1\,\sigma)^{+0.07}_{-0.07}(3\,\sigma)$ & 250.0 & 0.347 & $1.7 \cdot 10^{-21}$ \\[5pt]
HAQ0031+1328 & 1.02 & 0.28 $\pm$ 0.01 & 692.0 & 0.001 & 0.79$^{+0.01}_{-0.01}(1\,\sigma)^{+0.03}_{-0.04}(3\,\sigma)$ & 0.35$^{+0.02}_{-0.02}(1\,\sigma)^{+0.07}_{-0.07}(3\,\sigma)$ & 0.08$^{+0.01}_{-0.01}(1\,\sigma)^{+0.04}_{-0.04}(3\,\sigma)$ & 472.4 & 0.803 & $2.1 \cdot 10^{-48}$ \\[5pt] 
HAQ0051+1542 & 1.90 & 0.13 $\pm$ 0.01 & 362.4 & 0.002 & 1.24$^{+0.03}_{-0.03}(1\,\sigma)^{+0.10}_{-0.08}(3\,\sigma)$ & 0.39$^{+0.04}_{-0.03}(1\,\sigma)^{+0.11}_{-0.09}(3\,\sigma)$ & $-$0.15$^{+0.02}_{-0.03}(1\,\sigma)^{+0.07}_{-0.08}(3\,\sigma)$ & 211.8 & 0.750 & $2.0 \cdot 10^{-33}$ \\[5pt] 
HAQ0339+0420 & 1.80 & 0.23 $\pm$ 0.01 & 167.1 & 0.797 & 1.10$^{+0.06}_{-0.07}(1\,\sigma)^{+0.14}_{-0.18}(3\,\sigma)$ & 0.44$^{+0.07}_{-0.06}(1\,\sigma)^{+0.22}_{-0.17}(3\,\sigma)$ & 0.01$^{+0.03}_{-0.04}(1\,\sigma)^{+0.09}_{-0.13}(3\,\sigma)$ & 112.9 & 0.625 & $1.7 \cdot 10^{-12}$ \\[5pt] 
HAQ1248+2951 & 3.55 & 0.01 $\pm$ 0.01 & 298.9 & 0.148 & 2.36$^{+0.10}_{-0.09}(1\,\sigma)^{+0.26}_{-0.21}(3\,\sigma)$ & 0.20$^{+0.02}_{-0.02}(1\,\sigma)^{+0.10}_{-0.06}(3\,\sigma)$ & $-$0.03$^{+0.01}_{-0.01}(1\,\sigma)^{+0.02}_{-0.04}(3\,\sigma)$ & 184.8 & 0.806 & $1.7 \cdot 10^{-25}$ \\[5pt] 
HAQ1509+1214 & 2.80 & 0.09 $\pm$ 0.01 & 182.6 & 0.016 & 1.71$^{+0.06}_{-0.04}(1\,\sigma)^{+0.27}_{-0.11}(3\,\sigma)$ & 0.45$^{+0.13}_{-0.12}(1\,\sigma)^{+0.36}_{-0.28}(3\,\sigma)$ & $-$0.15$^{+0.05}_{-0.06}(1\,\sigma)^{+0.12}_{-0.16}(3\,\sigma)$ & 127.0 & 0.381 & $8.4 \cdot 10^{-13}$ \\[5pt]
HAQ2203-0052 & 1.24 & 0.13 $\pm$ 0.01 & 391.6 & 0.545 & 0.65$^{+0.04}_{-0.03}(1\,\sigma)^{+0.09}_{-0.08}(3\,\sigma)$ & 0.73$^{+0.07}_{-0.07}(1\,\sigma)^{+0.20}_{-0.22}(3\,\sigma)$ & $-$0.11$^{+0.02}_{-0.03}(1\,\sigma)^{+0.08}_{-0.08}(3\,\sigma)$ & 297.0 & 0.257 & $2.9 \cdot 10^{-21}$ \\[5pt]  
HAQ2343+0615 & 2.16 & 0.17 $\pm$ 0.01 & 193.2 & 0.203 & 1.48$^{+0.03}_{-0.03}(1\,\sigma)^{+0.08}_{-0.11}(3\,\sigma)$ & 0.29$^{+0.04}_{-0.04}(1\,\sigma)^{+0.13}_{-0.12}(3\,\sigma)$ & $-$0.05$^{+0.03}_{-0.03}(1\,\sigma)^{+0.09}_{-0.10}(3\,\sigma)$ & 144.2 & 0.404 & $2.3 \cdot 10^{-11}$ \\
\enddata
\tablecomments{Model parameters for the null model and the general model for the QSOs where the null model was rejected. $P_{\rm KS}$ refers to the $p$ value from the KS-test of the normalized residuals, and $p$ refers to the chance probability of the observed improvement in $\chi^2$ given the two extra free parameters in the general model.}
\end{deluxetable}

\clearpage

%%%%%%%%%%%%%%%%%%%%%%%%%%
%%  ROBUSTNESS OF LIKELIHOOD RATIO TEST

\section{Robustness of the Likelihood Ratio Test}
\label{appendix:simulations}

In order to test the robustness of the likelihood method, we generate a set of mock QSO data sets and analyze them with our algorithm the same way as we analyzed our data. Below, we summarize the details of how the data sets were generated.
The initial parameters were drawn randomly, following a uniform distribution, within the following limits. For the QSO redshift we use: $1 < z_{\rm QSO} < 3.5$; for the extinction at the QSO redshift we constrain the A(V) to be ${\rm A(V)_{QSO}} > -0.1$. The selection of candidates in terms of $g-r$ color implies that the extinction in the QSO's rest frame will be redshift dependent. Assuming the SMC extinction law in the QSO's rest frame, this means that less reddening is needed at higher redshifts in order to match the $g-r$ color criterion ($0.5<g-r<1$). We mimic this selection effect by invoking the following limit on the modeled extinction at the QSO redshift: ${\rm A(V)_{QSO}} < 1.5-0.37 \times z_{\rm QSO}~$. For the extinction at the absorber redshift we use $0 < {\rm A(V)_{abs}} < 1.0$; and for the absorber redshift we use $0.1 < z_{\rm abs} < 0.9 \times z_{\rm QSO}$. Here the upper limit is invoked to keep the absorption redshift well-defined in the fits. This also ensures that the absorption system will be physically separated from the QSO environment.

These four randomly drawn parameters are used to generate a QSO model assuming the combined QSO template by \citet{vandenBerk01, Glikman06}. We use the SMC extinction curve for the reddening at the QSO redshift and we generate two sets of models: one assuming SMC-type dust in the absorber and one assuming LMC-type dust. The QSO model is then smoothed to match the resolution of the ALFOSC instrument at the NOT and hereafter interpolated onto a wavelength grid similar to the spectral data from the aforementioned instrument. Then we generate synthetic photometric data by calculating the flux in each of the SDSS ($ugriz$) and UKIDSS ($YJHK_s$) bands by weighting the model with the appropriate filter curve. We denote the synthetic spectral and photometric data as a {\it synthetic data set}.\\
\indent We then add noise to the synthetic data set following a Gaussian noise model. We generate two sets for each assumed absorber extinction curve: one with high SNR and one with low SNR. High and low signal-to-noise ratios here refer to ${\rm SNR_{spec}=15,~SNR_{phot}=20}$ and
${\rm SNR_{spec}=5}$, ${\rm SNR_{phot}=10}$, respectively, where ${\rm SNR_{spec}}$ refers to the average SNR per pixel in the synthetic spectral data and ${\rm SNR_{phot}}$ refers to the SNR of each synthetic photometric band. For the spectral noise model, we further add a noise component mimicking the fringing in the red part of the CCD of the ALFOSC spectrograph. Specifically, this is $\sim20$\,\% (peak-to-peak fringe level) for wavelengths greater than $8000$\,{\AA}.

\begin{figure}
\plotone{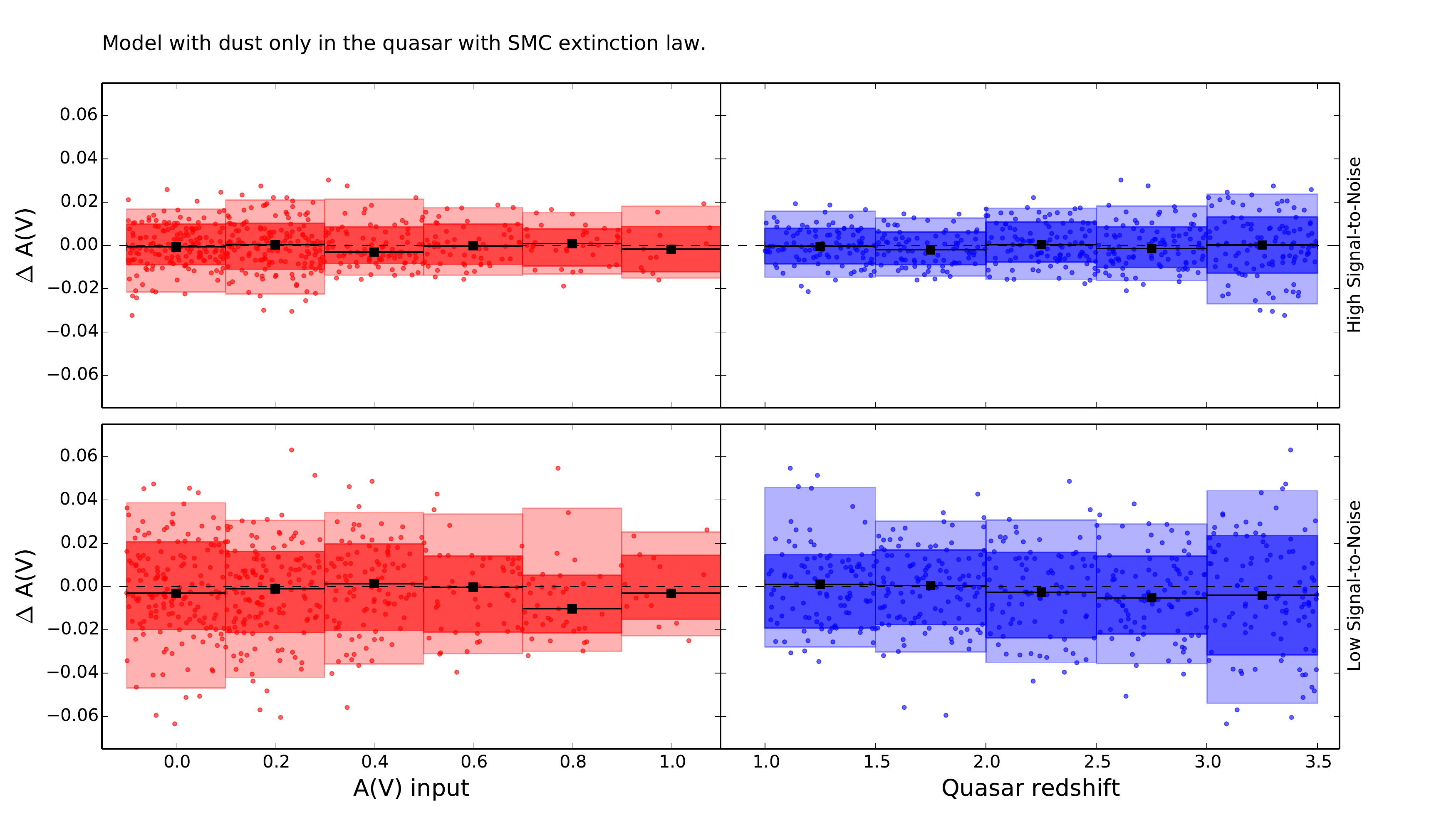}
\caption{Residuals of the recovered A(V) in the null model tests where we only add reddening at the QSO redshift. The top row shows the results of modeling at high SNR, while the bottom row shows the results obtained at low SNR. In each row, the left panel shows the residuals of the best-fit A(V) as function of the input A(V). The right panel shows the same residuals as function of QSO redshift. We observe no bias (i.e., no systematic offset in the residuals) and no significant dependence on input parameters.}
\label{fig:null_model}
\end{figure}

\begin{figure}
\plotone{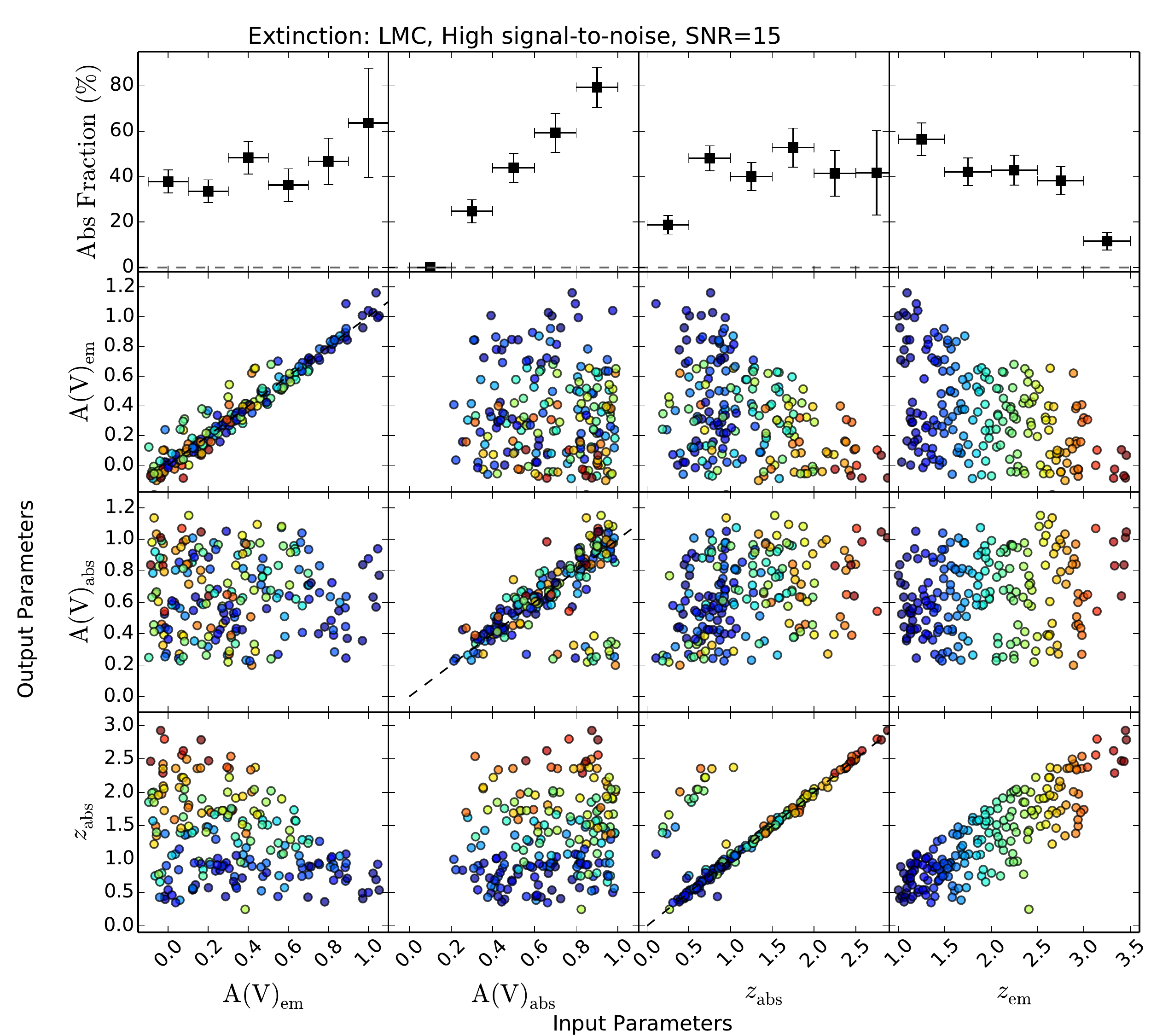}
\caption{The top row shows the recovered fraction of correctly identified absorbers using the likelihood ratio test and in the three subsequent rows, we show the best-fit output parameters as function of the input parameters. The data shown are for model runs with high SNR and assuming the LMC extinction curve. We only show the results for the correctly identified absorbers. The points are color-coded by redshift of the QSO (blue corresponds to $z=1$, and red indicates $z=3.5$.}
\label{fig:LMC_high_SNR}
\end{figure}

\begin{figure}
\plotone{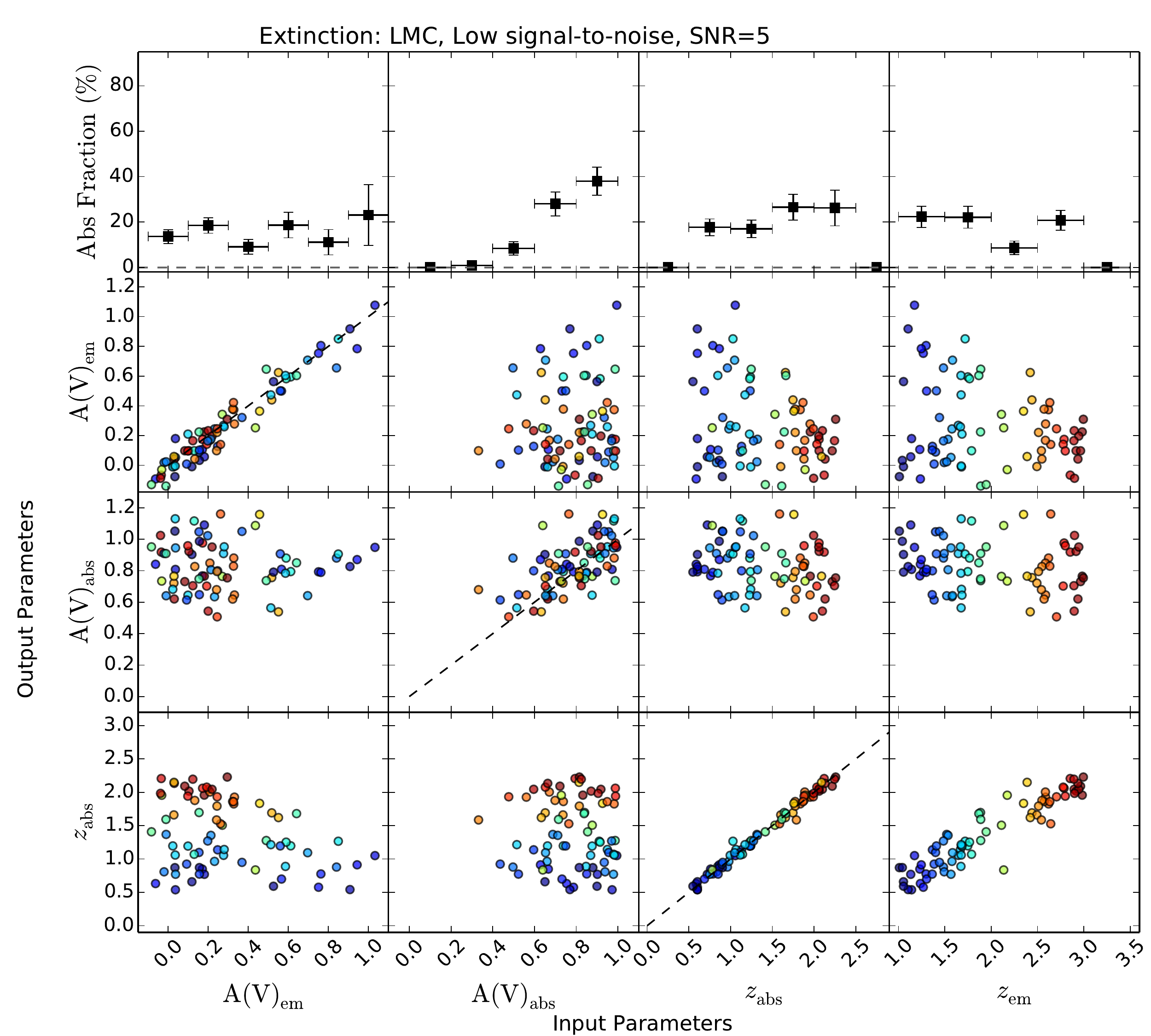}
\caption{Same as Fig.~\ref{fig:LMC_high_SNR} but for low SNR assuming the LMC extinction curve.}
\label{fig:LMC_low_SNR}
\end{figure}

\begin{figure}
\plotone{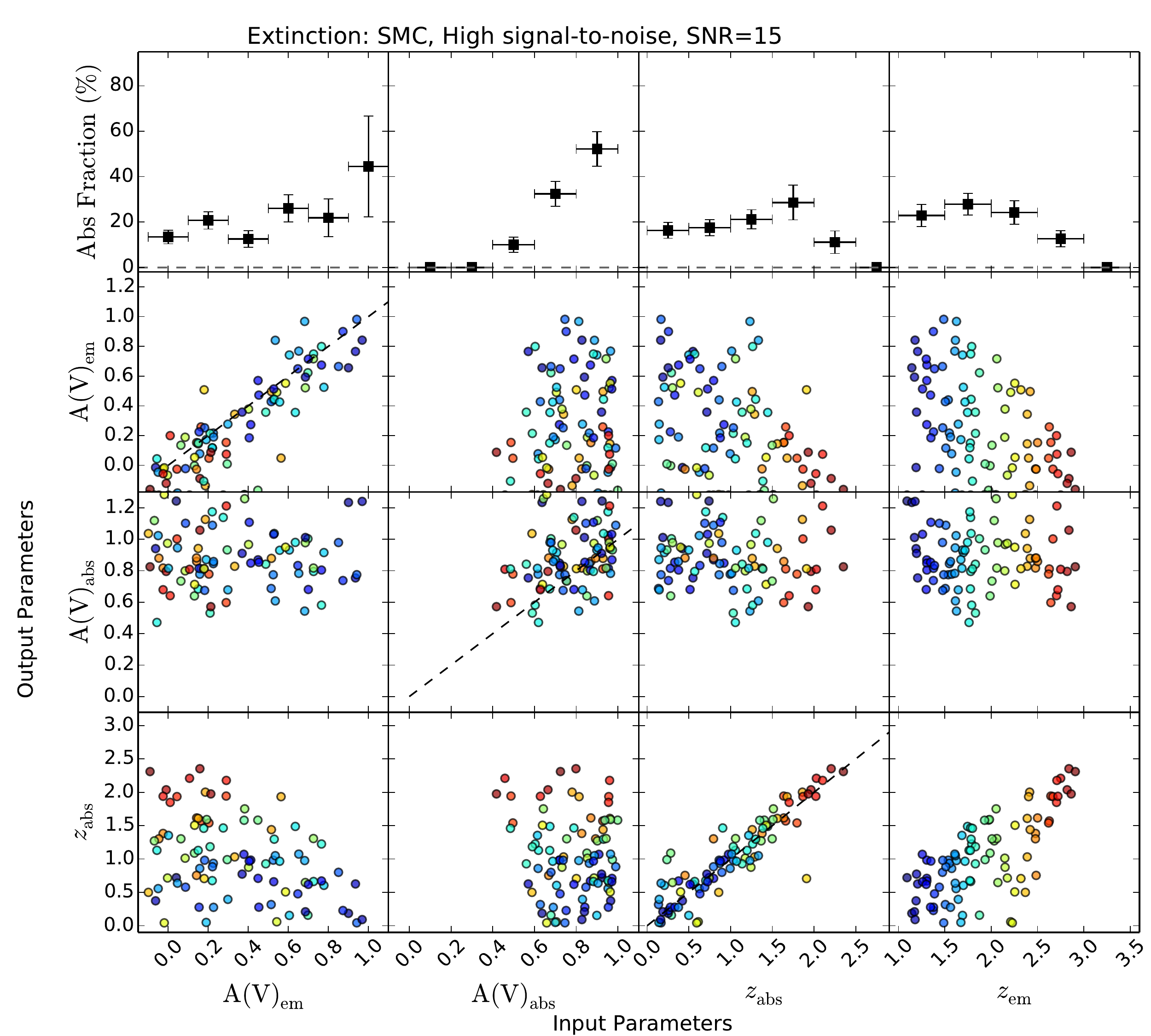}
\caption{Same as Fig.~\ref{fig:LMC_high_SNR} but for high SNR assuming the SMC extinction curve.}
\label{fig:SMC_low_SNR}
\end{figure}

In the following figures, we show the results of the analysis of our synthetic data sets. In Fig.~\ref{fig:null_model}, we show the analysis of the models with dust added only at the QSO redshift. We plot the difference between input parameter and the recovered best-fit parameter as function of input extinction and input QSO redshift. We observe no bias and the $1\,\sigma$ scatter in the recovered A(V) is $\pm0.010$ and $\pm0.019$ for high and low SNR, respectively.
In Fig.~\ref{fig:LMC_high_SNR}, \ref{fig:LMC_low_SNR}, and \ref{fig:SMC_low_SNR}, we show the best-fit parameters (output) as functions of the model parameters (input). We also show the fraction of correctly identified systems using the likelihood ratio test. We only show the results for SMC at high SNR since no absorbers are recovered at low SNR.

\clearpage

%%%%%%%%%%%%%%%%%%%%%%%%%%
%%  ALL SPECTRA

\section{Individual SEDs}
\label{appendix:SEDs}

In Figure C1 we show the individual spectral energy distributions for all targets in the sample. For QSOs in the sample we also shown the QSO continuum template from \citet{Richards06}
in blue, and the same continuum template reddened by the amount of dust inferred from the spectral modeling (shown in red). All available photometry from SDSS, UKIDSS, and WISE is shown.

\begin{figure}[h]
\plotone{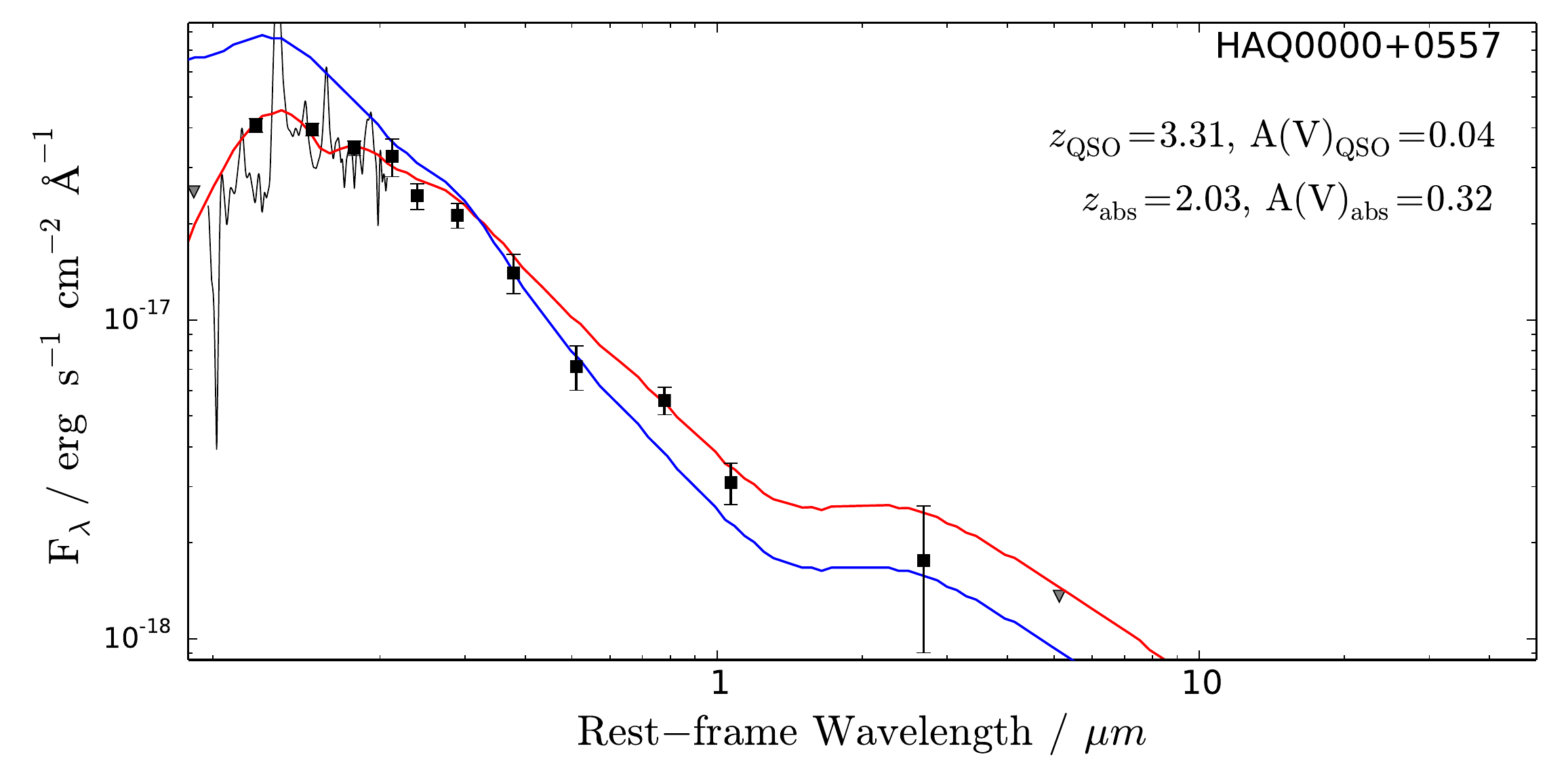}
\caption{Spectral energy distribution from photometry from SDSS, UKIDSS and WISE.
The NOT spectra have been smoothed for visual purposes. We show the
continuum template from \citet{Richards06} reddened by the amount of
reddening inferred from fitting the spectra ({\it red line}). In blue we show
the unreddened template. Upper limits (2\,$\sigma$) are shown as gray triangles.
(The full set of figures is available on the survey webpage http://www.dark-cosmology.dk/$\sim$krogager/redQSOs/data.html)}
\label{fig:SEDs}
\end{figure}

\clearpage

\section{QSOs with intervening absorbers from statistical modeling}
\label{appendix:intervening}

The remaining targets where an intervening dusty systems was identified with our statistical
algorithm (see Section~\ref{model_comparison}) are shown in Figure 15. We show both the simple model (in blue) with
dust only in the quasar itself and the general model (in red) with dust in both the quasar and the absorber.

\begin{figure}[h]
\plotone{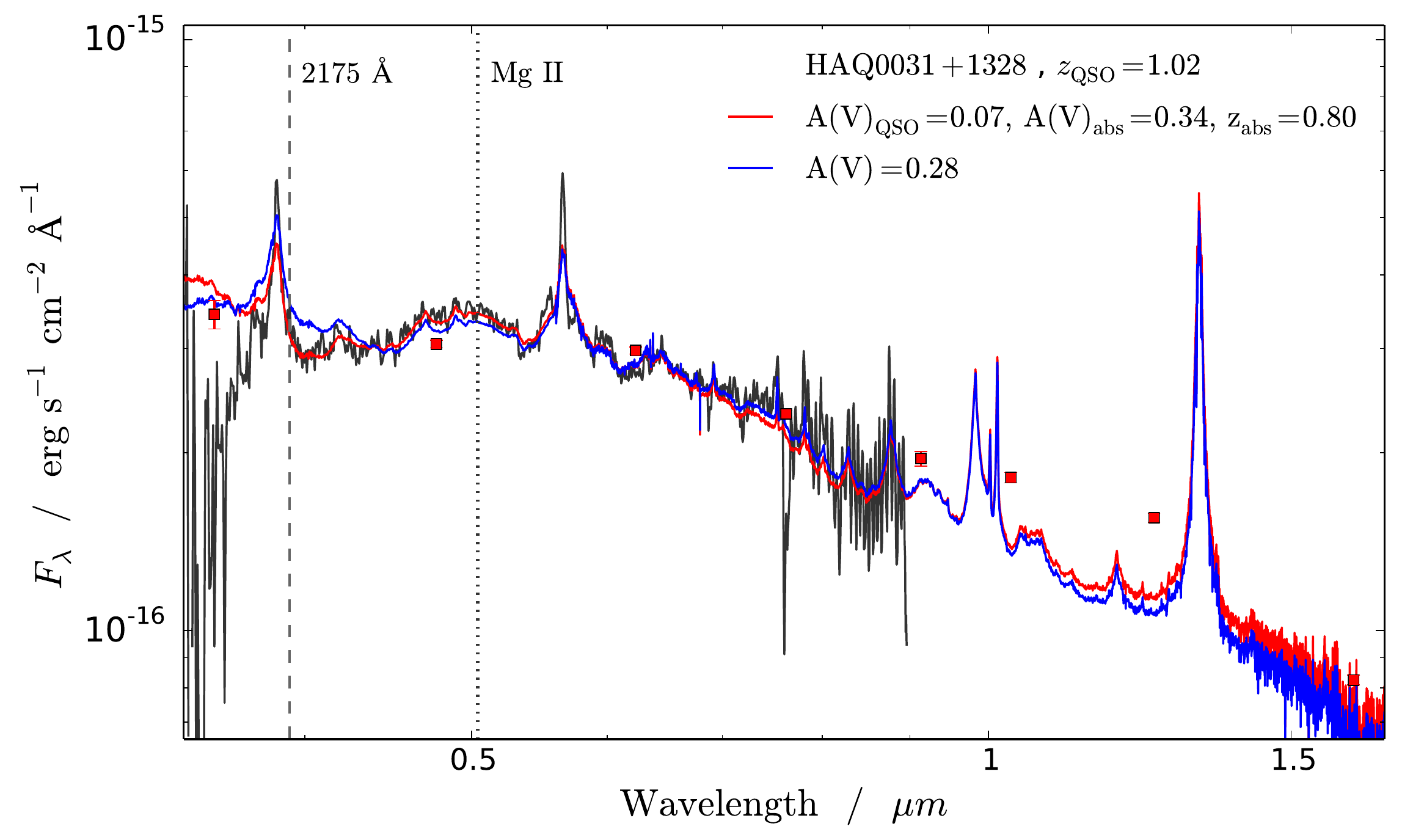}
\caption{Spectra and photometry for the QSOs with preferred dust in an intervening absorption system. The figure shows the spectral data in black and the SDSS and UKIDSS photometry as red squares. In cases where both SDSS and NOT spectra are available, we show the SDSS spectrum as black and the NOT spectrum in gray since the SDSS spectrum was used in the analysis. The SDSS spectrum, if available, has been smoothed with a 3-pixel Gaussian kernel for presentation purposes. The blue and red templates show the best-fit null model and general model, respectively. The dashed and dotted vertical lines indicate the locations of the 2175\,{\AA} bump and Mg\,{\sc ii} at the best-fit redshift for the absorber.
}
\end{figure}

\begin{figure}[b]
\epsscale{0.9}
\plotone{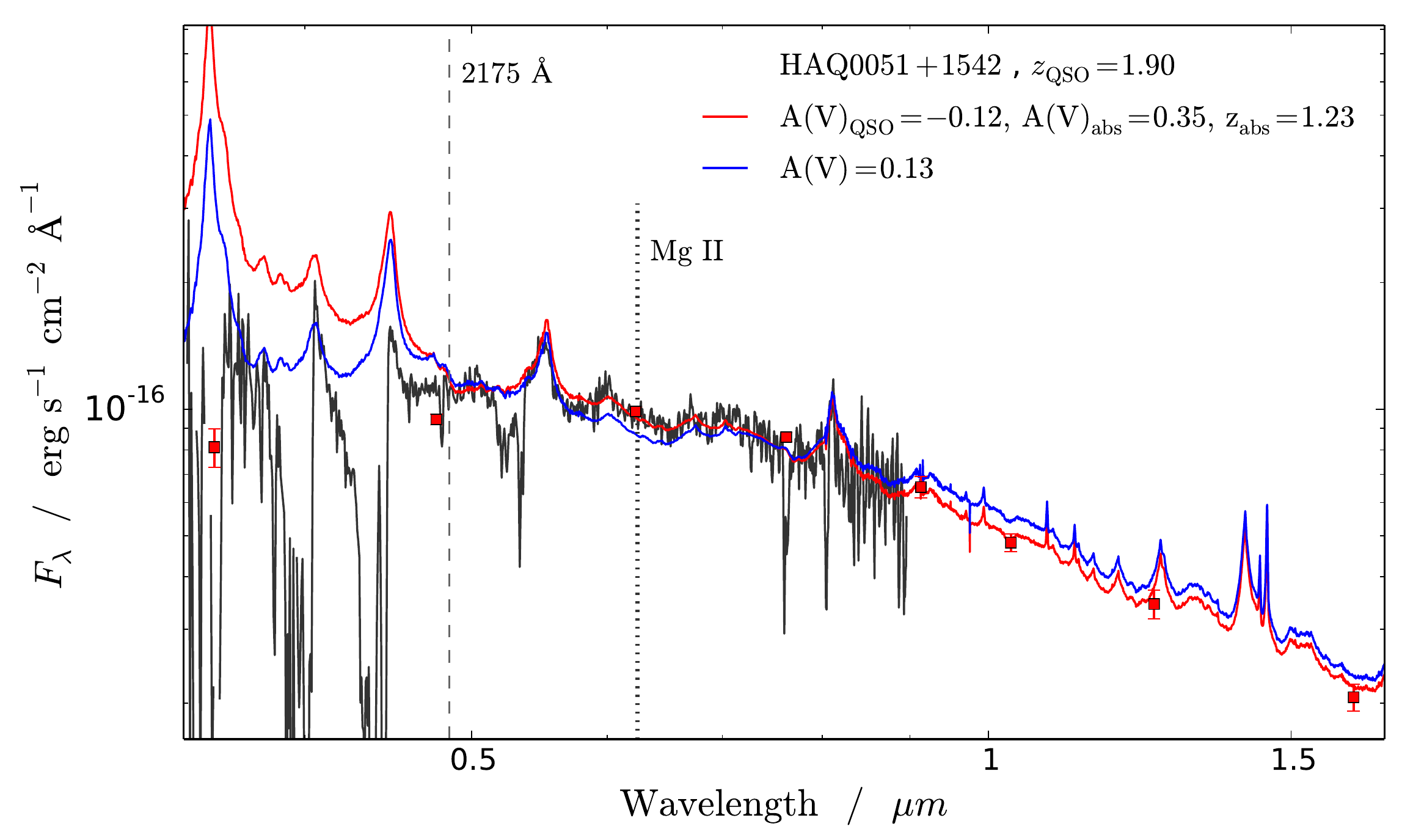}
\figurenum{15}

\end{figure}
\begin{figure}[b]
\plotone{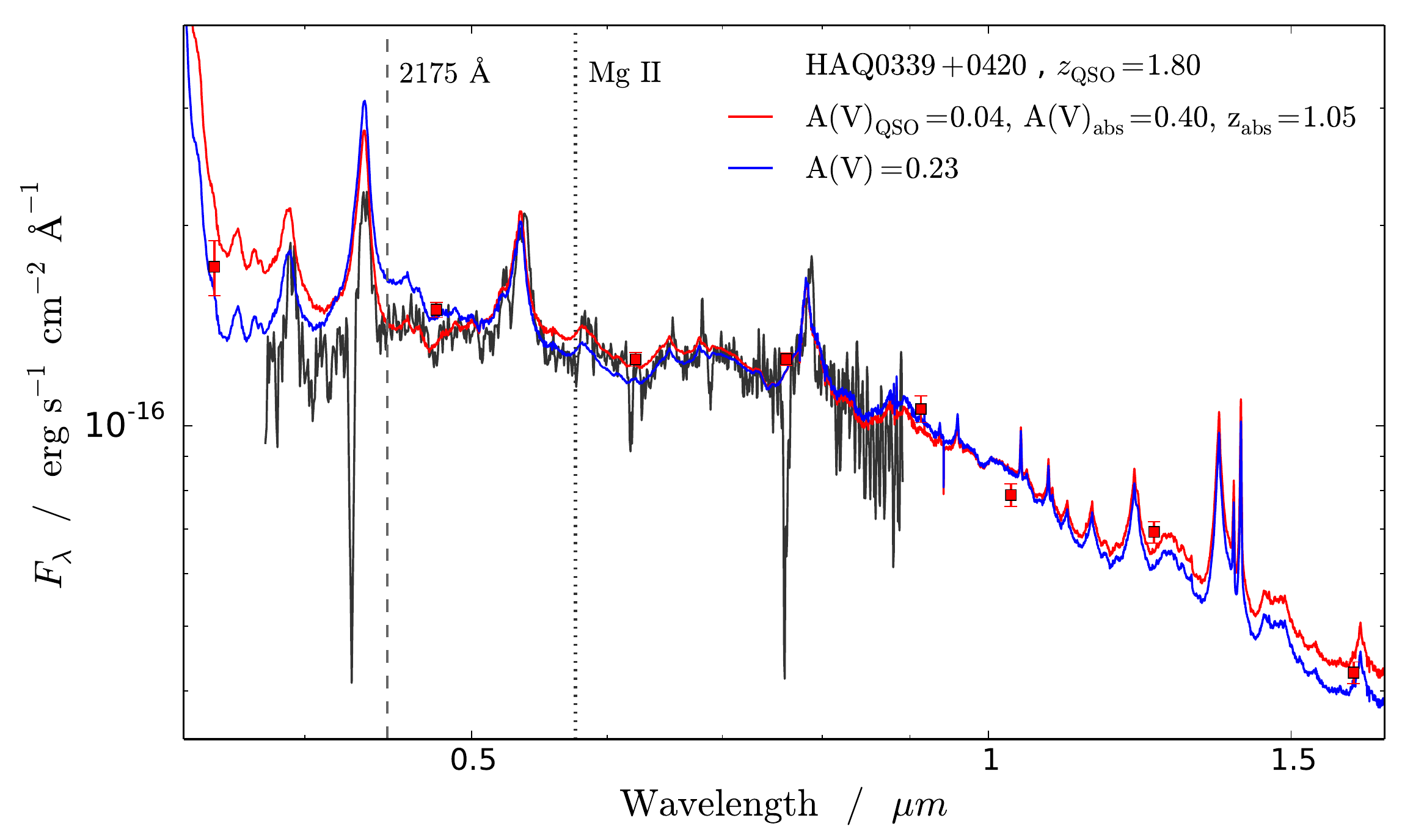}
\figurenum{15}
\caption{Continued.}
\end{figure}
\begin{figure}
\plotone{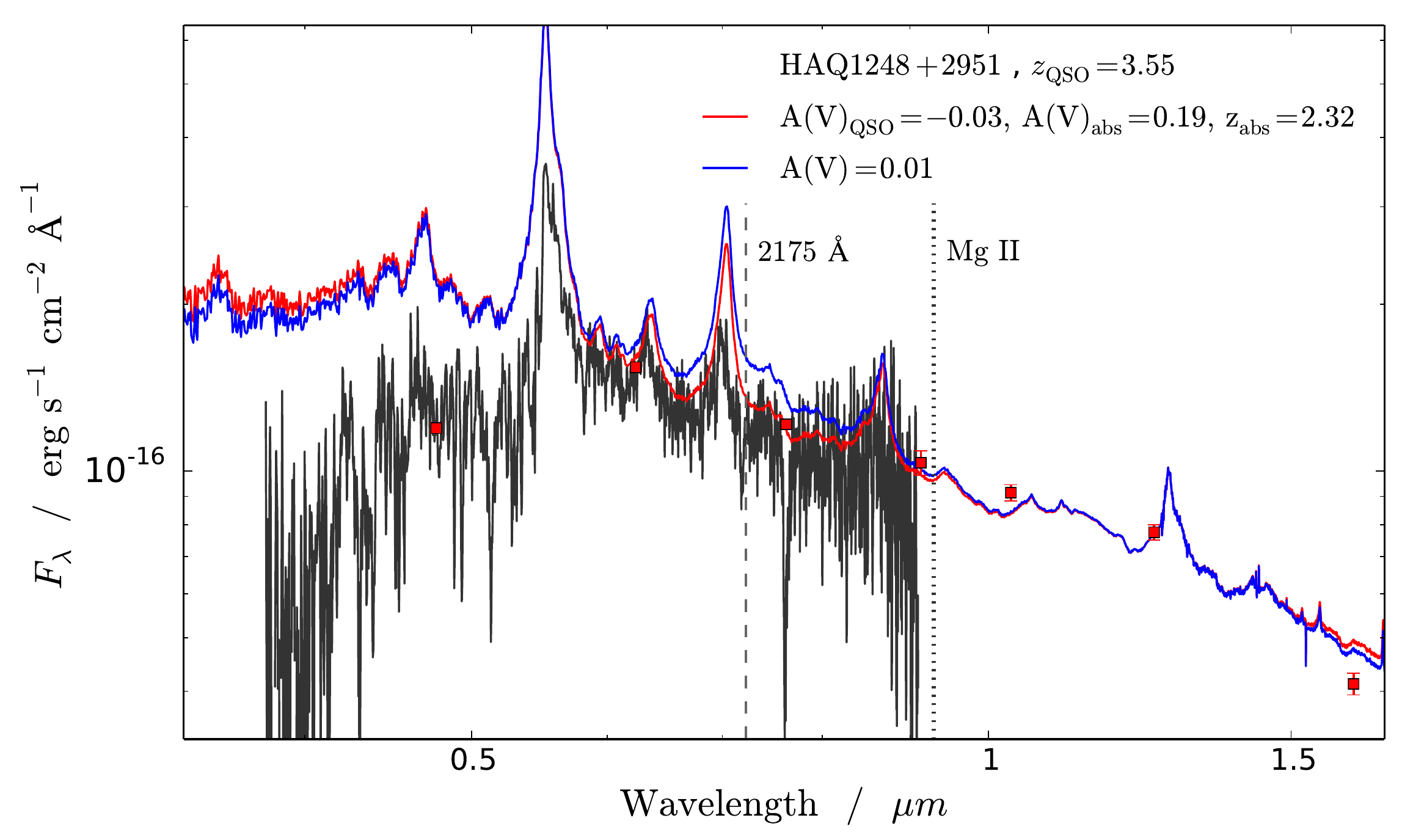}
\figurenum{15}

\end{figure}
\begin{figure}
\plotone{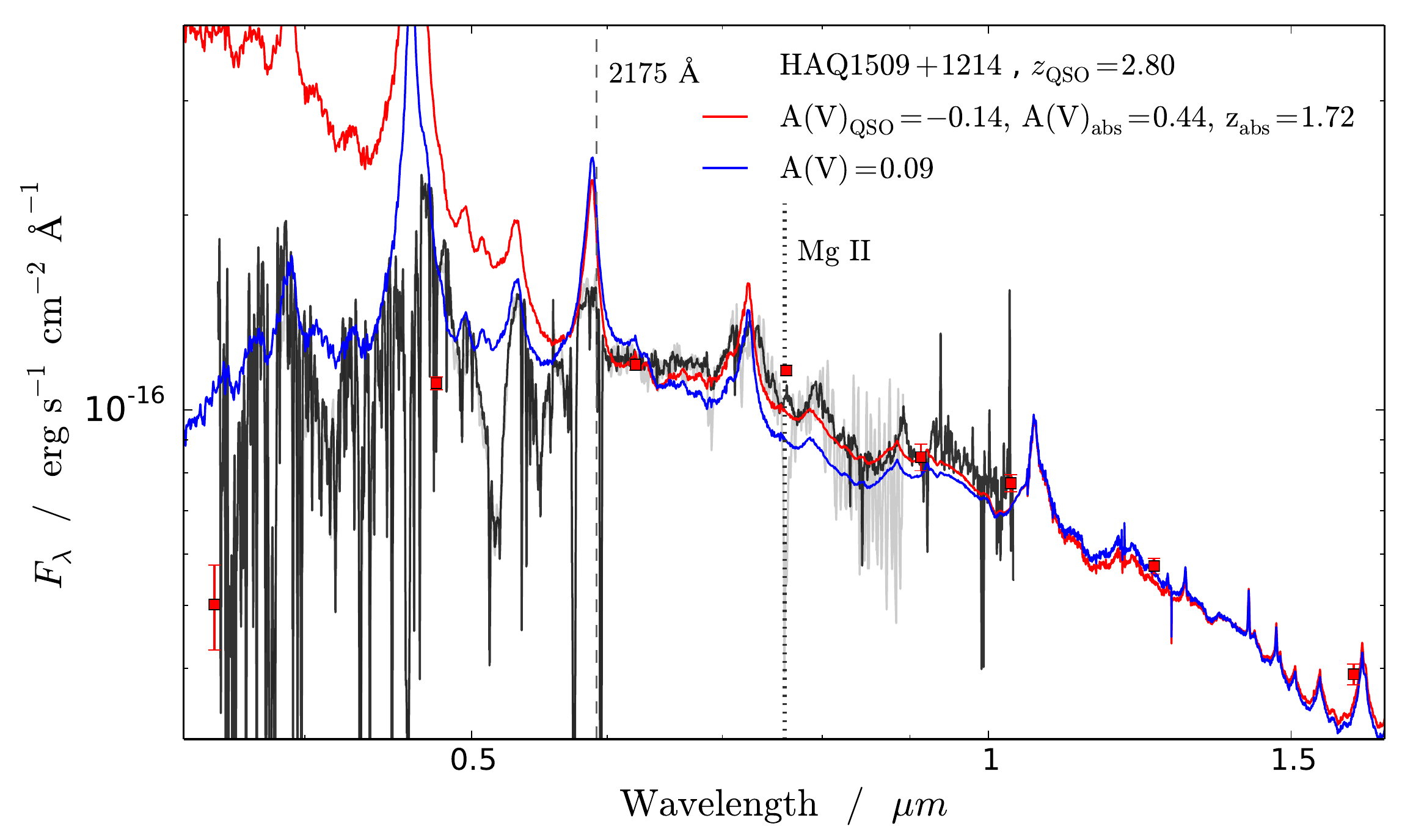}
\figurenum{15}
\caption{Continued.}
\end{figure}

\begin{figure}
\plotone{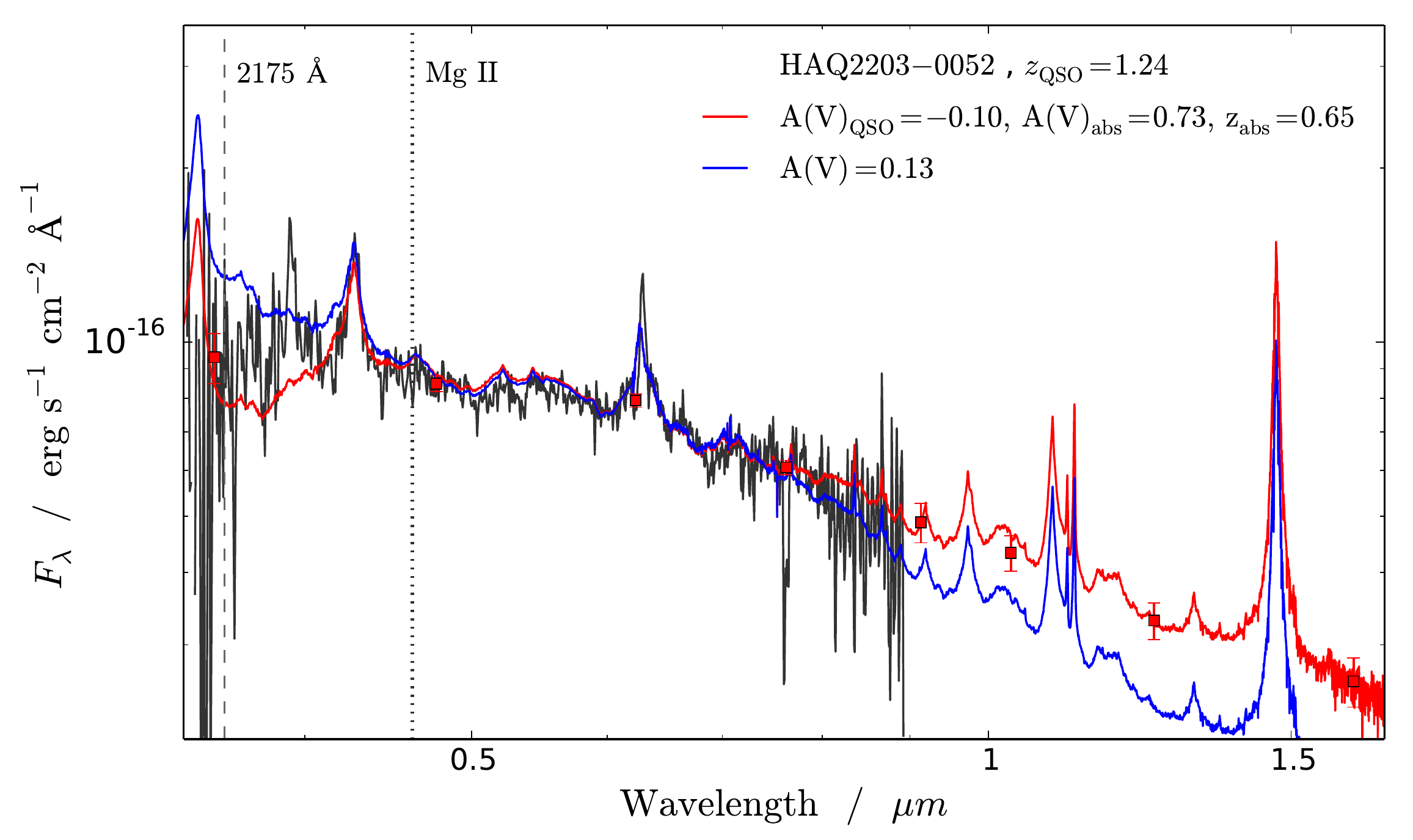}
\figurenum{15}

\end{figure}
\begin{figure}
\plotone{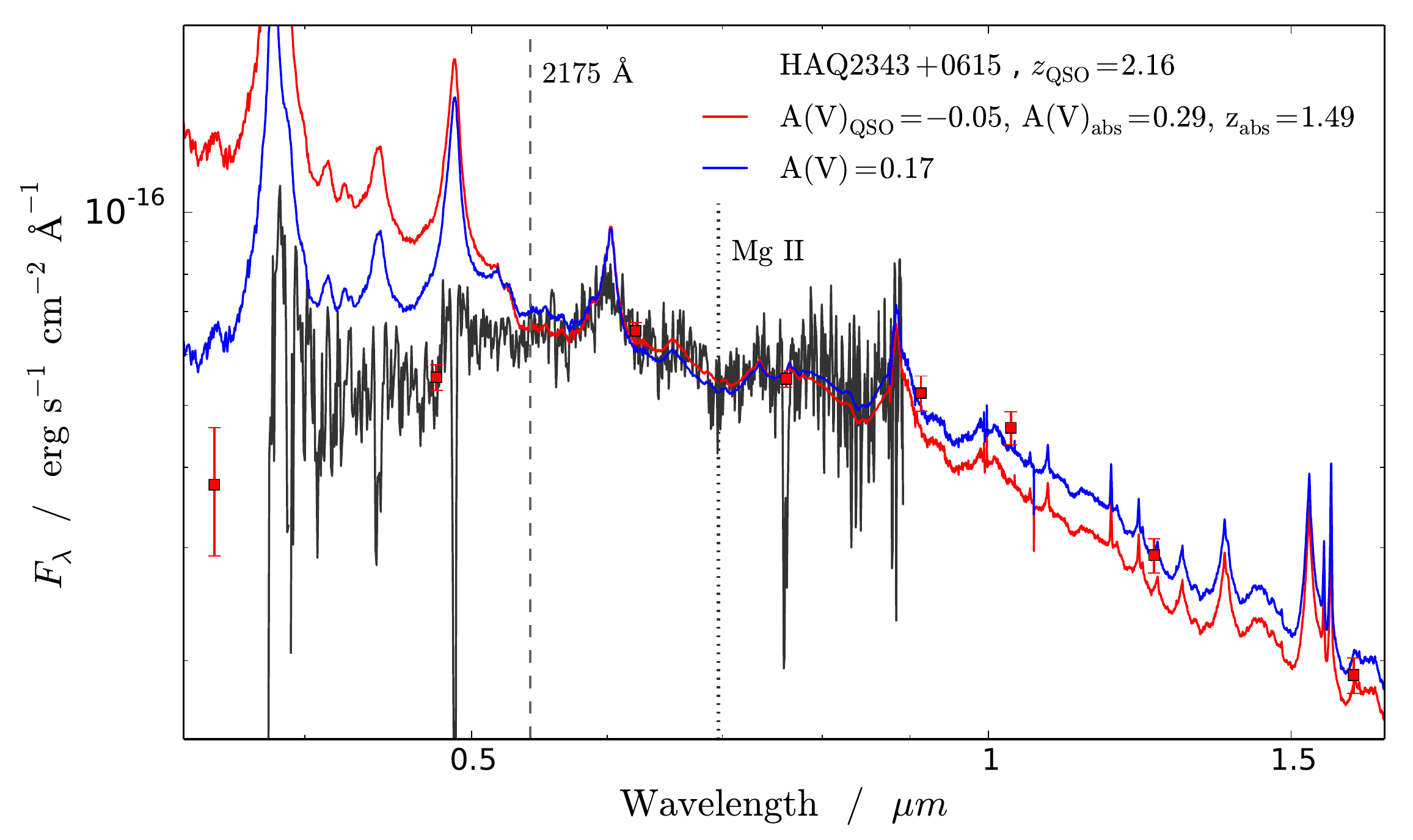}
\figurenum{15}
\caption{Continued.}
\end{figure}

\clearpage

\section{Spectra not classified as QSOs}
\label{appendix:contaminants}

The three targets that were not classified as QSOs are shown in Figures 16, 17, and 18.
The first two targets (Figures 16 and 17) are stellar spectra that entered the sample
by mistake. The first target (HAQ1524-0053) is a Galactic G dwarf, which was selected
due to an error in the $K_s$ band photometry. The second target (HAQ2254+0638) was observed by mistake, as the wrong target in the field was centred on the slit. The last target (HAQ1607+2611) remains unidentified.

\begin{figure}
\plotone{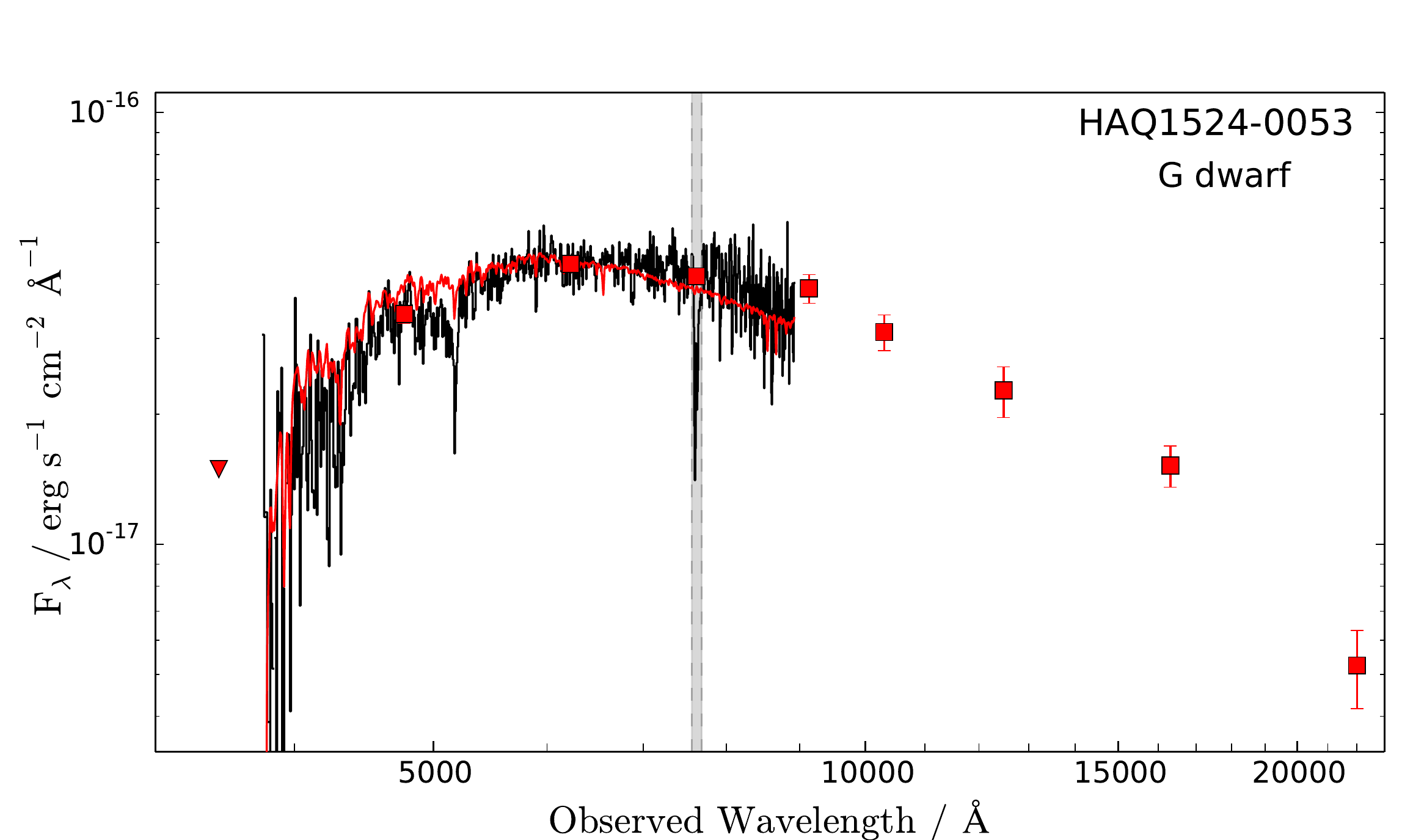}
\figurenum{16}
\caption{The observed spectrum is plotted as a solid black line. The red line shows a stellar template reddened by A(V)=0.52 mag (assuming Milky Way type extinction) to match the observed spectrum. The reddening is the measured extinction from the \citet{Schlegel98} dust map.
Overplotted with filled squares are the SDSS and UKIDSS photometric data points.
The NOT spectrum has been scaled to match the $r$-band photometric data point from SDSS. Note that the spectrum has not been corrected for telluric absorption (marked with a gray band at $\sim7600$~\AA).}
\end{figure}

\begin{figure}
\plotone{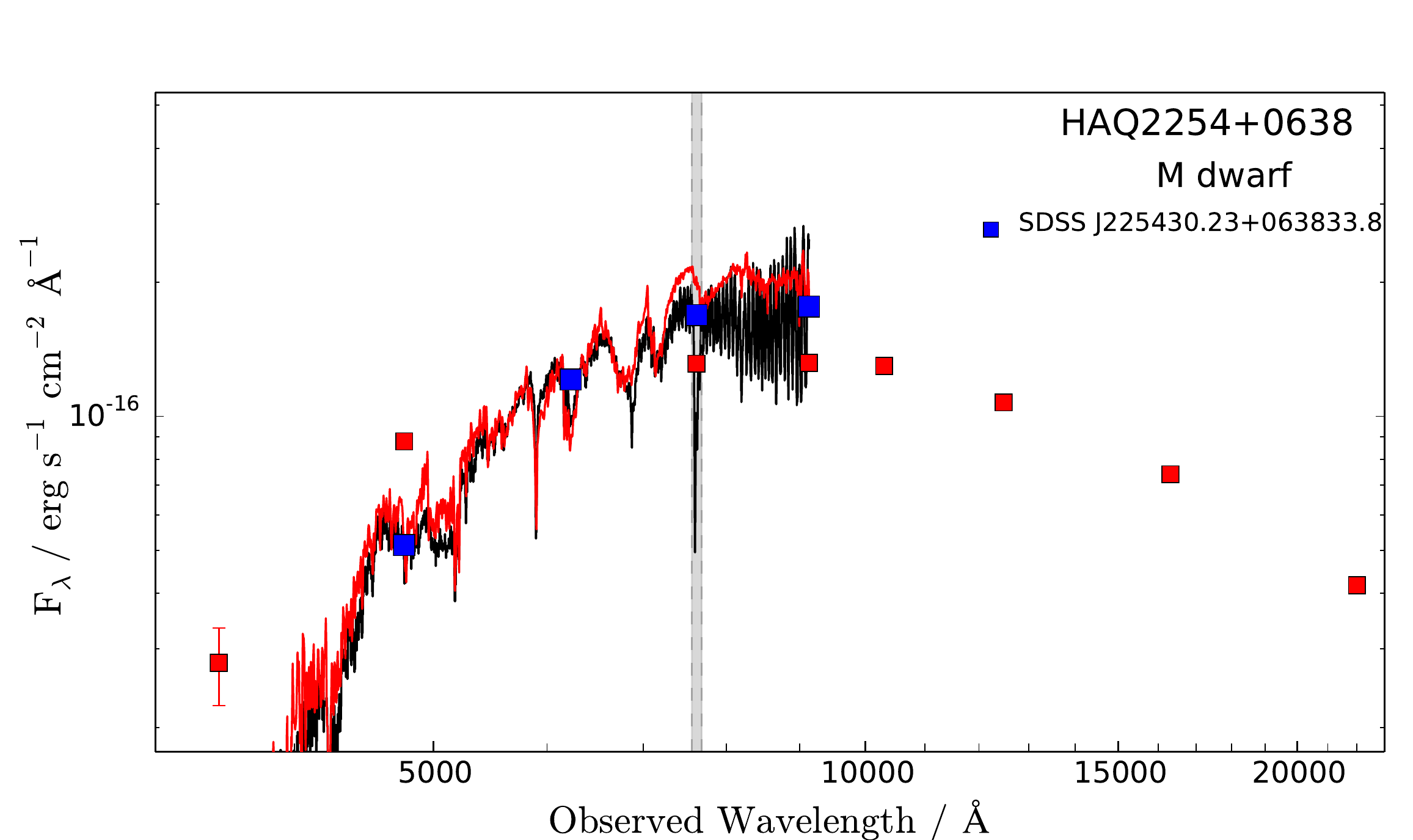}
\figurenum{17}
\caption{The observed spectrum is plotted as a solid black line. The red line shows a stellar template to match the observed spectrum. During the observations of HAQ2254+0638 a nearby star was accidentally placed on the slit instead of the QSO candidate. In red squares, we show the SDSS and UKIDSS photometric data points for the HAQ candidate and in blue squares we show the SDSS photometry of the actual star. The SDSS identifier of the star is given in the upper right corner. The mix up explains the apparent mismatch between the photometry of the candidate (in red squares) and the spectrum.
The NOT spectrum and the blue photometric points have been scaled to match the $r$-band photometric data point (red square) from SDSS, since both are almost 2 orders of magnitude brighter than the candidate photometry. Note that the spectrum has not been corrected for telluric absorption (marked with a gray band at $\sim7600$~\AA). 
}
\end{figure}

\begin{figure}
\plotone{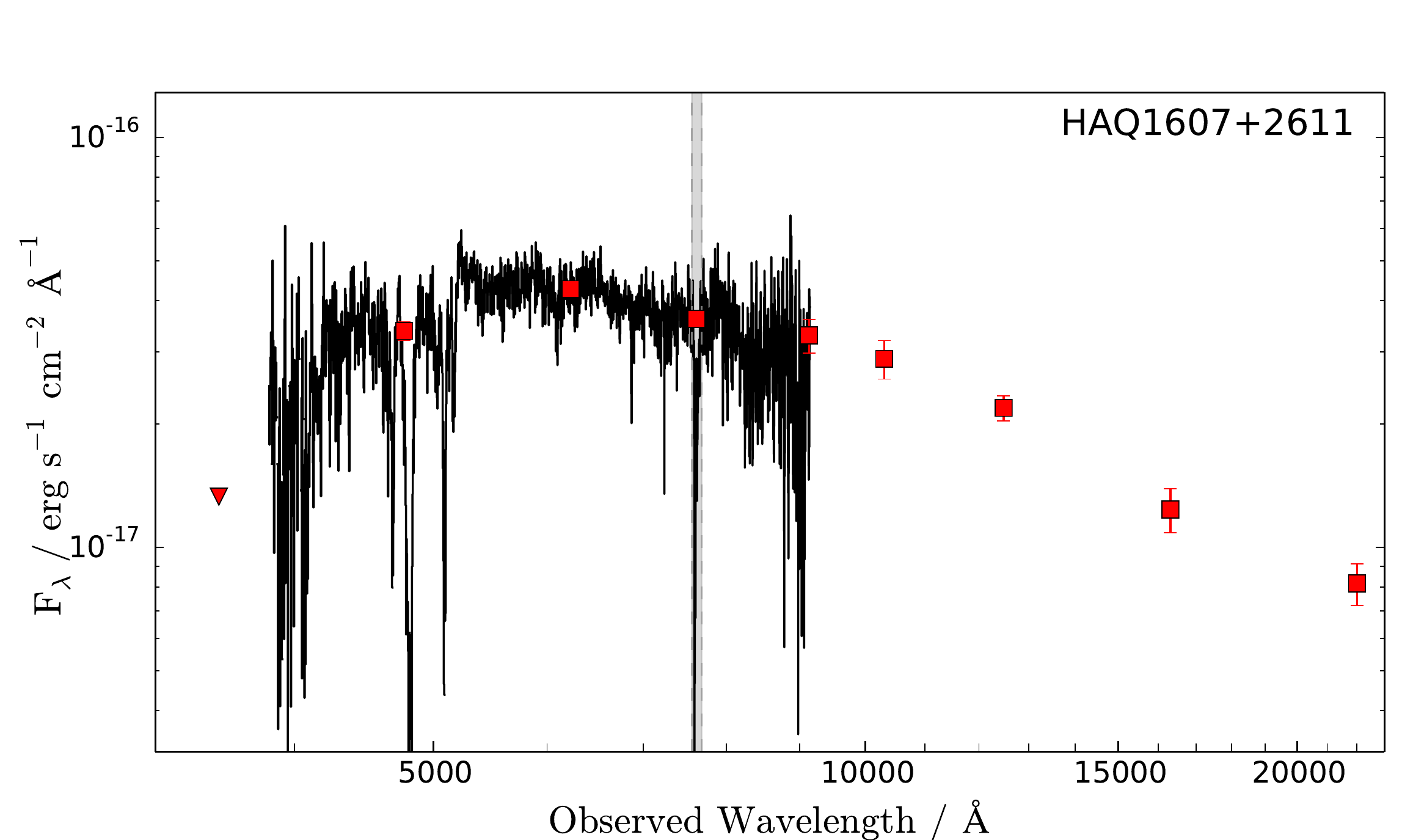}
\figurenum{18}
\caption{Unidentified target. The observed spectrum is plotted as a solid black line.
Overplotted with filled squares are the SDSS and UKIDSS photometric data points. In this case, we have not been able to securely classify the target.
The NOT spectrum has been scaled to match the $r$-band data from SDSS. Note that the spectrum has not been corrected for telluric absorption (marked with a gray band at $\sim7600$~\AA).
}
\end{figure}

\clearpage

\section{Candidates}
\label{candidates}

All the remaining candidates from our sample that were not followed-up with spectroscopy
are listed in Table~\ref{candidatetab}.

\begin{deluxetable}{lccc}
\tablecaption{Remaining Candidate QSOs.
\label{candidatetab}}
\tablewidth{0pt}
\tablehead{
\colhead{Target} & \colhead{RA (J\,2000)} & \colhead{Dec (J\,2000)} & \colhead{$r_{\rm SDSS}$ (AB)} \\ 
}
\startdata
HAQ0000+1250  &  00 00 08.483 & +12 50 33.48 &  18.84 \\ 
HAQ0001+0411  &  00 01 43.455 & +04 11 55.69 &  19.44 \\ 
HAQ0001+0431  &  00 01 42.854 & +04 31 38.89 &  19.44 \\ 
HAQ0003+0221  &  00 03 43.674 & +02 21 29.39 &  19.17 \\ 
HAQ0012+1535  &  00 12 15.986 & +15 35 51.08 &  19.71 \\ 
HAQ0014+1403  &  00 14 39.454 & +14 03 07.74 &  19.93 \\ 
HAQ0015+1213  &  00 15 10.919 & +12 13 53.30 &  19.85 \\ 
HAQ0019+0657  &  00 19 57.289 & +06 57 45.71 &  19.42 \\ 
HAQ0021+0633  &  00 21 02.642 & +06 33 25.69 &  19.50 \\ 
HAQ0022+1049  &  00 22 44.368 & +10 49 02.85 &  20.09 \\ 
\enddata
\tablecomments{Table \ref{candidatetab} is published in its entirety in the electronic edition of the journal. A portion is shown here for guidance regarding its form and content. The full table is available on the survey homepage: {\tt http://dark-cosmology.dk/$\sim$krogager/redQSOs/targets.html}
}
\end{deluxetable}

\end{document}

%% file: redQSOs_table1.tex
\begin{deluxetable}{@{}lccclr@{}}
\tablecaption{Full sample observed with the NOT.\label{sample}}
\tablewidth{0pt}
\tablehead{
\colhead{Target} & \colhead{RA (J\,2000)} & \colhead{Dec (J\,2000)} & \colhead{$r_{\rm{SDSS}}$} & \colhead{Telescope} & \colhead{Exptime} \\ 
\colhead{} & \colhead{} & \colhead{} & \colhead{(mag)} & \colhead{} & \colhead{(sec)} 
}
\startdata
HAQ\,0000+0557 & 00 00 33.979 & +05 57 53.77 & 19.75 & NOT, BOSS & 3$\times$780 \\ 
HAQ\,0001+0233 & 00 01 21.685 & +02 33 04.87 & 17.86 & NOT & 2$\times$600 \\ 
HAQ\,0008+0835 & 00 08 15.020 & +08 35 25.68 & 18.24 & NOT & 2$\times$450 \\ 
HAQ\,0008+0846 & 00 08 25.327 & +08 46 26.20 & 18.81 & NOT & 2$\times$600 \\ 
HAQ\,0011+0122 & 00 11 04.665 & +01 22 56.02 & 18.28 & NOT & 900 \\ 
HAQ\,0012+0651 & 00 12 55.339 & +06 51 22.76 & 18.70 & NOT & 2$\times$600 \\ 
HAQ\,0012+0657 & 00 12 43.499 & +06 57 36.74 & 18.98 & NOT & 2$\times$600 \\ 
HAQ\,0012+0944 & 00 12 01.587 & +09 44 02.21 & 18.13 & NOT & 2$\times$500 \\ 
HAQ\,0014+0444 & 00 14 05.324 & +04 44 09.35 & 17.75 & NOT & 2$\times$600 \\ 
HAQ\,0014+0939 & 00 14 13.425 & +09 39 06.47 & 18.77 & NOT, BOSS & 2$\times$600 \\ 
HAQ\,0015+0736 & 00 15 35.496 & +07 36 37.54 & 19.09 & NOT, BOSS & 3$\times$600 \\ 
HAQ\,0015+0811 & 00 15 31.492 & +08 11 36.96 & 19.00 & NOT & 2$\times$600 \\ 
HAQ\,0015+1129 & 00 15 22.046 & +11 29 59.89 & 17.89 & NOT & 2$\times$600 \\ 
HAQ\,0015+1340 & 00 15 58.255 & +13 40 05.37 & 19.61 & NOT & 3$\times$600 \\ 
HAQ\,0018+1133 & 00 18 25.423 & +11 33 01.62 & 19.03 & NOT & 2$\times$600 \\ 
HAQ\,0020+0259 & 00 20 20.401 & +02 59 13.88 & 19.03 & NOT & 2$\times$600 \\ 
HAQ\,0022+0147 & 00 22 24.417 & +01 47 31.23 & 17.52 & NOT & 3$\times$250 \\ 
HAQ\,0024+1037 & 00 24 06.099 & +10 37 58.03 & 17.06 & NOT & 3$\times$250 \\ 
HAQ\,0025+0220 & 00 25 10.345 & +02 20 06.35 & 18.93 & NOT & 2$\times$600 \\ 
HAQ\,0026+0640 & 00 26 12.462 & +06 40 36.80 & 18.83 & NOT & 2$\times$600 \\ 
HAQ\,0031+1328 & 00 31 16.660 & +13 28 48.44 & 17.60 & NOT & 2$\times$600, 300 \\ 
HAQ\,0033+0915 & 00 33 57.209 & +09 15 53.66 & 19.23 & NOT, BOSS & 3$\times$600 \\ 
HAQ\,0034+0950 & 00 34 28.096 & +09 50 20.64 & 18.33 & NOT & 900 \\ 
HAQ\,0038+1426 & 00 38 06.431 & +14 26 02.78 & 18.79 & NOT & 2$\times$600 \\ 
HAQ\,0042+1220 & 00 42 15.000 & +12 20 08.00 & 19.24 & NOT & 3$\times$600 \\ 
HAQ\,0043+0549 & 00 43 16.644 & +05 49 42.33 & 18.77 & NOT & 2$\times$600 \\ 
HAQ\,0043+1136 & 00 43 28.876 & +11 36 26.49 & 19.26 & NOT & 3$\times$540 \\ 
HAQ\,0044+0817 & 00 44 30.382 & +08 17 13.79 & 18.84 & NOT, BOSS & 3$\times$400 \\ 
HAQ\,0044+1250 & 00 44 45.699 & +12 50 19.86 & 19.17 & NOT & 2$\times$900 \\ 
HAQ\,0045+1217 & 00 45 43.339 & +12 17 11.83 & 17.43 & NOT & 3$\times$400 \\ 
HAQ\,0046+0839 & 00 46 33.895 & +08 39 13.78 & 19.71 & NOT, BOSS & 3$\times$720 \\ 
HAQ\,0047+0826 & 00 47 37.196 & +08 26 38.47 & 17.29 & NOT & 4$\times$200 \\ 
HAQ\,0051+1542 & 00 51 54.685 & +15 42 05.90 & 18.83 & NOT & 2$\times$600 \\ 
HAQ\,0053+0216 & 00 53 36.998 & +02 16 36.60 & 18.58 & NOT & 2$\times$600 \\ 
HAQ\,0056+1132 & 00 56 45.104 & +11 32 38.72 & 19.11 & NOT & 3$\times$600 \\ 
HAQ\,0057+1155 & 00 57 31.960 & +11 55 12.13 & 18.61 & NOT & 2$\times$600 \\ 
HAQ\,0059+1238 & 00 59 48.164 & +12 38 36.79 & 18.75 & NOT & 2$\times$600 \\ 
HAQ\,0102+0249 & 01 02 16.770 & +02 49 52.52 & 18.68 & NOT & 600 \\ 
HAQ\,0110+0303 & 01 10 13.499 & +03 03 56.17 & 18.91 & NOT & 2$\times$600 \\ 
HAQ\,0118+0323 & 01 18 13.583 & +03 23 34.30 & 18.62 & NOT & 2$\times$600 \\ 
HAQ\,0118+0700 & 01 18 57.315 & +07 00 29.15 & 19.16 & NOT, BOSS & 2$\times$600 \\ 
HAQ\,0119+0817 & 01 19 09.186 & +08 17 53.87 & 19.08 & NOT & 2$\times$900 \\ 
HAQ\,0120+0351 & 01 20 30.091 & +03 51 46.88 & 18.95 & NOT & 3$\times$480 \\ 
HAQ\,0121+0455 & 01 21 45.522 & +04 55 04.50 & 18.16 & NOT & 2$\times$900 \\ 
HAQ\,0122+0325 & 01 22 09.876 & +03 25 43.64 & 17.87 & NOT & 600 \\ 
HAQ\,0130+1439 & 01 30 16.520 & +14 39 53.71 & 19.93 & NOT & 4$\times$600 \\ 
HAQ\,0138+0124 & 01 38 02.071 & +01 24 24.47 & 18.26 & NOT, BOSS & 2$\times$600 \\ 
HAQ\,0138+0636 & 01 38 19.887 & +06 36 36.05 & 18.92 & NOT & 2$\times$600 \\ 
HAQ\,0143+1509 & 01 43 17.885 & +15 09 15.41 & 19.35 & NOT & 2$\times$1000 \\ 
HAQ\,0151+0618 & 01 51 36.733 & +06 18 31.67 & 19.16 & NOT & 3$\times$600 \\ 
HAQ\,0151+1453 & 01 51 57.714 & +14 53 08.34 & 19.53 & NOT & 3$\times$360 \\ 
HAQ\,0155+0438 & 01 55 03.728 & +04 38 30.42 & 18.38 & NOT & 900 \\ 
HAQ\,0201+0223 & 02 01 14.291 & +02 23 33.29 & 19.07 & NOT & 1080 \\ 
HAQ\,0204+0327 & 02 04 58.663 & +03 27 03.22 & 18.55 & NOT & 2$\times$600 \\ 
HAQ\,0206+0624 & 02 06 14.665 & +06 24 54.44 & 19.01 & NOT & 2$\times$600 \\ 
HAQ\,0208+0521 & 02 08 15.802 & +05 21 06.01 & 19.10 & NOT & 2$\times$900 \\ 
HAQ\,0211+1214 & 02 11 23.395 & +12 14 01.20 & 18.32 & NOT & 2$\times$600 \\ 
HAQ\,0226+0729 & 02 26 50.310 & +07 29 52.78 & 19.06 & NOT & 2$\times$900 \\ 
HAQ\,0236+0619 & 02 36 46.962 & +06 19 03.15 & 19.03 & NOT & 2$\times$600 \\ 
HAQ\,0243+0355 & 02 43 29.118 & +03 55 59.50 & 17.84 & NOT & 3$\times$240 \\ 
HAQ\,0318+0424 & 03 18 16.302 & +04 24 06.35 & 19.50 & NOT & 5$\times$400 \\ 
HAQ\,0318+0434 & 03 18 33.519 & +04 34 43.48 & 19.20 & NOT & 1500 \\ 
HAQ\,0319+0623 & 03 19 01.776 & +06 23 39.04 & 18.92 & NOT & 4$\times$450 \\ 
HAQ\,0329+0553 & 03 29 15.624 & +05 53 39.57 & 18.92 & NOT & 2$\times$600 \\ 
HAQ\,0329+0609 & 03 29 09.017 & +06 09 14.30 & 19.14 & NOT & 4$\times$500 \\ 
HAQ\,0337+0539 & 03 37 38.075 & +05 39 10.95 & 18.18 & NOT & 2$\times$600 \\ 
HAQ\,0339+0420 & 03 39 30.630 & +04 20 31.11 & 18.75 & NOT & 4$\times$360 \\ 
HAQ\,0340+0408 & 03 40 14.148 & +04 08 31.94 & 19.36 & NOT & 4$\times$600 \\ 
HAQ\,0345$-$0009 & 03 45 50.472 & $-$00 09 07.56 & 18.53 & NOT & 4$\times$375 \\ 
HAQ\,0347+0115 & 03 47 48.060 & +01 15 44.52 & 19.06 & NOT & 2$\times$600 \\ 
HAQ\,0355$-$0025 & 03 55 52.570 & $-$00 25 04.22 & 18.70 & NOT & 4$\times$400 \\ 
HAQ\,0355$-$0053 & 03 55 46.908 & $-$00 53 39.83 & 19.46 & NOT & 2$\times$780 \\ 
HAQ\,1106+0300 & 11 06 12.667 & +03 00 49.10 & 17.62 & NOT & 3$\times$300 \\ 
HAQ\,1114+1330 & 11 14 15.151 & +13 30 59.64 & 17.59 & NOT & 3$\times$400 \\ 
HAQ\,1115+0333 & 11 15 49.737 & +03 33 51.35 & 18.89 & NOT & 3$\times$600 \\ 
HAQ\,1148$-$0117 & 11 48 22.193 & $-$01 17 29.20 & 18.02 & NOT & 2$\times$500, 3$\times$400 \\ 
HAQ\,1207+1341 & 12 07 59.229 & +13 41 15.28 & 18.40 & NOT & 3$\times$500 \\ 
HAQ\,1233+1304 & 12 33 55.605 & +13 04 09.21 & 17.71 & NOT & 3$\times$250 \\ 
HAQ\,1247+3403 & 12 47 02.054 & +34 03 58.17 & 18.36 & NOT, BOSS & 2$\times$500 \\ 
HAQ\,1248+2951 & 12 48 48.423 & +29 51 06.73 & 18.19 & NOT & 2$\times$400 \\ 
HAQ\,1315+0440 & 13 15 21.006 & +04 40 00.56 & 18.70 & NOT & 3$\times$500 \\ 
HAQ\,1319+3214 & 13 19 02.712 & +32 14 51.29 & 19.50 & NOT & 4$\times$600 \\ 
HAQ\,1327+3206 & 13 27 57.361 & +32 06 50.63 & 18.55 & NOT & 2$\times$600 \\ 
HAQ\,1332+0052 & 13 32 54.515 & +00 52 50.63 & 18.35 & NOT, BOSS & 2$\times$600 \\ 
HAQ\,1339+3331 & 13 39 41.381 & +33 31 12.64 & 18.57 & NOT & 1200 \\ 
HAQ\,1355+3407 & 13 55 57.499 & +34 07 39.03 & 18.77 & NOT & 2$\times$800 \\ 
HAQ\,1358+2401 & 13 58 59.897 & +24 01 07.00 & 18.86 & NOT & 2$\times$900 \\ 
HAQ\,1400+0219 & 14 00 47.108 & +02 19 34.80 & 19.81 & NOT & 4$\times$600 \\ 
HAQ\,1409+0940 & 14 09 52.589 & +09 40 23.73 & 19.65 & NOT, BOSS & 3$\times$600 \\ 
HAQ\,1411$-$0104 & 14 11 59.671 & $-$01 04 42.60 & 19.28 & NOT & 3$\times$600 \\ 
HAQ\,1434+0448 & 14 34 15.006 & +04 48 46.83 & 19.77 & NOT & 4$\times$750 \\ 
HAQ\,1444+0752 & 14 44 43.539 & +07 52 24.28 & 19.45 & NOT, BOSS & 3$\times$600, 5$\times$750 \\ 
HAQ\,1451+3239 & 14 51 56.221 & +32 39 51.69 & 19.09 & NOT & 2$\times$600 \\ 
HAQ\,1506+0438 & 15 06 29.847 & +04 38 44.27 & 19.97 & NOT & 4$\times$600 \\ 
HAQ\,1509+1214 & 15 09 53.554 & +12 14 44.98 & 18.54 & NOT, BOSS & 2$\times$600 \\ 
HAQ\,1517+0817 & 15 17 55.714 & +08 17 27.66 & 19.15 & NOT & 2$\times$600 \\ 
HAQ\,1524$-$0053 & 15 24 44.652 & $-$00 53 09.70 & 19.50 & NOT & 3$\times$600 \\ 
HAQ\,1527+0250 & 15 27 10.942 & +02 50 19.20 & 17.06 & NOT & 5$\times$200 \\ 
HAQ\,1534+0013 & 15 34 52.677 & +00 13 17.71 & 18.82 & NOT & 2$\times$600 \\ 
HAQ\,1535+0157 & 15 35 53.854 & +01 57 11.36 & 18.86 & NOT & 2$\times$600 \\ 
HAQ\,1545$-$0130 & 15 45 49.013 & $-$01 30 09.26 & 19.34 & NOT & 3$\times$600 \\ 
HAQ\,1546+0005 & 15 46 58.586 & +00 05 38.33 & 19.01 & NOT, BOSS & 2$\times$750 \\ 
HAQ\,1600+2911 & 16 00 33.974 & +29 11 16.46 & 19.82 & NOT & 4$\times$900 \\ 
HAQ\,1603+2512 & 16 03 28.589 & +25 12 14.33 & 20.18 & NOT & 2$\times$1350 \\ 
HAQ\,1606+2902 & 16 06 01.112 & +29 02 18.85 & 19.58 & NOT & 4$\times$600 \\ 
HAQ\,1606+2903 & 16 06 28.064 & +29 03 33.80 & 17.53 & NOT, BOSS & 3$\times$480 \\ 
HAQ\,1607+2611 & 16 07 21.764 & +26 11 07.06 & 19.67 & NOT & 4$\times$600 \\ 
HAQ\,1611+2453 & 16 11 41.495 & +24 53 22.21 & 19.69 & NOT & 3$\times$600 \\ 
HAQ\,1620+2955 & 16 20 40.932 & +29 55 06.56 & 19.21 & NOT, BOSS & 1200 \\ 
HAQ\,1626+2517 & 16 26 21.102 & +25 17 14.00 & 19.33 & NOT, BOSS & 2$\times$1000 \\ 
HAQ\,1633+2851 & 16 33 50.413 & +28 51 56.77 & 19.45 & NOT & 2$\times$900 \\ 
HAQ\,1634+2811 & 16 34 39.605 & +28 11 38.28 & 19.25 & NOT & 2$\times$900 \\ 
HAQ\,1639+3157 & 16 39 57.963 & +31 57 26.71 & 19.60 & NOT, BOSS & 3$\times$600 \\ 
HAQ\,1643+2944 & 16 43 32.810 & +29 44 23.42 & 19.48 & NOT & 3$\times$600 \\ 
HAQ\,1645+3056 & 16 45 53.184 & +30 56 07.27 & 19.14 & NOT, BOSS & 3$\times$600 \\ 
HAQ\,1645+3130 & 16 45 47.802 & +31 30 03.28 & 19.50 & NOT, BOSS & 3$\times$600 \\ 
HAQ\,1655+3051 & 16 55 23.891 & +30 51 37.71 & 19.38 & NOT & 3$\times$600 \\ 
HAQ\,2159+0212 & 21 59 36.617 & +02 12 33.51 & 18.73 & NOT & 600, 2$\times$600 \\ 
HAQ\,2203$-$0052 & 22 03 08.631 & $-$00 52 34.54 & 19.05 & NOT & 2$\times$900 \\ 
HAQ\,2217+0359 & 22 17 40.515 & +03 59 33.78 & 19.43 & NOT & 2$\times$900 \\ 
HAQ\,2221+0145 & 22 21 43.557 & +01 45 37.37 & 19.05 & NOT, BOSS & 2$\times$600 \\ 
HAQ\,2222+0604 & 22 22 06.684 & +06 04 15.71 & 19.38 & NOT & 2$\times$900 \\ 
HAQ\,2225+0527 & 22 25 14.695 & +05 27 09.10 & 18.11 & NOT & 2$\times$600, 2$\times$450 \\ 
HAQ\,2229+0324 & 22 29 15.168 & +03 24 52.71 & 19.90 & NOT, BOSS & 3$\times$600 \\ 
HAQ\,2231+0509 & 22 31 15.996 & +05 09 48.62 & 19.10 & NOT & 2$\times$900 \\ 
HAQ\,2241+0818 & 22 41 51.844 & +08 18 59.09 & 19.06 & NOT & 2$\times$900 \\ 
HAQ\,2244+0335 & 22 44 53.750 & +03 35 23.29 & 18.66 & NOT, BOSS & 2$\times$600 \\ 
HAQ\,2245+0457 & 22 45 28.459 & +04 57 20.38 & 18.32 & NOT & 2$\times$600, 2$\times$600 \\ 
HAQ\,2246+0710 & 22 46 03.823 & +07 10 50.92 & 18.02 & NOT & 2$\times$600 \\ 
HAQ\,2247+0146 & 22 47 20.277 & +01 46 04.95 & 19.18 & NOT & 5$\times$600 \\ 
HAQ\,2252+0434 & 22 52 45.969 & +04 34 36.75 & 16.03 & NOT & 4$\times$90 \\ 
HAQ\,2253+1141 & 22 53 30.058 & +11 41 18.31 & 19.20 & NOT & 2$\times$600 \\ 
HAQ\,2254+0638 & 22 54 32.268 & +06 38 26.01 & 18.41 & NOT & 600 \\ 
HAQ\,2300+0914 & 23 00 56.010 & +09 14 03.85 & 18.86 & NOT & 2$\times$600 \\ 
HAQ\,2301+0832 & 23 01 22.475 & +08 32 01.43 & 17.26 & NOT & 2$\times$600 \\ 
HAQ\,2303+0238 & 23 03 17.785 & +02 38 09.60 & 18.75 & NOT & 2$\times$600 \\ 
HAQ\,2303+0630 & 23 03 12.032 & +06 30 14.21 & 18.67 & NOT & 2$\times$600 \\ 
HAQ\,2305+0117 & 23 05 48.832 & +01 17 41.63 & 19.55 & NOT & 3$\times$650 \\ 
HAQ\,2310+1117 & 23 10 46.942 & +11 17 21.55 & 20.06 & NOT & 4$\times$600 \\ 
HAQ\,2311+1444 & 23 11 38.547 & +14 44 36.72 & 19.18 & NOT, SDSS & 3$\times$600 \\ 
HAQ\,2313+0955 & 23 13 34.552 & +09 55 00.03 & 18.91 & NOT & 2$\times$600 \\ 
HAQ\,2318+0255 & 23 18 12.807 & +02 55 38.98 & 18.31 & NOT & 2$\times$600 \\ 
HAQ\,2326+0642 & 23 26 48.823 & +06 42 35.96 & 18.79 & NOT & 2$\times$600 \\ 
HAQ\,2326+1423 & 23 26 40.965 & +14 23 03.37 & 19.02 & NOT & 2$\times$450 \\ 
HAQ\,2330+1009 & 23 30 59.933 & +10 09 49.42 & 19.19 & NOT & 3$\times$500 \\ 
HAQ\,2333+0113 & 23 33 35.516 & +01 13 29.32 & 19.85 & NOT, BOSS & 4$\times$900 \\ 
HAQ\,2333+0619 & 23 33 11.434 & +06 19 31.18 & 19.31 & NOT & 2$\times$900 \\ 
HAQ\,2335+1407 & 23 35 12.070 & +14 07 31.56 & 19.71 & NOT & 4$\times$600 \\ 
HAQ\,2337+1343 & 23 37 18.332 & +13 43 06.02 & 19.00 & NOT & 2$\times$600 \\ 
HAQ\,2339+1232 & 23 39 38.008 & +12 32 01.49 & 18.40 & NOT & 2$\times$600 \\ 
HAQ\,2340+0121 & 23 40 54.272 & +01 21 41.34 & 18.81 & NOT & 2$\times$600 \\ 
HAQ\,2343+0615 & 23 43 44.229 & +06 15 00.60 & 19.39 & NOT & 3$\times$600 \\ 
HAQ\,2348+0716 & 23 48 44.464 & +07 16 58.34 & 18.49 & NOT & 900 \\ 
HAQ\,2351+1429 & 23 51 06.509 & +14 29 39.41 & 20.12 & NOT & 4$\times$750 \\ 
HAQ\,2352+0105 & 23 52 38.088 & +01 05 52.35 & 17.33 & NOT, BOSS & 4$\times$225 \\ 
HAQ\,2358+0339 & 23 58 33.476 & +03 39 55.79 & 17.91 & NOT & 2$\times$300 \\ 
HAQ\,2358+0359 & 23 58 16.016 & +03 59 44.69 & 19.95 & NOT & 4$\times$600 \\ 
HAQ\,2358+0520 & 23 58 46.404 & +05 20 52.00 & 19.47 & NOT & 3$\times$600 \\ 
HAQ\,2358+1436 & 23 58 19.661 & +14 36 42.03 & 19.45 & NOT & 3$\times$600
\enddata
\end{deluxetable}

%% file: redQSOs_table2.tex
\begin{deluxetable}{@{}lcclcl@{}}
\tablecaption{Results of the spectroscopic follow-up.
\label{followuptab}}
\tablewidth{400pt}
\tablehead{
\colhead{Target} & \colhead{Type} & \colhead{$z_{\rm QSO}$} & \colhead{A(V)} & \colhead{$F_{1.4\,GHz}$} & \colhead{Notes} \\ 
  &  &  &  &   mJy  &  
}
\startdata
HAQ\,0000+0557 & QSO & 3.31 & 0.32\tablenotemark{(a)} & $<$ 0.34 & {\small (BOSS: z=3.31)} \\ 
HAQ\,0001+0233 & BAL\,QSO & 1.89 & 0.06 & $<$ 0.40 &  \\ 
HAQ\,0008+0835 & QSO & 1.19 & 0.00 & $<$ 0.31 &  \\ 
HAQ\,0008+0846 & QSO & 1.23 & 0.04 & $<$ 0.36 &  \\ 
HAQ\,0011+0122 & QSO & 1.11 & 0.39\tablenotemark{(a)} & 5.03$\pm$0.14 &  \\ 
HAQ\,0012+0651 & QSO & 1.12 & 0.36 & $<$ 0.39 &  \\ 
HAQ\,0012+0657 & BAL\,QSO & 2.43 & 0.00 & $<$ 0.39 &  \\ 
HAQ\,0012+0944 & BAL\,QSO & 2.03 & 0.00 & $<$ 0.39 &  \\ 
HAQ\,0014+0444 & QSO & 1.05 & 0.48 & $<$ 0.36 &  \\ 
HAQ\,0014+0939 & BAL\,QSO & 3.19 & 0.00 & $<$ 0.39 & {\small (BOSS: z=3.23)} \\ 
HAQ\,0015+0736 & QSO & 3.63 & $-$0.05 & $<$ 0.39 & {\small (BOSS: z=3.67)} \\ 
HAQ\,0015+0811 & BAL\,QSO & 2.43 & 0.16 & $<$ 0.35 &  \\ 
HAQ\,0015+1129 & QSO & 0.87 & 0.78 & 1.64$\pm$0.16 &  \\ 
HAQ\,0015+1340 & QSO & 3.11 & 0.14 & 1.64$\pm$0.16 &  \\ 
HAQ\,0018+1133 & QSO & 1.90 & 0.18 & $<$ 0.46 &  \\ 
HAQ\,0020+0259 & BAL\,QSO & 2.47 & 0.15 & $<$ 0.40 &  \\ 
HAQ\,0022+0147 & QSO & 1.15 & 0.27 & $<$ 0.33 &  \\ 
HAQ\,0024+1037 & QSO & 1.22 & 0.27 & 0.77$\pm$0.15 &  \\ 
HAQ\,0025+0220 & QSO & 2.04 & 0.39 & $<$ 0.37 &  \\ 
HAQ\,0026+0640 & QSO & 1.20 & 0.22 & $<$ 0.36 &  \\ 
HAQ\,0031+1328 & QSO & 1.02 & 0.35\tablenotemark{(a)} & $<$ 0.36 &  \\ 
HAQ\,0033+0915 & QSO & 3.28 & 0.00 & $<$ 0.38 & {\small (BOSS: z=3.31)} \\ 
HAQ\,0034+0950 & QSO & 0.28 & 0.51 & $<$ 0.36 &  \\ 
HAQ\,0038+1426 & BAL\,QSO & 2.55 & 0.00 & $<$ 0.49 &  \\ 
HAQ\,0042+1220 & BAL\,QSO & 2.56 & 0.00 & $<$ 0.42 &  \\ 
HAQ\,0043+0549 & QSO & 1.22 & 0.47 & $<$ 0.33 &  \\ 
HAQ\,0043+1136 & QSO & 3.32 & 0.00 & $<$ 0.41 &  \\ 
HAQ\,0044+0817 & QSO & 3.35 & 0.00 & $<$ 0.26 & {\small (BOSS: z=3.31)} \\ 
HAQ\,0044+1250 & BAL\,QSO & 2.35 & 0.25 & $<$ 0.26 &  \\ 
HAQ\,0045+1217 & BL Lac & \nodata & \nodata & 83.81$\pm$0.15 &  \\ 
HAQ\,0046+0839 & BAL\,QSO & 2.85 & 0.28 & $<$ 0.35 & {\small (BOSS: z=2.84)} \\ 
HAQ\,0047+0826 & BAL\,QSO & 1.95 & 0.18 & $<$ 0.35 & Grism 6 \\ 
HAQ\,0051+1542 & BAL\,QSO & 1.90 & 0.39\tablenotemark{(a)} & $<$ 0.35 &  \\ 
HAQ\,0053+0216 & QSO & 0.99 & 0.71 & $<$ 0.36 &  \\ 
HAQ\,0056+1132 & QSO & 3.57 & 0.00 & $<$ 0.48 &  \\ 
HAQ\,0057+1155 & QSO & 0.60 & 1.20 & $<$ 0.47 &  \\ 
HAQ\,0059+1238 & QSO & 3.50 & 0.00 & $<$ 0.43 &  \\ 
HAQ\,0102+0249 & QSO & 3.49 & 0.01 & $<$ 0.37 &  \\ 
HAQ\,0110+0303 & QSO & 3.50 & 0.00 & $<$ 0.41 &  \\ 
HAQ\,0118+0323 & BAL\,QSO & 2.16 & 0.09 & $<$ 0.33 &  \\ 
HAQ\,0118+0700 & QSO & 3.50 & 0.00 & $<$ 0.34 & {\small (BOSS: z=3.50)} \\ 
HAQ\,0119+0817 & BAL\,QSO & 1.95 & 0.22 & $<$ 1.14 &  \\ 
HAQ\,0120+0351 & QSO & 3.09 & 0.00 & $<$ 0.34 &  \\ 
HAQ\,0121+0455 & QSO & 0.84 & 0.77 & $<$ 0.34 &  \\ 
HAQ\,0122+0325 & BAL\,QSO & 2.10 & 0.04 & $<$ 0.64 &  \\ 
HAQ\,0130+1439 & QSO & 1.84 & 0.40 & $<$ 0.64 &  \\ 
HAQ\,0138+0124 & BAL\,QSO & 2.53 & 0.00 & $<$ 0.45 & {\small (BOSS: z=2.61)} \\ 
HAQ\,0138+0636 & BAL\,QSO & 1.80 & 0.09 & $<$ 0.40 &  \\ 
HAQ\,0143+1509 & QSO & 3.76 & 0.00 & $<$ 0.40 &  \\ 
HAQ\,0151+0618 & QSO & 0.95 & 0.72 & 20.23$\pm$0.16 &  \\ 
HAQ\,0151+1453 & QSO & 1.18 & 0.05 & 20.23$\pm$0.16 & Grism 6 \\ 
HAQ\,0155+0438 & QSO & 1.13 & 0.46 & 167.63$\pm$0.29 &  \\ 
HAQ\,0201+0223 & QSO & 2.24 & 0.19 & 165.22$\pm$0.10 &  \\ 
HAQ\,0204+0327 & QSO & 0.83 & 0.66 & 7.53$\pm$0.13 &  \\ 
HAQ\,0206+0624 & QSO & 1.20 & 0.83 & 1.57$\pm$0.13 &  \\ 
HAQ\,0208+0521 & QSO & 1.07 & 0.40 & $<$ 0.37 & Grism 6 \\ 
HAQ\,0211+1214 & QSO & 2.11 & 0.05 & $<$ 0.37 &  \\ 
HAQ\,0226+0729 & QSO & 2.21 & 0.25 & $<$ 0.52 &  \\ 
HAQ\,0236+0619 & BAL\,QSO & 2.43 & 0.12 & $<$ 0.46 &  \\ 
HAQ\,0243+0355 & QSO & 3.30 & 0.00 & $<$ 0.33 &  \\ 
HAQ\,0318+0424 & BAL\,QSO & 3.06 & 0.14 & $<$ 0.42 &  \\ 
HAQ\,0318+0434 & BAL\,QSO & 2.38 & 0.00 & $<$ 0.54 &  \\ 
HAQ\,0319+0623 & QSO & 2.10 & 0.13 & $<$ 0.51 &  \\ 
HAQ\,0329+0553 & QSO & 1.11 & 0.15 & 23.72$\pm$0.24 &  \\ 
HAQ\,0329+0609 & QSO & 1.44 & 0.00 & 23.72$\pm$0.24 &  \\ 
HAQ\,0337+0539 & QSO & 3.28 & 0.03 & 7.73$\pm$0.15 &  \\ 
HAQ\,0339+0420 & QSO & 1.80 & 0.44\tablenotemark{(a)} & 1.28$\pm$0.14 &  \\ 
HAQ\,0340+0408 & QSO & 1.62 & 0.13 & $<$ 0.41 &  \\ 
HAQ\,0345$-$0009 & BAL\,QSO & 1.77 & 0.18 & $<$ 0.41 &  \\ 
HAQ\,0347+0115 & QSO & 0.99 & 0.40 & $<$ 0.40 &  \\ 
HAQ\,0355$-$0025 & QSO & 1.07 & 0.00 & $<$ 0.40 &  \\ 
HAQ\,0355$-$0053 & QSO & 2.15 & 0.00 & $<$ 0.40 &  \\ 
HAQ\,1106+0300 & QSO & 0.73 & 1.29 & $<$ 0.41 &  \\ 
HAQ\,1114+1330 & BAL\,QSO & 1.10 & 0.00 & $<$ 0.43 &  \\ 
HAQ\,1115+0333 & QSO & 3.10 & 0.15 & $<$ 0.45 & Grism 6, 7 \\ 
HAQ\,1148$-$0117 & BAL\,QSO & 2.70 & 0.12 & $<$ 0.45 &  \\ 
HAQ\,1207+1341 & BAL\,QSO & 2.37 & 0.12 & $<$ 0.41 &  \\ 
HAQ\,1233+1304 & BAL\,QSO & 2.34 & 0.02 & $<$ 0.44 &  \\ 
HAQ\,1247+3403 & QSO & 1.18 & 0.48 & $<$ 0.43 & {\small (BOSS: z=1.17)} \\ 
HAQ\,1248+2951 & QSO & 3.55 & 0.20\tablenotemark{(a)} & $<$ 0.41 &  \\ 
HAQ\,1315+0440 & BAL\,QSO & 2.17 & 0.13 & $<$ 0.43 &  \\ 
HAQ\,1319+3214 & QSO & 3.50 & 0.00 & $<$ 0.42 &  \\ 
HAQ\,1327+3206 & QSO & 2.48 & 0.01 & $<$ 1.64 &  \\ 
HAQ\,1332+0052 & QSO & 3.52 & 0.05 & $<$ 0.41 & {\small (BOSS: z=3.51)} \\ 
HAQ\,1339+3331 & QSO & 3.40 & 0.00 & $<$ 0.41 &  \\ 
HAQ\,1355+3407 & QSO & 3.16 & 0.00 & $<$ 0.44 &  \\ 
HAQ\,1358+2401 & QSO & 3.40 & 0.03 & $<$ 0.41 &  \\ 
HAQ\,1400+0219 & QSO & 0.86 & 0.85 & $<$ 0.46 &  \\ 
HAQ\,1409+0940 & QSO & 0.92 & 0.93 & $<$ 0.44 & {\small (BOSS: z=0.93)} \\ 
HAQ\,1411$-$0104 & QSO & 3.50 & 0.00 & $<$ 0.44 &  \\ 
HAQ\,1434+0448 & QSO & 1.20 & 0.57 & $<$ 0.45 &  \\ 
HAQ\,1444+0752 & BAL\,QSO & 2.42 & 0.05 & 6.17$\pm$0.14 & {\small (BOSS: z=2.45)} \\ 
HAQ\,1451+3239 & QSO & 3.55 & 0.00 & $<$ 0.41 &  \\ 
HAQ\,1506+0438 & QSO & 1.04 & 0.61 & $<$ 0.44 &  \\ 
HAQ\,1509+1214 & BAL\,QSO & 2.80 & 0.45\tablenotemark{(a)} & $<$ 0.45 & {\small (BOSS: z=2.89)} \\ 
HAQ\,1517+0817 & QSO & 0.66 & 0.57 & $<$ 0.40 &  \\ 
HAQ\,1524$-$0053 & G-dwarf & 0.00 & 0.52 & $<$ 0.40 & {\small MW type extinction} \\ 
HAQ\,1527+0250 & BAL\,QSO & 2.13 & 0.20 & $<$ 0.81 &  \\ 
HAQ\,1534+0013 & QSO & 3.44 & 0.00 & $<$ 0.42 &  \\ 
HAQ\,1535+0157 & QSO & 3.12 & 0.17 & $<$ 0.44 &  \\ 
HAQ\,1545$-$0130 & QSO & 3.49 & 0.00 & $<$ 0.44 &  \\ 
HAQ\,1546+0005 & QSO & 3.61 & 0.00 & $<$ 0.77 & {\small (BOSS: z=3.61)} \\ 
HAQ\,1600+2911 & QSO & 2.00 & 0.41 & $<$ 0.43 &  \\ 
HAQ\,1603+2512 & BAL\,QSO & 1.92 & 0.55 & $<$ 0.41 &  \\ 
HAQ\,1606+2902 & BAL\,QSO & 1.82 & 0.25 & $<$ 0.44 &  \\ 
HAQ\,1606+2903 & QSO\_CAP & 0.43 & 0.72 & $<$ 0.45 & {\small (BOSS: z=0.43)} \\ 
HAQ\,1607+2611 & Unknown & \nodata & \nodata & $<$ 0.43 &  \\ 
HAQ\,1611+2453 & QSO & 0.78 & 0.78 & $<$ 0.45 &  \\ 
HAQ\,1620+2955 & QSO & 3.36 & 0.11 & $<$ 0.43 & {\small (BOSS: z=3.36)} \\ 
HAQ\,1626+2517 & QSO & 1.25 & 0.30 & 6.72$\pm$0.14 & {\small (BOSS: z=1.26)} \\ 
HAQ\,1633+2851 & QSO & 1.14 & 0.52 & $<$ 0.43 &  \\ 
HAQ\,1634+2811 & QSO & 2.60 & 0.10 & $<$ 0.39 &  \\ 
HAQ\,1639+3157 & QSO & 0.81 & 0.79 & $<$ 0.41 & {\small (BOSS: z=0.81)} \\ 
HAQ\,1643+2944 & QSO & 1.08 & 0.44 & $<$ 0.45 &  \\ 
HAQ\,1645+3056 & BAL\,QSO & 2.50 & 0.10 & $<$ 0.41 & {\small (BOSS: z=2.53)} \\ 
HAQ\,1645+3130 & QSO & 0.93 & 0.69 & $<$ 0.40 & {\small (BOSS: z=0.93)} \\ 
HAQ\,1655+3051 & QSO & 1.27 & 0.53 & $<$ 0.43 &  \\ 
HAQ\,2159+0212 & QSO & 1.26 & 0.10 & $<$ 0.37 &  \\ 
HAQ\,2203$-$0052 & QSO & 1.24 & 0.74\tablenotemark{(a)} & $<$ 0.37 &  \\ 
HAQ\,2217+0359 & QSO & 0.98 & 0.52 & $<$ 0.37 &  \\ 
HAQ\,2221+0145 & QSO & 3.43 & $-$0.05 & $<$ 0.37 & {\small (BOSS: z=3.43)} \\ 
HAQ\,2222+0604 & BAL\,QSO & 2.45 & 0.25 & $<$ 0.46 &  \\ 
HAQ\,2225+0527 & QSO & 2.32 & 0.29 & 882.52$\pm$0.11 & Grism 6, 7 \\ 
HAQ\,2229+0324 & QSO & 2.64 & 0.28 & $<$ 0.40 & {\small (BOSS: z=2.66)} \\ 
HAQ\,2231+0509 & BAL\,QSO & 1.76 & 0.35 & $<$ 0.40 &  \\ 
HAQ\,2241+0818 & QSO & 2.43 & 0.04 & $<$ 0.33 &  \\ 
HAQ\,2244+0335 & QSO & 3.37 & 0.00 & $<$ 0.37 & {\small (BOSS: z=3.36)} \\ 
HAQ\,2245+0457 & BAL\,QSO & 2.10 & 0.12 & $<$ 0.36 &  \\ 
HAQ\,2246+0710 & QSO & 1.20 & 0.12 & $<$ 0.37 &  \\ 
HAQ\,2247+0146 & BAL\,QSO & $\sim$2.1 & $\sim$0.20 & $<$ 0.39 & Tentative identification \\ 
HAQ\,2252+0434 & QSO & 0.35 & 0.84 & 5.42$\pm$0.08 &  \\ 
HAQ\,2253+1141 & BAL\,QSO & 2.40 & 0.17 & $<$ 0.30 &  \\ 
HAQ\,2254+0638\tablenotemark{(b)} & M-dwarf & 0.00 & 0.00 & $<$ 0.39 & Wrong target on slit \\ 
HAQ\,2300+0914 & QSO & 1.99 & 0.29 & 64.79$\pm$0.13 & Grism 6 \\ 
HAQ\,2301+0832 & BAL\,QSO & 1.10 & 0.03 & $<$ 0.31 &  \\ 
HAQ\,2303+0238 & BAL\,QSO & 2.22 & 0.11 & $<$ 0.37 & Grism 6 \\ 
HAQ\,2303+0630 & QSO & 0.99 & 0.36 & $<$ 0.38 &  \\ 
HAQ\,2305+0117 & QSO & 2.67 & 0.13 & $<$ 0.40 &  \\ 
HAQ\,2310+1117 & QSO & 0.83 & 0.92 & $<$ 0.65 &  \\ 
HAQ\,2311+1444 & QSO\_HIZ & 3.31 & 0.00 & $<$ 0.65 & {\small (SDSS legacy: z=3.31)} \\ 
HAQ\,2313+0955 & QSO & 1.06 & 0.55 & $<$ 0.65 &  \\ 
HAQ\,2318+0255 & QSO & 3.50 & 0.00 & $<$ 0.37 &  \\ 
HAQ\,2326+0642 & QSO & 1.25 & 0.37 & $<$ 0.39 &  \\ 
HAQ\,2326+1423 & BAL\,QSO & 2.54 & 0.17 & $<$ 0.39 &  \\ 
HAQ\,2330+1009 & BAL\,QSO & 3.11 & 0.19 & 18.39$\pm$0.17 &  \\ 
HAQ\,2333+0113 & BAL\,QSO & 3.25 & 0.05 & $<$ 0.44 & {\small (BOSS: z=3.28)} \\ 
HAQ\,2333+0619 & BAL\,QSO & 2.72 & 0.13 & $<$ 0.53 &  \\ 
HAQ\,2335+1407 & QSO & 0.97 & 0.54 & $<$ 0.53 &  \\ 
HAQ\,2337+1343 & QSO & 3.57 & 0.00 & $<$ 0.53 &  \\ 
HAQ\,2339+1232 & QSO & 1.24 & 0.19 & $<$ 0.44 &  \\ 
HAQ\,2340+0121 & BAL\,QSO & 2.15 & 0.15 & $<$ 0.44 &  \\ 
HAQ\,2343+0615 & QSO & 2.16 & 0.29\tablenotemark{(a)} & $<$ 0.32 &  \\ 
HAQ\,2348+0716 & QSO & 0.88 & 0.22 & $<$ 0.36 &  \\ 
HAQ\,2351+1429 & BAL\,QSO & 2.96 & 0.23 & $<$ 0.40 &  \\ 
HAQ\,2352+0105 & QSO\_HIZ & 2.14 & 0.06 & $<$ 0.40 & {\small (BOSS: z=2.99)} \\ 
HAQ\,2358+0339 & BAL\,QSO & 2.09 & 0.06 & $<$ 0.36 &  \\ 
HAQ\,2358+0359 & BAL\,QSO & 2.89 & 0.20 & $<$ 0.35 &  \\ 
HAQ\,2358+0520 & QSO & 3.47 & 0.00 & $<$ 0.37 &  \\ 
HAQ\,2358+1436 & QSO & 1.05 & 0.65 & $<$ 0.26 &  \\ 
\enddata
\tablenotetext{(a)}{\, Marks systems with intervening dust. We here give the estimated A(V)$_{\rm abs}$ at the absorber redshift assuming the LMC extinction curve.}
\tablenotetext{(b)}{\, The spectrum for this target does not correspond to the photometry in SDSS, since a wrong target was accidentally put on the slit. The real SDSS identifier for this spectrum is J225430.23+063833.8.}
\tablecomments{
In cases where SDSS legacy had flagged the object as QSO for follow-up this
is indicated by 'QSO\_HIZ' or 'QSO\_CAP' as object 'Type'.
In the 'Notes' column, we list the targets with spectra available in BOSS or SDSS legacy along with their inferred redshift.
Other remarks such as complementary observations with different grisms are also listed here.\\
The classification of broad absorption line (BAL) QSOs does not follow a stringent methodology. For a detailed study and classification of BAL QSOs in our sample, we refer to Saturni et al. (2014), submitted.
}
\end{deluxetable}

%% file: HAQ_arXiv.bbl
\begin{thebibliography}{}
\expandafter\ifx\csname natexlab\endcsname\relax\def\natexlab#1{#1}\fi

\bibitem[{{Balokovi{\'c}} {et~al.}(2012){Balokovi{\'c}}, {Smol{\v c}i{\'c}},
  {Ivezi{\'c}}, {Zamorani}, {Schinnerer}, \& {Kelly}}]{Balokovic2012}
{Balokovi{\'c}}, M., {Smol{\v c}i{\'c}}, V., {Ivezi{\'c}}, {\v Z}., {et~al.}
  2012, \apj, 759, 30

\bibitem[{{Banerji} {et~al.}(2012){Banerji}, {McMahon}, {Hewett},
  {Alaghband-Zadeh}, {Gonzalez-Solares}, {Venemans}, \&
  {Hawthorn}}]{Banerji2012}
{Banerji}, M., {McMahon}, R.~G., {Hewett}, P.~C., {et~al.} 2012, \mnras, 427,
  2275

\bibitem[{{Becker} {et~al.}(1995){Becker}, {White}, \& {Helfand}}]{Becker1995}
{Becker}, R.~H., {White}, R.~L., \& {Helfand}, D.~J. 1995, \apj, 450, 559

\bibitem[{{Benn} {et~al.}(1998){Benn}, {Vigotti}, {Carballo},
  {Gonzalez-Serrano}, \& {S{\'a}nchez}}]{Benn98}
{Benn}, C.~R., {Vigotti}, M., {Carballo}, R., {Gonzalez-Serrano}, J.~I., \&
  {S{\'a}nchez}, S.~F. 1998, \mnras, 295, 451

\bibitem[{{Croom} {et~al.}(2004){Croom}, {Smith}, {Boyle}, {Shanks}, {Miller},
  {Outram}, \& {Loaring}}]{Croom2004}
{Croom}, S.~M., {Smith}, R.~J., {Boyle}, B.~J., {et~al.} 2004, \mnras, 349,
  1397

\bibitem[{{Dawson} {et~al.}(2013){Dawson}, {Schlegel}, {Ahn}, {Anderson},
  {Aubourg}, {Bailey}, {Barkhouser}, {Bautista}, {Beifiori}, {Berlind},
  {Bhardwaj}, {Bizyaev}, {Blake}, {Blanton}, {Blomqvist}, {Bolton}, {Borde},
  {Bovy}, {Brandt}, {Brewington}, {Brinkmann}, {Brown}, {Brownstein}, {Bundy},
  {Busca}, {Carithers}, {Carnero}, {Carr}, {Chen}, {Comparat}, {Connolly},
  {Cope}, {Croft}, {Cuesta}, {da Costa}, {Davenport}, {Delubac}, {de Putter},
  {Dhital}, {Ealet}, {Ebelke}, {Eisenstein}, {Escoffier}, {Fan}, {Filiz Ak},
  {Finley}, {Font-Ribera}, {G{\'e}nova-Santos}, {Gunn}, {Guo}, {Haggard},
  {Hall}, {Hamilton}, {Harris}, {Harris}, {Ho}, {Hogg}, {Holder}, {Honscheid},
  {Huehnerhoff}, {Jordan}, {Jordan}, {Kauffmann}, {Kazin}, {Kirkby}, {Klaene},
  {Kneib}, {Le Goff}, {Lee}, {Long}, {Loomis}, {Lundgren}, {Lupton}, {Maia},
  {Makler}, {Malanushenko}, {Malanushenko}, {Mandelbaum}, {Manera}, {Maraston},
  {Margala}, {Masters}, {McBride}, {McDonald}, {McGreer}, {McMahon}, {Mena},
  {Miralda-Escud{\'e}}, {Montero-Dorta}, {Montesano}, {Muna}, {Myers},
  {Naugle}, {Nichol}, {Noterdaeme}, {Nuza}, {Olmstead}, {Oravetz}, {Oravetz},
  {Owen}, {Padmanabhan}, {Palanque-Delabrouille}, {Pan}, {Parejko},
  {P{\^a}ris}, {Percival}, {P{\'e}rez-Fournon}, {P{\'e}rez-R{\`a}fols},
  {Petitjean}, {Pfaffenberger}, {Pforr}, {Pieri}, {Prada}, {Price-Whelan},
  {Raddick}, {Rebolo}, {Rich}, {Richards}, {Rockosi}, {Roe}, {Ross}, {Ross},
  {Rossi}, {Rubi{\~n}o-Martin}, {Samushia}, {S{\'a}nchez}, {Sayres}, {Schmidt},
  {Schneider}, {Sc{\'o}ccola}, {Seo}, {Shelden}, {Sheldon}, {Shen}, {Shu},
  {Slosar}, {Smee}, {Snedden}, {Stauffer}, {Steele}, {Strauss}, {Streblyanska},
  {Suzuki}, {Swanson}, {Tal}, {Tanaka}, {Thomas}, {Tinker}, {Tojeiro},
  {Tremonti}, {Vargas Maga{\~n}a}, {Verde}, {Viel}, {Wake}, {Watson}, {Weaver},
  {Weinberg}, {Weiner}, {West}, {White}, {Wood-Vasey}, {Yeche}, {Zehavi},
  {Zhao}, \& {Zheng}}]{BOSS}
{Dawson}, K.~S., {Schlegel}, D.~J., {Ahn}, C.~P., {et~al.} 2013, \aj, 145, 10

\bibitem[{{El{\'{\i}}asd{\'o}ttir} {et~al.}(2009){El{\'{\i}}asd{\'o}ttir},
  {Fynbo}, {Hjorth}, {Ledoux}, {Watson}, {Andersen}, {Malesani}, {Vreeswijk},
  {Prochaska}, {Sollerman}, \& {Jaunsen}}]{Eliasdottir2009}
{El{\'{\i}}asd{\'o}ttir}, {\'A}., {Fynbo}, J.~P.~U., {Hjorth}, J., {et~al.}
  2009, \apj, 697, 1725

\bibitem[{{Ellison} {et~al.}(2006){Ellison}, {Vreeswijk}, {Ledoux}, {Willis},
  {Jaunsen}, {Wijers}, {Smette}, {Fynbo}, {M{\o}ller}, {Hjorth}, \&
  {Kaufer}}]{Ellison2006}
{Ellison}, S.~L., {Vreeswijk}, P., {Ledoux}, C., {et~al.} 2006, \mnras, 372,
  L38

\bibitem[{{Elvis} {et~al.}(2002){Elvis}, {Marengo}, \& {Karovska}}]{Elvis2002}
{Elvis}, M., {Marengo}, M., \& {Karovska}, M. 2002, \apjl, 567, L107

\bibitem[{{Fitzpatrick} \& {Massa}(2005)}]{FM2005}
{Fitzpatrick}, E.~L., \& {Massa}, D. 2005, \aj, 130, 1127

\bibitem[{{Foreman-Mackey} {et~al.}(2013){Foreman-Mackey}, {Hogg}, {Lang}, \&
  {Goodman}}]{emcee}
{Foreman-Mackey}, D., {Hogg}, D.~W., {Lang}, D., \& {Goodman}, J. 2013, \pasp,
  125, 306

\bibitem[{{Fynbo} {et~al.}(2013){Fynbo}, {Krogager}, {Venemans}, {Noterdaeme},
  {Vestergaard}, {M{\o}ller}, {Ledoux}, \& {Geier}}]{Fynbo2013a}
{Fynbo}, J.~P.~U., {Krogager}, J.-K., {Venemans}, B., {et~al.} 2013, \apjs,
  204, 6

\bibitem[{{Fynbo} {et~al.}(2011){Fynbo}, {Ledoux}, {Noterdaeme}, {Christensen},
  {M{\o}ller}, {Durgapal}, {Goldoni}, {Kaper}, {Krogager}, {Laursen}, {Maund},
  {Milvang-Jensen}, {Okoshi}, {Rasmussen}, {Thorsen}, {Toft}, \&
  {Zafar}}]{Fynbo11}
{Fynbo}, J.~P.~U., {Ledoux}, C., {Noterdaeme}, P., {et~al.} 2011, \mnras, 413,
  2481

\bibitem[{{Fynbo} {et~al.}(2014){Fynbo}, {Kr{\"u}hler}, {Leighly}, {Ledoux},
  {Vreeswijk}, {Schulze}, {Noterdaeme}, {Watson}, {Wijers}, {Bolmer}, {Cano},
  {Christensen}, {Covino}, {D'Elia}, {Flores}, {Friis}, {Goldoni}, {Greiner},
  {Hammer}, {Hjorth}, {Jakobsson}, {Japelj}, {Kaper}, {Klose}, {Knust},
  {Leloudas}, {Levan}, {Malesani}, {Milvang-Jensen}, {M{\o}ller}, {Nicuesa
  Guelbenzu}, {Oates}, {Pian}, {Schady}, {Sparre}, {Tagliaferri}, {Tanvir},
  {Th{\"o}ne}, {de Ugarte Postigo}, {Vergani}, {Wiersema}, {Xu}, \&
  {Zafar}}]{Fynbo2014}
{Fynbo}, J.~P.~U., {Kr{\"u}hler}, T., {Leighly}, K., {et~al.} 2014, \aap, 572,
  A12

\bibitem[{{Glikman} {et~al.}(2006){Glikman}, {Helfand}, \& {White}}]{Glikman06}
{Glikman}, E., {Helfand}, D.~J., \& {White}, R.~L. 2006, \apj, 640, 579

\bibitem[{{Glikman} {et~al.}(2007){Glikman}, {Helfand}, {White}, {Becker},
  {Gregg}, \& {Lacy}}]{Glikman07}
{Glikman}, E., {Helfand}, D.~J., {White}, R.~L., {et~al.} 2007, \apj, 667, 673

\bibitem[{{Glikman} {et~al.}(2012){Glikman}, {Urrutia}, {Lacy}, {Djorgovski},
  {Mahabal}, {Myers}, {Ross}, {Petitjean}, {Ge}, {Schneider}, \&
  {York}}]{Glikman2012}
{Glikman}, E., {Urrutia}, T., {Lacy}, M., {et~al.} 2012, \apj, 757, 51

\bibitem[{{Glikman} {et~al.}(2013){Glikman}, {Urrutia}, {Lacy}, {Djorgovski},
  {Urry}, {Croom}, {Schneider}, {Mahabal}, {Graham}, \& {Ge}}]{Glikman2013}
{Glikman}, E., {Urrutia}, T., {Lacy}, M., {et~al.} 2013, \apj, 778, 127

\bibitem[{{Gordon} {et~al.}(2003){Gordon}, {Clayton}, {Misselt}, {Landolt}, \&
  {Wolff}}]{Gordon03}
{Gordon}, K.~D., {Clayton}, G.~C., {Misselt}, K.~A., {Landolt}, A.~U., \&
  {Wolff}, M.~J. 2003, \apj, 594, 279

\bibitem[{{Graham} {et~al.}(2014){Graham}, {Djorgovski}, {Drake}, {Mahabal},
  {Chang}, {Stern}, {Donalek}, \& {Glikman}}]{Graham2014}
{Graham}, M.~J., {Djorgovski}, S.~G., {Drake}, A.~J., {et~al.} 2014, \mnras,
  439, 703

\bibitem[{{Gregg} {et~al.}(2002){Gregg}, {Lacy}, {White}, {Glikman}, {Helfand},
  {Becker}, \& {Brotherton}}]{Gregg02}
{Gregg}, M.~D., {Lacy}, M., {White}, R.~L., {et~al.} 2002, \apj, 564, 133

\bibitem[{{Hall} {et~al.}(2002){Hall}, {Anderson}, {Strauss}, {York},
  {Richards}, {Fan}, {Knapp}, {Schneider}, {Vanden Berk}, {Geballe}, {Bauer},
  {Becker}, {Davis}, {Rix}, {Nichol}, {Bahcall}, {Brinkmann}, {Brunner},
  {Connolly}, {Csabai}, {Doi}, {Fukugita}, {Gunn}, {Haiman}, {Harvanek},
  {Heckman}, {Hennessy}, {Inada}, {Ivezi{\'c}}, {Johnston}, {Kleinman},
  {Krolik}, {Krzesinski}, {Kunszt}, {Lamb}, {Long}, {Lupton}, {Miknaitis},
  {Munn}, {Narayanan}, {Neilsen}, {Newman}, {Nitta}, {Okamura}, {Pentericci},
  {Pier}, {Schlegel}, {Snedden}, {Szalay}, {Thakar}, {Tsvetanov}, {White}, \&
  {Zheng}}]{Hall2002}
{Hall}, P.~B., {Anderson}, S.~F., {Strauss}, M.~A., {et~al.} 2002, \apjs, 141,
  267

\bibitem[{{Hopkins} {et~al.}(2004){Hopkins}, {Strauss}, {Hall}, {Richards},
  {Cooper}, {Schneider}, {Vanden Berk}, {Jester}, {Brinkmann}, \&
  {Szokoly}}]{Hopkins04}
{Hopkins}, P.~F., {Strauss}, M.~A., {Hall}, P.~B., {et~al.} 2004, \aj, 128,
  1112

\bibitem[{{Ivezi{\'c}} {et~al.}(2002){Ivezi{\'c}}, {Menou}, {Knapp}, {Strauss},
  {Lupton}, {Vanden Berk}, {Richards}, {Tremonti}, {Weinstein}, {Anderson},
  {Bahcall}, {Becker}, {Bernardi}, {Blanton}, {Eisenstein}, {Fan},
  {Finkbeiner}, {Finlator}, {Frieman}, {Gunn}, {Hall}, {Kim}, {Kinkhabwala},
  {Narayanan}, {Rockosi}, {Schlegel}, {Schneider}, {Strateva}, {SubbaRao},
  {Thakar}, {Voges}, {White}, {Yanny}, {Brinkmann}, {Doi}, {Fukugita},
  {Hennessy}, {Munn}, {Nichol}, \& {York}}]{Ivezic2002}
{Ivezi{\'c}}, {\v Z}., {Menou}, K., {Knapp}, G.~R., {et~al.} 2002, \aj, 124,
  2364

\bibitem[{{Jiang} {et~al.}(2010){Jiang}, {Ge}, {Prochaska}, {Wang}, {Zhou}, \&
  {Wang}}]{Jiang2010}
{Jiang}, P., {Ge}, J., {Prochaska}, J.~X., {et~al.} 2010, \apj, 724, 1325

\bibitem[{{Jiang} {et~al.}(2013){Jiang}, {Zhou}, {Ji}, {Shu}, {Liu}, {Wang},
  {Dong}, {Bai}, {Wang}, \& {Wang}}]{Jiang2013}
{Jiang}, P., {Zhou}, H., {Ji}, T., {et~al.} 2013, \aj, 145, 157

\bibitem[{{Junkkarinen} {et~al.}(2004){Junkkarinen}, {Cohen}, {Beaver},
  {Burbidge}, {Lyons}, \& {Madejski}}]{Junkkarinen2004}
{Junkkarinen}, V.~T., {Cohen}, R.~D., {Beaver}, E.~A., {et~al.} 2004, \apj,
  614, 658

\bibitem[{{Kaplan} {et~al.}(2010){Kaplan}, {Prochaska}, {Herbert-Fort},
  {Ellison}, \& {Dessauges-Zavadsky}}]{Kaplan10}
{Kaplan}, K.~F., {Prochaska}, J.~X., {Herbert-Fort}, S., {Ellison}, S.~L., \&
  {Dessauges-Zavadsky}, M. 2010, \pasp, 122, 619

\bibitem[{{Khare} {et~al.}(2012){Khare}, {vanden Berk}, {York}, {Lundgren}, \&
  {Kulkarni}}]{Khare2012}
{Khare}, P., {vanden Berk}, D., {York}, D.~G., {Lundgren}, B., \& {Kulkarni},
  V.~P. 2012, \mnras, 419, 1028

\bibitem[{{Lacy} {et~al.}(2007){Lacy}, {Petric}, {Sajina}, {Canalizo},
  {Storrie-Lombardi}, {Armus}, {Fadda}, \& {Marleau}}]{Lacy2007}
{Lacy}, M., {Petric}, A.~O., {Sajina}, A., {et~al.} 2007, \aj, 133, 186

\bibitem[{{Larson} {et~al.}(1996){Larson}, {Whittet}, \& {Hough}}]{Larson1996}
{Larson}, K.~A., {Whittet}, D.~C.~B., \& {Hough}, J.~H. 1996, \apj, 472, 755

\bibitem[{{Leighly} {et~al.}(2014){Leighly}, {Terndrup}, {Baron}, {Lucy},
  {Dietrich}, \& {Gallagher}}]{Leighly2014}
{Leighly}, K.~M., {Terndrup}, D.~M., {Baron}, E., {et~al.} 2014, \apj, 788, 123

\bibitem[{{Maddox} {et~al.}(2012){Maddox}, {Hewett}, {P{\'e}roux}, {Nestor}, \&
  {Wisotzki}}]{Maddox2012}
{Maddox}, N., {Hewett}, P.~C., {P{\'e}roux}, C., {Nestor}, D.~B., \&
  {Wisotzki}, L. 2012, \mnras, 424, 2876

\bibitem[{{Maddox} {et~al.}(2008){Maddox}, {Hewett}, {Warren}, \&
  {Croom}}]{Maddox2008}
{Maddox}, N., {Hewett}, P.~C., {Warren}, S.~J., \& {Croom}, S.~M. 2008, \mnras,
  386, 1605

\bibitem[{{Malhotra}(1997)}]{Malhotra1997}
{Malhotra}, S. 1997, \apjl, 488, L101

\bibitem[{{Matthews} \& {Sandage}(1963)}]{Matthews1963}
{Matthews}, T.~A., \& {Sandage}, A.~R. 1963, \apj, 138, 30

\bibitem[{{Meusinger} {et~al.}(2012){Meusinger}, {Schalldach}, {Scholz}, {in
  der Au}, {Newholm}, {de Hoon}, \& {Kaminsky}}]{Meusinger2012}
{Meusinger}, H., {Schalldach}, P., {Scholz}, R.-D., {et~al.} 2012, \aap, 541,
  A77

\bibitem[{{Nishiyama} {et~al.}(2008){Nishiyama}, {Nagata}, {Tamura}, {Kandori},
  {Hatano}, {Sato}, \& {Sugitani}}]{Nishiyama2008}
{Nishiyama}, S., {Nagata}, T., {Tamura}, M., {et~al.} 2008, \apj, 680, 1174

\bibitem[{{Nishiyama} {et~al.}(2009){Nishiyama}, {Tamura}, {Hatano}, {Kato},
  {Tanab{\'e}}, {Sugitani}, \& {Nagata}}]{Nishiyama2009}
{Nishiyama}, S., {Tamura}, M., {Hatano}, H., {et~al.} 2009, \apj, 696, 1407

\bibitem[{{Noterdaeme} {et~al.}(2009{\natexlab{a}}){Noterdaeme}, {Ledoux},
  {Srianand}, {Petitjean}, \& {Lopez}}]{Noterdaeme09b}
{Noterdaeme}, P., {Ledoux}, C., {Srianand}, R., {Petitjean}, P., \& {Lopez}, S.
  2009{\natexlab{a}}, \aap, 503, 765

\bibitem[{{Noterdaeme} {et~al.}(2010){Noterdaeme}, {Petitjean}, {Ledoux},
  {L{\'o}pez}, {Srianand}, \& {Vergani}}]{Noterdaeme10}
{Noterdaeme}, P., {Petitjean}, P., {Ledoux}, C., {et~al.} 2010, \aap, 523, A80+

\bibitem[{{Noterdaeme} {et~al.}(2009{\natexlab{b}}){Noterdaeme}, {Petitjean},
  {Ledoux}, \& {Srianand}}]{Noterdaeme09a}
{Noterdaeme}, P., {Petitjean}, P., {Ledoux}, C., \& {Srianand}, R.
  2009{\natexlab{b}}, \aap, 505, 1087

\bibitem[{{Noterdaeme} {et~al.}(2012){Noterdaeme}, {Laursen}, {Petitjean},
  {Vergani}, {Maureira}, {Ledoux}, {Fynbo}, {L{\'o}pez}, \&
  {Srianand}}]{Noterdaeme12}
{Noterdaeme}, P., {Laursen}, P., {Petitjean}, P., {et~al.} 2012, \aap, 540, A63

\bibitem[{{Peth} {et~al.}(2011){Peth}, {Ross}, \& {Schneider}}]{Peth11}
{Peth}, M.~A., {Ross}, N.~P., \& {Schneider}, D.~P. 2011, \aj, 141, 105

\bibitem[{{Pitman} {et~al.}(2000){Pitman}, {Clayton}, \& {Gordon}}]{Pitman2000}
{Pitman}, K.~M., {Clayton}, G.~C., \& {Gordon}, K.~D. 2000, \pasp, 112, 537

\bibitem[{{Polletta} {et~al.}(2006){Polletta}, {Wilkes}, {Siana}, {Lonsdale},
  {Kilgard}, {Smith}, {Kim}, {Owen}, {Efstathiou}, {Jarrett}, {Stacey},
  {Franceschini}, {Rowan-Robinson}, {Babbedge}, {Berta}, {Fang}, {Farrah},
  {Gonz{\'a}lez-Solares}, {Morrison}, {Surace}, \& {Shupe}}]{Polletta2006}
{Polletta}, M.~d.~C., {Wilkes}, B.~J., {Siana}, B., {et~al.} 2006, \apj, 642,
  673

\bibitem[{{Pontzen} \& {Pettini}(2009)}]{Pontzen2009}
{Pontzen}, A., \& {Pettini}, M. 2009, \mnras, 393, 557

\bibitem[{{Richards} {et~al.}(2001){Richards}, {Fan}, {Schneider}, {Vanden
  Berk}, {Strauss}, {York}, {Anderson}, {Anderson}, {Annis}, {Bahcall},
  {Bernardi}, {Briggs}, {Brinkmann}, {Brunner}, {Burles}, {Carey}, {Castander},
  {Connolly}, {Crocker}, {Csabai}, {Doi}, {Finkbeiner}, {Friedman}, {Frieman},
  {Fukugita}, {Gunn}, {Hindsley}, {Ivezi{\'c}}, {Kent}, {Knapp}, {Lamb},
  {Leger}, {Long}, {Loveday}, {Lupton}, {McKay}, {Meiksin}, {Merrelli}, {Munn},
  {Newberg}, {Newcomb}, {Nichol}, {Owen}, {Pier}, {Pope}, {Richmond},
  {Rockosi}, {Schlegel}, {Siegmund}, {Smee}, {Snir}, {Stoughton}, {Stubbs},
  {SubbaRao}, {Szalay}, {Szokoly}, {Tremonti}, {Uomoto}, {Waddell}, {Yanny}, \&
  {Zheng}}]{Richards2001}
{Richards}, G.~T., {Fan}, X., {Schneider}, D.~P., {et~al.} 2001, \aj, 121, 2308

\bibitem[{{Richards} {et~al.}(2003){Richards}, {Hall}, {Vanden Berk},
  {Strauss}, {Schneider}, {Weinstein}, {Reichard}, {York}, {Knapp}, {Fan},
  {Ivezi{\'c}}, {Brinkmann}, {Budav{\'a}ri}, {Csabai}, \&
  {Nichol}}]{Richards03}
{Richards}, G.~T., {Hall}, P.~B., {Vanden Berk}, D.~E., {et~al.} 2003, \aj,
  126, 1131

\bibitem[{{Richards} {et~al.}(2006){Richards}, {Lacy}, {Storrie-Lombardi},
  {Hall}, {Gallagher}, {Hines}, {Fan}, {Papovich}, {Vanden Berk}, {Trammell},
  {Schneider}, {Vestergaard}, {York}, {Jester}, {Anderson}, {Budav{\'a}ri}, \&
  {Szalay}}]{Richards06}
{Richards}, G.~T., {Lacy}, M., {Storrie-Lombardi}, L.~J., {et~al.} 2006, \apjs,
  166, 470

\bibitem[{{Rodr{\'{\i}}guez-Ardila} \& {Mazzalay}(2006)}]{Rodriguez-Ardila2006}
{Rodr{\'{\i}}guez-Ardila}, A., \& {Mazzalay}, X. 2006, \mnras, 367, L57

\bibitem[{{Ross} {et~al.}(2012){Ross}, {Myers}, {Sheldon}, {Y{\`e}che},
  {Strauss}, {Bovy}, {Kirkpatrick}, {Richards}, {Aubourg}, {Blanton}, {Brandt},
  {Carithers}, {Croft}, {da Silva}, {Dawson}, {Eisenstein}, {Hennawi}, {Ho},
  {Hogg}, {Lee}, {Lundgren}, {McMahon}, {Miralda-Escud{\'e}},
  {Palanque-Delabrouille}, {P{\^a}ris}, {Petitjean}, {Pieri}, {Rich}, {Roe},
  {Schiminovich}, {Schlegel}, {Schneider}, {Slosar}, {Suzuki}, {Tinker},
  {Weinberg}, {Weyant}, {White}, \& {Wood-Vasey}}]{Ross2012}
{Ross}, N.~P., {Myers}, A.~D., {Sheldon}, E.~S., {et~al.} 2012, \apjs, 199, 3

\bibitem[{{Sanders} {et~al.}(1989){Sanders}, {Phinney}, {Neugebauer}, {Soifer},
  \& {Matthews}}]{Sanders1989}
{Sanders}, D.~B., {Phinney}, E.~S., {Neugebauer}, G., {Soifer}, B.~T., \&
  {Matthews}, K. 1989, \apj, 347, 29

\bibitem[{{Schlegel} {et~al.}(1998){Schlegel}, {Finkbeiner}, \&
  {Davis}}]{Schlegel98}
{Schlegel}, D.~J., {Finkbeiner}, D.~P., \& {Davis}, M. 1998, \apj, 500, 525

\bibitem[{{Srianand} {et~al.}(2008){Srianand}, {Gupta}, {Petitjean}, \&
  {Saikia}}]{Srianand08}
{Srianand}, R., {Gupta}, N., {Petitjean}, P., \& {Saikia}, D.~J. 2008, \mnras,
  391, L69

\bibitem[{{Stein} {et~al.}(1976){Stein}, {Odell}, \&
  {Strittmatter}}]{Stein1976}
{Stein}, W.~A., {Odell}, S.~L., \& {Strittmatter}, P.~A. 1976, \araa, 14, 173

\bibitem[{{Sumi}(2004)}]{Sumi2004}
{Sumi}, T. 2004, \mnras, 349, 193

\bibitem[{{Urrutia} {et~al.}(2009){Urrutia}, {Becker}, {White}, {Glikman},
  {Lacy}, {Hodge}, \& {Gregg}}]{Urrutia09}
{Urrutia}, T., {Becker}, R.~H., {White}, R.~L., {et~al.} 2009, \apj, 698, 1095

\bibitem[{{van Dokkum}(2001)}]{vanDokkum01}
{van Dokkum}, P.~G. 2001, \pasp, 113, 1420

\bibitem[{{Vanden Berk} {et~al.}(2001){Vanden Berk}, {Richards}, {Bauer},
  {Strauss}, {Schneider}, {Heckman}, {York}, {Hall}, {Fan}, {Knapp},
  {Anderson}, {Annis}, {Bahcall}, {Bernardi}, {Briggs}, {Brinkmann}, {Brunner},
  {Burles}, {Carey}, {Castander}, {Connolly}, {Crocker}, {Csabai}, {Doi},
  {Finkbeiner}, {Friedman}, {Frieman}, {Fukugita}, {Gunn}, {Hennessy},
  {Ivezi{\'c}}, {Kent}, {Kunszt}, {Lamb}, {Leger}, {Long}, {Loveday}, {Lupton},
  {Meiksin}, {Merelli}, {Munn}, {Newberg}, {Newcomb}, {Nichol}, {Owen}, {Pier},
  {Pope}, {Rockosi}, {Schlegel}, {Siegmund}, {Smee}, {Snir}, {Stoughton},
  {Stubbs}, {SubbaRao}, {Szalay}, {Szokoly}, {Tremonti}, {Uomoto}, {Waddell},
  {Yanny}, \& {Zheng}}]{vandenBerk01}
{Vanden Berk}, D.~E., {Richards}, G.~T., {Bauer}, A., {et~al.} 2001, \aj, 122,
  549

\bibitem[{{Vladilo} {et~al.}(2008){Vladilo}, {Prochaska}, \&
  {Wolfe}}]{Vladilo08}
{Vladilo}, G., {Prochaska}, J.~X., \& {Wolfe}, A.~M. 2008, \aap, 478, 701

\bibitem[{{Wang} {et~al.}(2004){Wang}, {Hall}, {Ge}, {Li}, \&
  {Schneider}}]{Wang2004}
{Wang}, J., {Hall}, P.~B., {Ge}, J., {Li}, A., \& {Schneider}, D.~P. 2004,
  \apj, 609, 589

\bibitem[{{Wang} {et~al.}(2012){Wang}, {Zhou}, {Ge}, {Jiang}, {Lu},
  {Prochaska}, {Hamann}, {Wang}, {Wang}, \& {Yuan}}]{Wang2012}
{Wang}, J.-G., {Zhou}, H.-Y., {Ge}, J., {et~al.} 2012, \apj, 760, 42

\bibitem[{{Warren} {et~al.}(2007){Warren}, {Hambly}, {Dye}, {Almaini}, {Cross},
  \& {Edge}}]{Warren07}
{Warren}, S.~J., {Hambly}, N.~C., {Dye}, S., {et~al.} 2007, \mnras, 375, 213

\bibitem[{{Warren} {et~al.}(2000){Warren}, {Hewett}, \& {Foltz}}]{Warren00}
{Warren}, S.~J., {Hewett}, P.~C., \& {Foltz}, C.~B. 2000, \mnras, 312, 827

\bibitem[{{Welty} \& {Fowler}(1992)}]{Welty1992}
{Welty}, D.~E., \& {Fowler}, J.~R. 1992, \apj, 393, 193

\bibitem[{{White} {et~al.}(2007){White}, {Helfand}, {Becker}, {Glikman}, \& {de
  Vries}}]{White2007}
{White}, R.~L., {Helfand}, D.~J., {Becker}, R.~H., {Glikman}, E., \& {de
  Vries}, W. 2007, \apj, 654, 99

\bibitem[{{York} {et~al.}(2000){York}, {Adelman}, {Anderson}, {Anderson},
  {Annis}, {Bahcall}, {Bakken}, {Barkhouser}, {Bastian}, {Berman}, {Boroski},
  {Bracker}, {Briegel}, {Briggs}, {Brinkmann}, {Brunner}, {Burles}, {Carey},
  {Carr}, {Castander}, {Chen}, {Colestock}, {Connolly}, {Crocker}, {Csabai},
  {Czarapata}, {Davis}, {Doi}, {Dombeck}, {Eisenstein}, {Ellman}, {Elms},
  {Evans}, {Fan}, {Federwitz}, {Fiscelli}, {Friedman}, {Frieman}, {Fukugita},
  {Gillespie}, {Gunn}, {Gurbani}, {de Haas}, {Haldeman}, {Harris}, {Hayes},
  {Heckman}, {Hennessy}, {Hindsley}, {Holm}, {Holmgren}, {Huang}, {Hull},
  {Husby}, {Ichikawa}, {Ichikawa}, {Ivezi{\'c}}, {Kent}, {Kim}, {Kinney},
  {Klaene}, {Kleinman}, {Kleinman}, {Knapp}, {Korienek}, {Kron}, {Kunszt},
  {Lamb}, {Lee}, {Leger}, {Limmongkol}, {Lindenmeyer}, {Long}, {Loomis},
  {Loveday}, {Lucinio}, {Lupton}, {MacKinnon}, {Mannery}, {Mantsch}, {Margon},
  {McGehee}, {McKay}, {Meiksin}, {Merelli}, {Monet}, {Munn}, {Narayanan},
  {Nash}, {Neilsen}, {Neswold}, {Newberg}, {Nichol}, {Nicinski}, {Nonino},
  {Okada}, {Okamura}, {Ostriker}, {Owen}, {Pauls}, {Peoples}, {Peterson},
  {Petravick}, {Pier}, {Pope}, {Pordes}, {Prosapio}, {Rechenmacher}, {Quinn},
  {Richards}, {Richmond}, {Rivetta}, {Rockosi}, {Ruthmansdorfer}, {Sandford},
  {Schlegel}, {Schneider}, {Sekiguchi}, {Sergey}, {Shimasaku}, {Siegmund},
  {Smee}, {Smith}, {Snedden}, {Stone}, {Stoughton}, {Strauss}, {Stubbs},
  {SubbaRao}, {Szalay}, {Szapudi}, {Szokoly}, {Thakar}, {Tremonti}, {Tucker},
  {Uomoto}, {Vanden Berk}, {Vogeley}, {Waddell}, {Wang}, {Watanabe},
  {Weinberg}, {Yanny}, {Yasuda}, \& {SDSS Collaboration}}]{York2000}
{York}, D.~G., {Adelman}, J., {Anderson}, Jr., J.~E., {et~al.} 2000, \aj, 120,
  1579

\bibitem[{{York} {et~al.}(2006){York}, {Khare}, {Vanden Berk}, {Kulkarni},
  {Crotts}, {Lauroesch}, {Richards}, {Schneider}, {Welty}, {Alsayyad}, {Kumar},
  {Lundgren}, {Shanidze}, {Smith}, {Vanlandingham}, {Baugher}, {Hall},
  {Jenkins}, {Menard}, {Rao}, {Tumlinson}, {Turnshek}, {Yip}, \&
  {Brinkmann}}]{York2006}
{York}, D.~G., {Khare}, P., {Vanden Berk}, D., {et~al.} 2006, \mnras, 367, 945

\bibitem[{{Zafar} {et~al.}(2012){Zafar}, {Watson}, {El{\'{\i}}asd{\'o}ttir},
  {Fynbo}, {Kr{\"u}hler}, {Schady}, {Leloudas}, {Jakobsson}, {Th{\"o}ne},
  {Perley}, {Morgan}, {Bloom}, \& {Greiner}}]{Zafar2012}
{Zafar}, T., {Watson}, D., {El{\'{\i}}asd{\'o}ttir}, {\'A}., {et~al.} 2012,
  \apj, 753, 82

\bibitem[{{Zhou} {et~al.}(2010){Zhou}, {Ge}, {Lu}, {Wang}, {Yuan}, {Jiang}, \&
  {Shan}}]{Zhou2010}
{Zhou}, H., {Ge}, J., {Lu}, H., {et~al.} 2010, \apj, 708, 742

\end{thebibliography}
